\documentclass[]{interact}

\usepackage{subfigure}% Support for small, `sub' figures and tables

\usepackage{natbib}% Citation support using natbib.sty
\bibpunct[, ]{(}{)}{;}{a}{}{,}% Citation support using natbib.sty
\theoremstyle{plain}% Theorem-like structures provided by amsthm.sty
\newtheorem{theorem}{Theorem}[section]

\newtheorem{corollary}[theorem]{Corollary}
\newtheorem{proposition}[theorem]{Proposition}

\theoremstyle{definition}
\newtheorem{definition}[theorem]{Definition}

\newtheorem{assumption}[theorem]{Assumption} % Custom

\theoremstyle{remark}
\newtheorem{remark}{Remark}

\newcommand\Alt{j}
\newcommand\Altt{{j'}}
\newcommand\Ind{i}

\newcommand\SocUtSym{b}
\newcommand\SocWelfareSym{B}

\newcommand\IndUtSym{V}
\newcommand\IndUtAfterSym{U}
\newcommand\IncSym{y} % incentive
\newcommand\IncSymMaj{Y}
\newcommand\TrSym{z} % transfer
\newcommand\TotIncSym{Y}
\newcommand\WeightSym{w} % incentive

\newcommand\NumOfInd{m}
\newcommand\SplitInd{s}
\newcommand\SplitAlt{t}

\newcommand\DecSym{x}

\newcommand\Budget{Q}
\newcommand\BudgetUsed{\widetilde{\Budget}}

\newcommand\NonDomSetSym{\mathcal{R}}
\newcommand\NumOfNonDomSym{r}
\newcommand\EffSym{e}

\newcommand\Curve{\mathcal{C}}

\newcommand\IndSet{\mathcal{I}}

\newcommand\ProbSym{\mathbb{P}}
\newcommand\ExpSym{\mathbb{E}}

\newcommand\AccProbSym{\pi}

\newcommand\defeq{\equiv}

\newcommand\InitSetSym{\mathcal{N} }

\newcommand\OfInd[1]{_{#1}}
\newcommand\OfIndDef{\OfInd{\Ind}}

\newcommand\OfIndAlt[2]{_{#1,#2}}
\newcommand\OfIndAltDef{\OfIndAlt{\Ind}{\Alt}}

\newcommand\InitSet[1]{\InitSetSym\OfInd{#1}}
\newcommand\InitSetDef{\InitSet{\Ind}}
\newcommand\IndUt[2]{\IndUtSym\OfIndAlt{#1}{#2}}
\newcommand\IndUtDef{\IndUtSym\OfIndAltDef}

\newcommand\IndUtAfterFunc[3]{\IndUtAfterSym\OfIndAlt{#1}{#2}(#3)}
\newcommand\IndUtAfterDef{\IndUtAfterSym\OfIndAltDef}
\newcommand\IndUtAfterFuncDef{\IndUtAfterFunc{\Ind}{\Alt}{\Policy}}
\newcommand\SocUt[2]{\SocUtSym\OfIndAlt{#1}{#2}}
\newcommand\SocUtDef{\SocUtSym\OfIndAltDef}
\newcommand\Inc[2]{\IncSym\OfIndAlt{#1}{#2}}
\newcommand\IncDef{\IncSym\OfIndAltDef}
\newcommand\Tr[2]{\TrSym\OfIndAlt{#1}{#2}}
\newcommand\TrDef{\TrSym\OfIndAltDef}
\newcommand\IncPolicy{\mathbf{\IncSym}}
\newcommand\IncPolicySet{\mathcal{\IncSymMaj}}
\newcommand\Policy{\mathbf{\TrSym}}
\newcommand\Weight[2]{\WeightSym\OfIndAlt{#1}{#2}}

\newcommand\WeightDef{\WeightSym\OfIndAltDef}

\newcommand\AltPair[2]{[#1,#2]}
\newcommand\AltPairDef{\AltPair{\Ind}{\Alt}}
\newcommand\DefAlt[1]{\AltEq{#1}{0}} % default alternative
\newcommand\DefAltDef{\DefAlt{\Ind}}

\newcommand\Dec[2]{\DecSym\OfIndAlt{#1}{#2}} % decision variable
\newcommand\DecDef{\Dec{\Ind}{\Alt}}
\newcommand\MaxSocUt[1]{\SocWelfareSym^*(#1)}
\newcommand\MaxSocUtDef{\MaxSocUt{\Budget}}
\newcommand\MaxSocUtDeff{\MaxSocUt{\TotIncSym}}
\newcommand\AlgSocUt[1]{\SocWelfareSym(#1)}
\newcommand\AlgSocUtDef{\AlgSocUt{\TotIncSym}}
\newcommand\SocWelfare[1]{\SocWelfareSym(#1)}

\newcommand\NonDomSet[1]{\NonDomSetSym\OfInd{#1}}
\newcommand\NonDomSetDef{\NonDomSet{\Ind}}
\newcommand\NumOfNonDom[1]{\NumOfNonDomSym\OfInd{#1}}
\newcommand\NumOfNonDomDef{\NumOfNonDom{\Ind}}
\newcommand\IncrSocUt[2]{\tilde{\SocUtSym}\OfIndAlt{#1}{#2}}
\newcommand\IncrSocUtDef{\IncrSocUt{\Ind}{\Alt}}
\newcommand\IncrWeight[2]{\tilde{\WeightSym}\OfIndAlt{#1}{#2}}
\newcommand\IncrWeightDef{\IncrWeight{\Ind}{\Alt}}
\newcommand\IncrEffSym{\tilde{\EffSym}}
\newcommand\Eff[2]{\EffSym\OfIndAlt{#1}{#2}}
\newcommand\EffDef{\Eff{\Ind}{\Alt} }
\newcommand\IncrEff[2]{\IncrEffSym\OfIndAlt{#1}{#2}}
\newcommand\IncrEffDef{\IncrEff{\Ind}{\Alt}}
\newcommand\TotInc{\TotIncSym}

\newcommand\TotNonDomSet{\NonDomSetSym}
\newcommand\TotSocUt{{\SocWelfareSym}}

\newcommand\NumOfAlt[1]{|\InitSet{#1}|}
\newcommand\NumOfAltDef{\NumOfAlt{\Ind}}
\newcommand\NumOfConvHull[1]{|\NonDomSet{#1}|}
\newcommand\NumOfConvHullDef{\NumOfConvHull{\Ind}}
\newcommand\TotNumOfConvHull{|\TotNonDomSet|}

\newcommand\Tax{\tau}
\newcommand\Rev{R}
\newcommand\VarUtSym{\Delta U}
\newcommand\VarUt[1]{\VarUtSym(#1)}

\newcommand\ExpensesSym{Y}
\newcommand\Exp[1]{\ExpensesSym(#1)}

\newcommand\DisutSym{\delta}
\newcommand\Disut[1]{\DisutSym(#1)}
\newcommand\DisutDef{\Disut{\Policy}}
\newcommand\ThrSym{A} % tax threshold level
\newcommand\RandomSym{\epsilon}
\newcommand\Random[2]{\RandomSym\OfIndAlt{#1}{#2}} % random term
\newcommand\RandomDef{\Random{\Ind}{\Alt}}
\newcommand\AltEq[2]{\Alt^{*}_{#1}(#2)} % equilibrium alternative
\newcommand\AltEqDef{\AltEq{\Ind}{\Policy}}
\newcommand\IndUtDet[2]{\hat{\IndUtSym}\OfIndAlt{#1}{#2}}
\newcommand\IndUtDetDef{\IndUtDet{\Ind}{\Alt}}
\newcommand\IncDet[2]{\hat{\IncSym}\OfIndAlt{#1}{#2}}
\newcommand\IncDetDef{\IncDet{\Ind}{\Alt}}

\newcommand\AccProb[3]{\AccProbSym_{#1, #2}(#3)}
\newcommand\AccProbDef{\AccProb{\Ind}{\Alt}{\IncDef}}

\newcommand\IterSym{k}
\newcommand\OfIter[1]{^{[#1]}}
\newcommand\OfIterDef{\OfIter{\IterSym}}

\newcommand\Thr{\ThrSym\OfIndDef}
\newcommand\vGroup[2]{\underset{#2}{\underbrace{#1} } }

\newcommand{\cotwo}{CO\textsubscript{2}}
\usepackage[ruled,linesnumbered]{algorithm2e}
\usepackage{siunitx}
\usepackage[normalem]{ulem}

\usepackage{hyperref}

\usepackage[colorinlistoftodos, textsize=scriptsize]{todonotes} % todo

\DeclareMathOperator*{\argmax}{arg\,max}

\newcommand\fakesubsec[1]{\underline{\textbf{#1} } }

\begin{document}

% \articletype{ARTICLE TEMPLATE}

\title{Personalized Incentives with Constrained Regulator's Budget}

\author{
\name{Lucas Javaudin\textsuperscript{a}\thanks{CONTACT Lucas Javaudin. Email: lucas.javaudin@cyu.fr}, Andrea Araldo\textsuperscript{b} and André de Palma\textsuperscript{a}}
\affil{\textsuperscript{a}THEMA, CY Cergy Paris Université; \textsuperscript{b}Telecom SudParis, Institut Polytechnique de Paris}
}

\maketitle

\begin{abstract}
    We consider a regulator driving individual choices towards increasing social welfare by providing personal incentives.
    We formalize and solve this problem by maximizing social welfare under a budget constraint.
    The personalized incentives depend on the alternatives available to each individual and on her preferences. 
    A polynomial time approximation algorithm computes a policy within few seconds.
    We analytically prove that it is boundedly close to the optimum.
    We efficiently calculate the Maximum Social Welfare Curve to achieve for a range of incentive budgets.
    This curve provides the right incentive budget to invest.
    We extend our formulation to enforcement, taxation and non-personalized-incentive policies.
    We analytically show that our personalized-incentive policy is also optimal within this class of policies and construct close-to-optimal enforcement and proportional tax-subsidy policies.
    We then compare analytically and numerically our policy with other state-of-the-art policies. Finally, we simulate a large-scale application to mode choice to reduce \cotwo{} emissions.
\end{abstract}

\begin{keywords}
    Personalized incentives; Knapsack problem; Tax policy; \cotwo{} emissions; Modal shift
\end{keywords}

\begin{jelcode}
    C61, H2, Q58, R41
\end{jelcode}

\textbf{\textcolor{red}{This is an original manuscript of an article published by Taylor \& Francis in Transportmetrica A: Transport Science on 23 Nov 2023, available at:} \url{https://doi.org/10.1080/23249935.2023.2284353}.}

\section{Introduction}

%We consider a population of individuals, each of them facing a specific choice set.
%The individuals choose their preferred alternative which, does not correspond necessarily to the socially optimal choice.
%In this section, we present the mathematical formulation of the problem used throughout this paper.
%The framework that we propose could potentially be adapted to many real world situations in transportation area, in marketing and in several other fields, in particular, where externalities are prevalent.

Taxes and subsidies in transportation are often perceived by the population as unfair, since they neglect the alternatives actually available to each individual and the individual preferences.%
\footnote{A preliminary version of this work was presented at the 2022 IEEE 25th International Conference on Intelligent Transportation Systems (ITSC), and a conference paper appears in the proceedings of that conference \citep{javaudin2022large}.}
On the other hand, with the increase in information available to governments \citep{clarke2014governments}, economic policies can be improved to consider the peculiarities of each individual.
Customized policies could be used to align the individual cost with the social cost in the individuals' decisions, without penalizing anyone.
We do not discuss here the legal dimension of such policies (which should be debated in the political arena).

We propose a policy of personalized incentives in a framework where individuals choose between multiple alternatives (or options).
A benevolent regulator has a limited budget that he can use to propose monetary incentives, with the goal to induce individuals to change their choice toward socially-better ones.
The policy we present is fair in the sense that no individual increases or decreases her utility.
This is a clear advantage over road pricing, the most deployed demand management scheme, which usually decreases the utility of some individuals.

Consider a regulator aiming to induce car buyers to choose more environmentally-friendly car models.
Suppose that two buyers, A and B, both consider buying a car with high \cotwo{} emissions and suppose that buyer A (resp.\ buyer B) can be convinced to buy instead a car with low \cotwo{} emissions if she gets a discount of \SI{2000}[\$]{} (resp.\ of \SI{5000}[\$]{}).
With the policy of personalized incentives envisaged in this paper, the regulator could give \SI{2000}[\$]{} (resp.\ \SI{5000}[\$]{}) to buyer A (resp.\ to buyer B) if she accepts to buy the low-emission car.
%This policy is similar to the bonus-malus policy in France, where car buyers receive up to \SI{7000}[\$]{} when buying an electric car.
%However, the bonus-malus policy is not personalized and thus, buyers receive \SI{7000}[\$]{} even if only \SI{2000}[\$]{} was needed to convince them to buy a more environmentally friendly car (or if they would have bought it even without subvention).
In this simple example, the regulator could convince the two buyers to choose the low-emission car for only \SI{7000}[\$]{} (\SI{2000}[\$]{} to buyer A and \SI{5000}[\$]{} to buyer B), while, with a non-personalized subsidy policy, the regulator would need to spend at least \SI{10000}[\$]{} to convince both buyers (each buyer receives \SI{5000}[\$]{}).
Hence, a personalized-incentive policy allows to reduce the average \cotwo{} emissions by the same amount than a non-personalized policy, while spending less.

We define the optimal personalized-incentive policy as the allocation of incentives that maximizes social welfare (defined as the reduction of \cotwo{} emissions in the example above), for a given budget.
With two cars and two buyers, the optimal policy is easy to compute by simple enumeration.
However, the problem is combinatorial so, with a large number of heterogeneous buyers and a large number of car models to choose from, we need more sophisticated methods.

We formalize the problem of finding a personalized-incentive policy maximizing social welfare under the regulator's budget constraint and show that it reduces to the well-known Multiple-Choice Knapsack Problem (MCKP -- see Section~\ref{sec:policy}).
The MCKP consists in packing `items' of different `classes' into a knapsack of a certain `capacity'.
We show that the MCKP provides a natural formalization of an optimal incentive policy, where a class is an individual, an item is an alternative and the knapsack capacity is the budget of incentives.
To approximate the optimal policy in polynomial time, we adapt a greedy algorithm from the Operations Research literature and we analyse some of its analytical (e.g., suboptimality gap bound) and economic (e.g., diminishing returns) properties (Section~\ref{sec:algorithm}).

We then frame personalized-incentive policies into a larger family of demand management policies, including enforcement, tax and non-personalized-incentives (Section~\ref{sec:policies_comparison}).
These policies aim to maximize social welfare subject to a disutility constraint, where the disutility is the total loss of surplus for both the regulator and the individuals.
We find that personalized-incentive policies are optimal within this larger family of policies.
Moreover, they are `fair', since they guarantee that the utility of each individual remains unchanged, and thus no one is penalized.
We also compute a theoretical lower bound on the gap between state-of-the-art incentive policies, which are not personalized, and our personalized-incentive policy.
 Furthermore, we show that our greedy algorithm can not only construct incentive policies, but also enforcement and proportional tax-subsidy policies. We show that also in this case, the social welfare is boundedly close to the theoretical optimum.
 While in most of the paper we assume that the regulator knows exactly the preferences of each individual, we also study the case of imperfect information (Section~\ref{sec:imperfect-information}).

Using data from the French census, we evaluate the \cotwo{} reduction achieved via the policy computed with our algorithm in a large-scale use-case, where individuals are incentivized to shift toward greener transportation modes for their commute to work, at the scale of a French department (Section~\ref{sec:case_study}).
The results confirm the theoretical findings, showing in particular that our personalized incentives achieve the same \cotwo{} reduction as flat subsidies, but with a considerably smaller amount of incentives spent.
Our code is available as open source.\footnote{\url{https://github.com/LucasJavaudin/individualized-incentives-algorithm}}

Even though the case study is about modal shift, the proposed methods can be applied in various contexts.
For example, consider the marketing department of a large firm selling mutually exclusive goods.
To increase the profits of the firm, the marketing department could use its budget to propose price discounts to some consumers in order to convince them to shift to goods with higher margins.
Another potential example of application is in the telecommunications management context.
In recent years, governments are planning to subsidize local organizations to improve the access of rural population to the Internet (France alone will spend 3 billions euros in 10 years, \citealp{THD2021}).
With our methods, governments could allocate optimally these subsidies.

%The paper is organized as follows.
%We discuss the related work in Section~\ref{sec:related-work}.
%We formalize the problem of finding the optimal personalized-incentive policy and Maximum Social Welfare Curve in Section~\ref{sec:policy}.
%Section~\ref{sec:algorithm} presents a polynomial time algorithm to approximate the optimal incentive policy and show its analytical properties.
%Section~\ref{sec:policies_comparison} shows that the algorithm can also be used to approximate the optimal enforcement and tax-subsidy policy.
%Section~\ref{sec:imperfect-information} extends the model to the case where the regulator has imperfect information on the individuals' preferences.
%Section~\ref{sec:case_study} proposes an application of the algorithm to mode choice in France.
%Section~\ref{sec:conclusion} concludes the paper.

\section{Related Work}%
\label{sec:related-work}

We first discuss the literature on incentive policies (Section~\ref{sec:incentives}), in particular in transportation, which is the main application domain we envision.
We then review the applications of the Multiple Choice Knapsack Problem, on which our optimization is based, in economics and transportation (Section~\ref{sec:mckp-transportation}) and, finally, in other domains (Section~\ref{sec:mckp-other}).

\subsection{Incentive Policies in Transportation}
\label{sec:incentives}

Earlier studies of welfare analysis in a discrete-choice framework have been conducted by \citet{Small1981} and \citet{Anderson1992}.
\citet{DeBorger2001} studies the optimal taxation in a discrete-choice framework with externalities.
His model is close to ours but he does not consider incentive policies.
Some papers conduct an empirical study of an incentive policy in the transportation context (e.g., \citealp{Merugu2009}, \citealp{Ettema2010}, \citealp{Yue2015}, \citealp{Hu2015a}) but they do not carry out a theoretical study of the optimal policy.

Incentives are a promising tool for policy makers to trigger a transition toward greener transportation.
\citet{Mirhedayatian2018} model the reaction of a single logistics company to several incentives for buying and adopting electric vehicles.
We are interested instead in calculating optimal incentives for a large plethora of individuals. 
A vast literature exists on time-varying incentives and/or surcharges to shift departure times, in order to reduce congestion.
To this aim, \citet{Sun2020} adopt a bottleneck model of a road segment.
\citet{HaiYang2020} propose an optimization model to calculate transit surcharges and incentives, during peak and off-peak, respectively, to avoid over-crowding.
In the two aforementioned works, the incentives are not personalized, in that they do not depend on the individual's profile and all the individuals go from the same origin to the same destination.
We instead consider a large set of individuals, each with a different set of alternatives, resulting in different contributions to the social welfare and individual perceived utility.
Our incentives are personalized, in that we encourage social-welfare maximizing alternatives with an incentive that compensate for the reduction in individual utility loss, which changes from an individual to another.

Closer to our work, \citet{Araldo2019} devise Tripod, a simulation-based optimization method to calculate incentives to encourage energy efficient transportation alternatives.
However, the incentives do not depend on individual specificities.
Indeed, the system computes a unique `Token Energy Efficiency' (TEE) value, and computes the incentive for each alternative by simply multiplying the TEE by the estimated energy savings achieved with that alternative.
Such approach is pertinent when the regulator has no information on the individual preferences. However, when perfect information is available, our approach is able to achieve the same social welfare of Tripod with less incentives spent or, equivalently, to achieve a larger social welfare with the same incentives spent. We show these findings both analytically (Section~\ref{sec:comparison-proportional-tripod}) and numerically (Section~\ref{sec:comparison-with-other-policies}). 

\subsection{Multiple Choice Knapsack Problem in Economics and Transportation}
\label{sec:mckp-transportation}
The Multiple Choice Knapsack Problem (MCKP -- \citealp[chap.~11]{Kellerer2004}) can be used to model a decision maker willing to optimally invest a limited budget in order to increase an objective function.
The possible investments options are divided into separate groups, and the decision maker has to choose at most one option for each group.

We now review the few examples of applications of MCKP in Economics and Transportation. 
\citet{Zhong2010} study the decision of a transportation planner willing to select a subset of candidate projects for funding.
They do not propose any resolution algorithm and solve the problem in an exact way using a solver.
Later, \citet{Colorni2017} use MCKP as a subroutine of a more general multi-criteria project-selection problem.
Since the problem is NP hard \citep[chap.~11]{Kellerer2004}, the aforementioned exact approach would require an unfeasibly large computation time in the large-scale applications we target.
For this reason, we resort instead to a polynomial time approximation algorithm.
\citet[Sec.~6 and 7]{Zoltners1979} use MCKP as a subroutine for a problem where a sales representative with a finite time-budget has to optimally allocate a call frequency to each accounts.
They solve such a subroutine with an algorithm similar in spirit to our Algorithm~\ref{alg:greedy-max-soc-ut-curve}, but in a more complicated setting, due to iterating decisions over multiple time-slots.

\subsection{Multiple Choice Knapsack Problem in Computer Science and other Domains}
\label{sec:mckp-other}

MKCP is widely adopted in the Computer Science community, where a certain resource must be allocated among different entities.
In the work of \citet{Varshney2015}, a central information aggregator receives information from several selfish sensors, which can transmit it at several precision levels: the more the precision level the higher the sensor cost in terms of energy.
The aggregator needs to select one precision level (or none) per sensor and compensates the corresponding loss of energy of each sensor via payments.
In \citet{Fielder2016}, a manager of an information system invests in security controls.
Per each control, it has to select a certain `level': the higher the level, the higher the protection of the organization, but also the higher the investment.
\citet{AraldoSAC2020} allocate computational resources among different service providers; the owner of the resources selects one configuration for each of them in order to increase the overall system utility.
To this aim, they use Multi-Dimensional Multiple-Node MCKP.
 \citet{Geunes2018} solve the problem of a seller, who needs to decide the amount of product to buy from several providers, each proposing a different pricing scheme, in order to maximize its overall utility.

\subsection{Position with respect to the Related Work}
To the best of our knowledge, we are the first to formalize the problem of computing optimal personalized-incentives with MCKP.
By finding the assumptions that enable such a formalization, we show in this paper that MCKP describes naturally such a problem, since it manages to represent the different alternatives of each individual.
The adoption of MCKP also allows us to devise an efficient algorithm for large-scale applications, adapting existing solutions from Operations Research.

%%%%%%%%%%%%%%%%%%%%%%%%%%%%%%%%%%%%%%%%%%%%%%%%%%%%%%%%%%%%%%%%
%%%%%%%%%%%%%%%%%%%%%%%%% FRAMEWORK AND INCENTIVE POLICY %%%%%%%
%%%%%%%%%%%%%%%%%%%%%%%%%%%%%%%%%%%%%%%%%%%%%%%%%%%%%%%%%%%%%%%%

\section{Framework and Personalized-Incentive Policy}
\label{sec:policy}

In this section, we first formalize the model studied and present the underlying assumptions (Section~\ref{sec:model-and-assumptions}).
We also characterize the personalized-incentive policy that will be studied throughout this paper (Section~\ref{sec:personalized-incentive-policy}).
We then present the Maximum Social Welfare Problem, which consists in finding the optimal incentive policy under a budget constraint (Section~\ref{sec:maximum-social-welfare-problem}).
Finally, we present the Maximum Social Welfare Curve problem, which solves the previous problem for a range of budget values (Section~\ref{sec:maximum-social-welfare-curve-problem}).

The notations used throughout this paper are summarized in Table~\ref{tab:notation}.
All the proofs are relegated to Appendix \ref{sec:proofs}.

\begin{table}
    \centering
    \begin{footnotesize}
        \caption{Notation used throughout the paper.}%
        \label{tab:notation}
        \begin{tabular}{cl}
            \toprule
            $\Ind$ & Index to denote an individual
            \\
            $\Alt$ & Index to denote an alternative
            \\
            $\AltPairDef$ & Alternative $\Alt$ of individual $\Ind$
            \\
            $\IndSet$, $\NumOfInd$ & Set of individuals and number of individuals \\
            $\InitSetDef$ & Set of alternatives available to individual $\Ind$ \\
            $\IndUtDef$ & Intrinsic utility of individual $\Ind$ when choosing alternative $\Alt$ \\
            $\SocUtDef$ & Social indicator of alternative $\Alt$ of individual $\Ind$ 
            \\
            $\TrDef$ & Monetary transfer received (or paid) by individual $\Ind$ when choosing alternative $\Alt$ \\
            $\Policy$ & General policy (set $\{\TrDef\}\OfIndAltDef$ of monetary transfers) \\
            $\IncDef$ & Personalized incentive proposed to individual $\Ind$, conditional on choosing alternative $\Alt$ \\
            $\IncPolicy$ & Personalized-incentive policy (set $\{\IncDef\}\OfIndAltDef$ of incentives) \\
            $\IncPolicySet$ & Set of all personalized-incentive policies \\
            $\IndUtAfterFuncDef$ & Utility of individual $\Ind$ when choosing alternative $\Alt$, given policy $\Policy$~\eqref{eq:utility_def} \\
            $\AltEqDef$ & Alternative chosen by individual $\Ind$, given policy $\Policy$ \\
            $\DefAltDef$ & Default alternative, i.e., alternative chosen by individual $\Ind$, in the absence of policy~\eqref{eq:default_alt}\\
            $\WeightDef$ & Weight of alternative $\Alt$ of individual $\Ind$, equation \eqref{eq:weight_def}
            \\
            $\EffDef$ & Efficiency of alternative $\Alt$ of individual $\Ind$ (Definition~\ref{def:efficiency})
            \\
            $\Budget$ & Maximum budget available to the regulator \\
            $\BudgetUsed$ & Budget actually spent by the policy computed by Algorithm~\ref{alg:greedy-max-soc-ut-curve} \\
            $\MaxSocUtDeff$ & Maximum social welfare reachable with a personalized-incentive policy,\\
                            & with a total incentive expenditure of $\TotIncSym$ \\
            $\AlgSocUtDef$ & Social welfare obtained with the personalized-incentive policy produced by our algorithm, \\
                           & with a total incentive expenditure of $\TotIncSym$ \\
            $\AlgSocUt{\Policy}$ & Social welfare achieved with a policy $\Policy$, equation \eqref{eq:policy-social-welfare}
            \\
            $\Exp{\Policy}$ & Expenses of the regulator for a policy $\Policy$, equation \eqref{eq:expenses}
            \\
            $\VarUt{\Policy}$ & Total variation in individual utility of a policy $\Policy$, equation \eqref{eq:utility_loss}
            \\
            $\Disut{\Policy}$ & Disutility of policy $\Policy$, equation~\eqref{eq:disutility}
            \\
            $\NonDomSetDef\subseteq\InitSetDef$, $\NumOfNonDomDef=|\NonDomSetDef|$ & Set of LP-extremes alternatives of individual $\Ind$, and its cardinality \\
            $\IncrSocUtDef$ & Incremental social indicator of the alternative $\Alt$ of individual $\Ind$ \\
            $\IncrWeightDef$ & Incremental incentive of the alternative $\Alt$ of individual $\Ind$ \\
            $\IncrEffDef$ & Incremental efficiency of the alternative $\Alt$ of individual $\Ind$ 
            \\
            $\IncrEff{\SplitInd}{\SplitAlt}$ & Incremental efficiency of the split item (Algorithm~\ref{alg:greedy-max-soc-ut-curve})
            \\
            $\IterSym$ & Iteration index of Algorithm~\ref{alg:greedy-max-soc-ut-curve}
            \\
            $\EffSym\OfIterDef,\IncrEffSym\OfIterDef$ & Overall and incremental efficiency of Algorithm~\ref{alg:greedy-max-soc-ut-curve} at iteration $\IterSym$ (Definition~\ref{def:marginal-and-overall-efficiency}) 
            \\
            $\Tax$ & Tax level (Section~\ref{sec:finding-the-optimal-tax-subsidy-policy})
            \\
            $\Thr$ & Baseline social-indicator of individual $\Ind$ (Section~\ref{sec:finding-the-optimal-tax-subsidy-policy})
            \\
            \bottomrule
        \end{tabular}
    \end{footnotesize}
\end{table}

%In this section, we introduce a simple discrete-choice framework and define our incentive policy to maximize social utility under a budget constraint. 
%We then express the optimization problem that we want to solve.
%Furthermore, we will show analytically that the social utility obtained from the algorithm is boundedly close to the theoretical maximum.}

\subsection{Model and Assumptions}
\label{sec:model-and-assumptions}

We consider a population $\IndSet \equiv \{1, \dots, \NumOfInd\}$ of $\NumOfInd$ individuals.
Each individual $\Ind\in\IndSet$ chooses an alternative $\Alt$ among an \emph{individual-specific choice-set} $\InitSetDef$.
For example, we can consider individuals choosing a mode of transportation to commute to their work.
In this case, the choice set could be $\InitSetDef=\{\text{car}, \text{walk}, \text{bike}, \text{public transit}\}$.
The choice set can be individual-specific so that if individual $\Ind$ owns a car but individual $\Ind'$ does not, we could have $\InitSetDef=\{\text{car}, \text{walk}, \text{bike}, \text{public transit}\}$ and $\InitSet{\Ind'}=\{\text{walk}, \text{bike}, \text{public transit}\}$.
The mode-choice example is studied extensively in Section \ref{sec:case_study}.
As another example, $\IndSet$ could be a set of individuals purchasing a car. In this case, the set of alternatives $\InitSetDef$ of individual $\Ind$ would include the models of cars available in the market.

Let $\TrDef\in\mathbb{R}$ be a monetary transfer, from the regulator to individual~$\Ind$, induced when she chooses alternative~$\Alt$.
This monetary transfer can be an incentive, if positive, or a tax, if negative.
Any \emph{policy} can thus be described by a set of monetary transfers proposed to all the individuals for any of their alternatives, which we compactly denote with $\Policy \equiv \{\TrDef\}\OfIndAltDef$.

A policy influences the individual choice since the proposed monetary transfers change her utilities.

The \emph{utility} $\IndUtAfterFuncDef$ of individual $\Ind$ when choosing alternative $\Alt\in\InitSetDef$ is given by
\begin{equation}\label{eq:utility_def}
    \IndUtAfterFuncDef = \IndUtDef + \TrDef,
\end{equation}
where $\IndUtDef\in\mathbb{R}$ is the intrinsic utility (in the absence of policy).
We implicitly assumed that utility is quasi-linear with respect to income, which means that both $\IndUtDef$ and $\TrDef$ are expressed in the same unit as income and that $\TrDef$ has an additive effect on utility, hence equation \eqref{eq:utility_def}.
%\todo[inline]{aa: Why is the observation that the utility is quasi-linear with respect to the income is useful for our paper? Please, try to comment more on this observation.\\
%A note for myself: in previous discussions, I was pinpointing that the choices of individuals depend on their income (as the income influences $\IndUtDef$). The implications of quasi-linearity are in Sec. 10.3 of~\cite{Varian1992}.}

Given a policy $\Policy$, each individual $\Ind$ chooses an alternative $\AltEqDef$ which maximizes her utility:
\begin{equation*}
    %\label{eq:max-ind-ut}
    \AltEqDef \in \argmax_{\Alt} \IndUtAfterFuncDef.
\end{equation*}

We consider a regulator aiming to maximize a social welfare indicator, whose value depends on the individuals' choices.
More formally, each alternative $\Alt$ of individual $\Ind$ is characterized by a \emph{social indicator} $\SocUtDef\in\mathbb{R}$.
In the mode-choice example, the social indicator could be the opposite of \cotwo{} emissions induced by the commutes.

The goal of the regulator is to find a policy $\Policy$ which maximizes the global social indicator, or \emph{social welfare} indicator, defined by
\begin{equation}
	\label{eq:policy-social-welfare}
    \AlgSocUt{\Policy} \defeq \sum_{\Ind=1}^\NumOfInd \SocUt{\Ind}{\AltEqDef},
\end{equation}
i.e., the sum of the social indicators of the alternatives chosen by the individuals.
Intuitively, a policy $\Policy$ which maximizes welfare could be
\begin{equation*}
    \TrDef =
    \left\{
        \begin{array}{ll}
             0 & \text{if}\:\Alt\in\argmax_\Altt \SocUt{\Ind}{\Altt} \\
            -\infty & \text{else}
        \end{array}
    \right.
    ,\quad\forall \Ind,\Alt,
\end{equation*}
which is equivalent to a ban of all alternatives that do not maximize the social indicator for each individual.
However, in practice, the regulator is affected by some political constraints and such extreme policy is not feasible.

\begin{definition}[Expenses]
    For any policy $\Policy$, we define the expenses $\Exp{\Policy}$ of the regulator (or his revenues $-\Exp{\Policy}$) as
    \begin{equation}
        \label{eq:expenses}
        \Exp{\Policy} \defeq \sum^{\NumOfInd}_{\Ind=1} \Tr{\Ind}{\AltEq{\Ind}{\Policy}},
    \end{equation}
    i.e., the sum of the monetary transfers paid or received for the alternative $\AltEq{\Ind}{\Policy}$ chosen by each individual $\Ind$.
\end{definition}

The following assumptions are made.
First, we assume that individuals cannot affect each other's intrinsic utility.
\begin{assumption}[Independent intrinsic utilities]
    \label{ass:independant_indiv}
    Given a policy $\Policy$, for each individual $\Ind$ and each alternative $\Alt\in\InitSetDef$, the intrinsic utility $\IndUtDef$ is independent of the alternative $\AltEq{\Ind'}{\Policy}$ chosen by any other individual $\Ind'\neq\Ind$.
\end{assumption}
%Note that this assumption does not prevent externalities, in the sense that an individual's choice can still affect the other individuals' utility, as long as it affect all their alternatives uniformly.
%For example, the \cotwo{} emissions of an individual $\Ind$ can impact the utility of another individual $\Ind'$ but the relative difference between the utility of all her alternatives must not vary.
Similarly, we assume that the social indicator of the alternatives is independent of the choices of the individuals.
\begin{assumption}[Independent social indicators]
    \label{ass:independant_social}
    Given a policy $\Policy$, for each individual $\Ind$ and each alternative $\Alt\in\InitSetDef$, the social indicator $\SocUtDef$ is independent of the alternative $\AltEq{\Ind'}{\Policy}$ chosen by any other individual $\Ind'\neq\Ind$.
\end{assumption}
Note that Assumptions \ref{ass:independant_indiv} and \ref{ass:independant_social} hold in many practical situations.
For instance, in the numerical scenario on transport mode choice (Section \ref{sec:case_study}), we achieve relevant social welfare improvement (CO2 reduction), while inducing only few individuals to change their modes, with a negligible impact on the utilities of the other individuals.

%The consequences of these two assumptions depend on the context.
We further assume that the utilities and social indicators are known to the regulator.
\begin{assumption}[Perfect information]
    \label{ass:information}
    The regulator has perfect information: it knows exactly the intrinsic utilities $\{\IndUtDef\}\OfIndAltDef$ and social indicators $\{\SocUtDef\}\OfIndAltDef$ of all the alternatives, for all the individuals.
\end{assumption}
%If the pair $(\IndUtSet, \SocUtSet)$ is imperfectly known to the regulator, he would not be able to compute the optimal policy.

In Sections~\ref{sec:imperfect-information} and~\ref{sec:case_study} we discuss the implications of assumption \ref{ass:information} in the context of mode choice and we show how to relax it by more realistically assuming that the intrinsic utilities $\{\IndUtDef\}\OfIndAltDef$ are imperfectly known to the regulator.
Developing our theoretical framework under Assumption~\ref{ass:information} allows us to develop optimization algorithms that can then be applied, \emph{mutatis mutandis}, also to the realistic case when Assumption~\ref{ass:information} does not hold, as we show in Section~\ref{sec:modal-choice-imperfect-information}.

With no loss of generality, we rule out identical alternatives.
\begin{assumption}[No identical alternatives]
    \label{ass:no-identical-alternatives}
    We assume that, for any individual $\Ind$, there are no identical alternatives $\Alt,\Altt\in\InitSetDef$, i.e., such that $\IndUtDef=\IndUt{\Ind}{\Altt}$ and $\SocUtDef=\SocUt{\Ind}{\Altt}$.
\end{assumption}

We need to characterize more precisely the behaviour of individuals when multiple alternatives maximize their utility.
\begin{assumption}[Tie-breaking rule]
\label{ass:behavior}
For any policy $\Policy$, if the set $\argmax_\Alt \IndUtAfterFuncDef$ contains more than one alternative, individual $\Ind$ chooses the alternative $\Altt$ with the largest social indicator, i.e.,
\begin{equation} 
\label{eq:behavior}
\AltEqDef = \argmax_{\Altt\in\argmax_j\IndUtAfterFuncDef} \SocUt{\Ind}{\Altt}. 
\end{equation}
%Thanks to Assumption~\ref{ass:no-identical-alternatives}, the set $\underset{\Altt\in\argmax_j\IndUtAfterFuncDef}{ \argmax{} } \SocUt{\Ind}{\Altt}$ is a singleton, and we can write
%\begin{equation*} 
%\AltEqDef = \argmax_{\Altt\in\argmax_j\IndUtAfterFuncDef} \SocUt{\Ind}{\Altt}. 
%\end{equation*}
%\todoi{lj: You said previously that Assumption \ref{ass:no-identical-alternatives} does not guarantee that it is a singleton since we can have some a policy such that there is a tie. I suggest that we do not confuse the reader by either keeping the notation $\in$ (although the proofs assume that it is a singleton, but it is always a singleton with our policy) or the notation $=$ (although it is not correct for some policies).}
\end{assumption}
The previous assumption is merely a technical assumption that could be relaxed by proposing incentives infinitesimally larger to ensure that the set $\argmax_\Alt \IndUtAfterFuncDef$ is always a singleton.

The alternative chosen by each individual $\Ind$ in the absence of policy (i.e., where $\TrDef=0$, $\forall\Ind, \Alt$) is called \emph{default alternative}, and denoted $\DefAltDef$.
Under Assumption~\ref{ass:behavior}, the default alternative is given by
\begin{equation}
    \label{eq:default_alt}
    \DefAltDef \defeq \argmax_{\Altt\in\argmax_\Alt\IndUtDef} \SocUt{\Ind}{\Altt}.
\end{equation}

\subsection{Personalized-Incentive Policies}
\label{sec:personalized-incentive-policy}

We assume that the space of policies available to the regulator is limited to policies such that $\TrDef\geq0$, for each alternative $\Alt$ and individual $\Ind$.
In other words, the regulator never taxes alternatives, for some political reasons.
Note that we allow the regulator to give different monetary transfers to different individuals for the same alternative (e.g., some individuals might receive \SI{2}[\$]{} for commuting by foot, while others may only receive \SI{1}[\$]{}).
Hence, this space of policies is referred to as the set of \emph{personalized-incentive policies}, denoted $\IncPolicySet$.
To distinguish personalized-incentive policies from more general policies, we denote them with $\IncPolicy = \{ \IncDef \}\OfIndAltDef$, where $\IncDef$ is the incentive given to individual $\Ind$, conditional on her choosing alternative $\Alt$, and thus,
\begin{equation*}
    \IncPolicySet = \big\{ \IncPolicy = \{ \IncDef \}\OfIndAltDef\::\:\IncDef \geq 0,\: \forall \Ind, \Alt \big\}.
\end{equation*}

The incentive $\IncDef$ reduces the budget of the regulator only if individual $\Ind$ chooses alternative $\Alt$.
Therefore, if the regulator wants to spend at most a budget $\Budget$, the budget constraint can be written as
\begin{equation*}
    %\label{eq:budget-constraint}
    \TotInc(\IncPolicy) = 
    \sum_{\Ind=1}^\NumOfInd \Inc{\Ind}{\AltEq{\Ind}{\IncPolicy}} \leq \Budget,
\end{equation*}
where $\AltEq{\Ind}{\IncPolicy}$ is the alternative chosen by individual $\Ind$ under the personalized-incentive policy $\AltEq{\Ind}{\IncPolicy}$.

In the rest of this subsection, we characterize more precisely the set of policies we consider, discarding `inefficient' policies.
\begin{proposition}
    \label{propo:incentive_amount}
    The regulator can induce any individual $\Ind\in\IndSet$ to shift from her default alternative $\DefAltDef$ to any alternative $\Alt\in\InitSetDef$, with a higher social indicator (i.e., $\SocUtDef > \SocUt{\Ind}{\DefAltDef}$), by proposing the following incentives
    \begin{equation}
        \label{eq:incentive_amount}
        \begin{array}{rll}
            \IncDef & = \IndUt{\Ind}{\DefAltDef} - \IndUtDef, &\\
            \Inc{\Ind}{\Altt} & =0, & \text{for any other alternative }\Altt\neq\Alt
        \end{array}
    \end{equation} 
    Additionally, $\IncDef$, defined above, is the minimum incentive required to induce individual $\Ind$ to shift to alternative $\Alt$.
\end{proposition}

Such a proposition tells us that it suffices to incentivize only one alternative per individual and no more than that.
Therefore, we can limit the space of the studied personalized-incentive policies as in the following assumption, with no loss of generality.

\begin{assumption}
	\label{ass:types-of-policies}
    We only study in this paper personalized-incentive policies that propose incentives in the form of~\eqref{eq:incentive_amount}, i.e., only one alternative $\Alt$ per individual $\Ind$ is incentivized, with an incentive equal to $\IncDef = \IndUt{\Ind}{\DefAltDef} - \IndUtDef$.
\end{assumption}

%\begin{remark}
    %From the definition of the default alternative $\DefAltDef$, the incentive amount computed in \eqref{eq:incentive_amount} is non-negative, i.e., $\IncDef \geq 0$.
%\end{remark}

\begin{remark} 
    \label{rem:unchanged-utility}
    If an individual $\Ind$ is given an incentive $\IncDef$, given by \eqref{eq:incentive_amount}, to shift to alternative $\Alt$, then her utility $\IndUtAfterFunc{\Ind}{\Alt}{\IncPolicy}$ remains unchanged since, from equation \eqref{eq:utility_def},
    \begin{equation*}
        \IndUtAfterFunc{\Ind}{\Alt}{\IncPolicy} = \IndUtDef + \vGroup{\IndUt{\Ind}{\DefAltDef} - \IndUt{\Ind}{\Alt} }{\IncDef}= \IndUt{\Ind}{\DefAltDef}.
    \end{equation*}
    In other words, the incentive amount is such that the utility of the individual does not change.
    In this sense, our approach is fully equitable.
\end{remark}

%Assumption~\ref{ass:behavior} is a technical assumption needed to characterize the incentives: as we will see below, the optimal incentive policy is such that there is a tie between the utility of the default alternative $\DefAltDef$ and another alternative $\Alt$ such that $\IncDef>0$ and $\SocUtDef>\SocUt{\Ind}{\DefAltDef}$; Assumption~\ref{ass:behavior} ensures that the socially superior alternative $\Alt$ will be chosen.
%Alternatively, we could assume that incentive $\IncDef$ is infinitesimally larger to prevent a tie.

%To develop our formulation, we remove Pareto-dominated alternatives since, given Assumption~\ref{ass:behavior}, they are not relevant for our analysis.
%Pareto dominance is defined as follows.
With no loss of generality, we can remove from any individual choice-set the alternatives that are never chosen, as the ones defined below.

\begin{proposition}[Pareto-dominance]
\label{prop:pareto}
Let us consider individual $\Ind$ facing two alternatives $\Alt,\Altt\in\InitSetDef$. Alternative $\Alt$ is said to be \emph{Pareto-dominated} by $\Altt$ if $\SocUt{\Ind}{\Altt} \geq \SocUtDef$ and $\IndUt{\Ind}{\Altt} > \IndUtDef$.
Alternative $\Alt$ is \emph{Pareto-dominated}, if it is Pareto-dominated by some other alternative.

A personalized-incentive policy $\IncPolicy$ that incentivizes a Pareto-dominated alternative can be discarded, since there always exists another policy that obtains at least the same social welfare, by spending less budget.
\end{proposition}

We thus exclude Pareto-dominated alternatives from the choice-set $\InitSetDef$ of each individual $\Ind$, as they would never be chosen by individuals, under the considered policies.
\begin{assumption}[No Pareto-dominated alternatives]
\label{ass:no-pareto}
For any individual $\Ind$, there are no Pareto-dominated alternatives in her set of alternatives $\InitSetDef$.
\end{assumption}

\subsection{Maximum Social Welfare Problem}
\label{sec:maximum-social-welfare-problem}

We can now formally define the optimization problem of the regulator, who chooses the personalized-incentive policy $\IncPolicy=\{\IncDef\}\OfIndAltDef$ which maximizes social welfare under his budget constraint.
We refer to this problem as \emph{Maximum Social Welfare Problem}:
\begin{equation}
    \label{eq:optimization_problem}
    \left\{
        \begin{array}{cll}
            \displaystyle\max_{\IncPolicy\in\IncPolicySet} & \displaystyle\SocWelfare{\IncPolicy} & \\
            \text{s.t.} & \displaystyle\Exp{\IncPolicy} \leq \Budget & \\
            & \IncDef \ge 0, & \forall \Ind\in\IndSet, \Alt\in\InitSetDef
                        %& \displaystyle\AltEq{\Ind}{\IncPolicy} = \argmax_{k\in\argmax_j \IndUtAfterFunc{\Ind}{\Alt}{\IncPolicy}} \SocUt{\Ind}{k} & \forall \Ind\in\InitSetDef
        \end{array}
    \right..
\end{equation}

\begin{definition}[Optimal personalized-incentive policy]
    \label{def:optimal_incentive_policy}
    An optimal personalized-incentive policy $\IncPolicy$, for a budget $\Budget$, is a solution of Problem~\ref{eq:optimization_problem}.
\end{definition}

We denote with $\{\Alt_1, \dots, \Alt_{\NumOfInd}\}$ a \emph{chosen-alternative set}, where $\Alt_\Ind$ denotes the alternative chosen by individual $\Ind \in \IndSet$.
According to Proposition~\ref{propo:incentive_amount}, the regulator can induce any chosen-alternative set $\{\Alt_1, \dots, \Alt_{\NumOfInd}\}$ by proposing to any individual $\Ind$ the incentives $\Inc{\Ind}{\Alt_{\Ind}}=\IndUt{\Ind}{\DefAltDef} - \IndUt{\Ind}{\Alt_\Ind}$ and $\IncDef=0$, for any $\Alt\neq\Alt_{\Ind}$.
Thanks to the same proposition, the regulator cannot induce this set of alternatives by spending less.
Therefore, the optimization problem of the regulator \eqref{eq:optimization_problem} amounts to finding the chosen-alternative set $\{\Alt_1, \dots, \Alt_{\NumOfInd}\}$ which maximizes social welfare $\sum_{\Ind=1}^{\NumOfInd} \SocUt{\Ind}{\Alt_{\Ind}}$, subject to the constraint that the corresponding spendings $\sum_{\Ind=1}^{\NumOfInd} \Inc{\Ind}{\Alt_{\Ind}}$ must not exceed the budget $\Budget$.

Such a problem can be expressed as an Integer Linear Program (ILP).
In order to do so, we introduce a \emph{weight} $\WeightDef$ for any alternative $\Alt\in\InitSetDef$ of individual $\Ind$.
The weight is defined as the incentive amount that would be proposed to individual $\Ind$ if the regulator were to induce her to choose alternative $\Alt$, which is, according to Proposition~\ref{propo:incentive_amount},
\begin{equation}
    \label{eq:weight_def}
    \WeightDef \defeq \IndUt{\Ind}{\DefAltDef} - \IndUtDef, \quad \forall \Ind, \Alt.
\end{equation}
Note that $\WeightDef$ is a fixed value that we used to compute the optimal policy, while $\IncDef$ represents the incentive amount chosen by the regulator.
The personalized-incentive policy is such that $\IncDef = \WeightDef$, if individual $\Ind$ is induced to choose alternative $\Alt$, and $\IncDef = 0$ otherwise.

We introduce the binary decision variable $\DecDef$ that is equal to $1$ if the regulator wants to make individual $\Ind$ choose alternative $\Alt$, and that is equal to $0$ otherwise, with the natural constraint that $\sum_{\Alt\in\InitSetDef} \DecDef = 1 $ (only one alternative is chosen).
The Maximum Social Welfare problem \eqref{eq:optimization_problem} can be written as
\begin{equation}
\label{eq:max-soc-ut}
\left\{
	\begin{array}{cll}
            \max_{\{\DecDef\}\OfIndAltDef} & \sum_{\Ind\in\IndSet} \sum_{\Alt\in\InitSetDef} \SocUtDef \DecDef & \\
            \text{s.t.} & \sum_{\Ind\in\IndSet} \sum_{\Alt\in\InitSetDef} \WeightDef \DecDef 
            \le \Budget & \\
                        & \sum_{\Alt\in\InitSetDef} \DecDef = 1, & \Ind\in\IndSet \\
                        & \DecDef \in \{0,1\}, &\Ind\in\IndSet,\  \Alt\in\InitSetDef\\
                        & \WeightDef = \IndUt{\Ind}{\DefAltDef} - \IndUtDef, & \Ind\in\IndSet, \ \Alt\in\InitSetDef
	\end{array}
\right. ,
\end{equation}
which is a Multiple-Choice Knapsack Problem (MCKP) with weights $\WeightDef$ and profits $\SocUtDef$ \citep[chap.~11]{Kellerer2004}.

Observe that the solution $\{\DecDef\}\OfIndAltDef$ of problem~\eqref{eq:max-soc-ut} corresponds to the personalized-incentive policy $\IncPolicy$, solution of~\eqref{eq:optimization_problem}, where
\begin{equation*}
    %\label{eq:incentive-policy}
    \IncDef=\DecDef\cdot \WeightDef, \quad \forall \Ind, \Alt.
\end{equation*}

For any budget $\Budget$, we indicate with $\MaxSocUt{\Budget}$ the maximum of the social welfare, solution of problem~\eqref{eq:max-soc-ut}.

\subsection{Maximum Social Welfare Curve Problem}
\label{sec:maximum-social-welfare-curve-problem}

Suppose now that the regulator is endowed with a maximum budget $\Budget$ and that he can spend any budget in the interval $\TotIncSym\in[0,\Budget]$. To decide the exact amount of budget that is convenient to spend, it is useful to obtain the \emph{Maximum Social Welfare Curve} $\Curve^*_\Budget$, representing the maximum social welfare reachable, $\MaxSocUt{\TotIncSym}$, for any budget $\TotIncSym\in[0, \Budget]$, i.e.
\begin{equation}
    \Curve^*_\Budget = \left\{ \big(\TotIncSym, \MaxSocUt{\TotIncSym} \big) \:\middle|\: \TotIncSym \in [0, \Budget] \right\}.
    \label{eq:max-soc-ut-curve}
\end{equation}
The \emph{Maximum Social Welfare Curve Problem} consists in finding the curve $\Curve^*_\Budget$, for a given maximum budget $\Budget$.
It is easy to show that it is monotone non-decreasing (the larger the budget spent, the larger the social welfare reached).
Observe that, although a maximum budget $\Budget$ is available, the regulator may not want to indiscriminately spend it all, but may choose the actual budget to invest in incentives, based on several criteria.
For instance, the regulator may use the above curve to find the minimum budget needed to reach a certain social-welfare target.
Moreover, in many practical cases, the social welfare is converted into money metric.
The coefficient of conversion is usually fixed based on political considerations.
For instance, in our numerical results (Section~\ref{sec:case_study}), we convert \cotwo{} emission reduction into money, using the cost of 100 euros per ton of \cotwo{}.
After converting social welfare in money metric, the regulator may choose to invest an incentive budget such that the gain of social welfare equals the incentive spent.
Such a value can be found on the Maximum Social Welfare Curve.

%%%%%%%%%%%%%%%%%%%%%%%%%%%%%%%%%%%%%%%%%%%%%
%%%%%%%%%%%%%% APPROX ALGO %%%%%%%%%%%%%%%%%%
%%%%%%%%%%%%%%%%%%%%%%%%%%%%%%%%%%%%%%%%%%%%%
\section{Approximation Algorithm}
\label{sec:algorithm}

\citet{Kellerer2004} shows that the MCKP problem, and thus the Maximum Social Welfare problem \eqref{eq:max-soc-ut}, is NP-hard.
Therefore, for large instances of such problems, finding the optimal solution is unfeasible and we need to resort to heuristics.
We provide in this section a polynomial time algorithm based on greedy algorithms from the Operations Research literature, which gives us solutions boundedly close to the optimum.
We then discuss some relevant properties of such an algorithm, e.g., its diminishing returns behaviour and the fact that it is an anytime algorithm (explained in Remark~\ref{rem:anytime}).

In the following subsection, we introduce some preliminary mathematical concepts.

\subsection{Preliminary Steps}
\label{sec:preliminary-steps}
Before presenting the proposed algorithm, we need to `clean' the input of the problem, removing some irrelevant alternatives from the set $\InitSetDef$ of the alternatives of any individual $\Ind$.
In broad terms, irrelevant alternatives are the ones that do not provide enough social indicator compared to the incentive amount needed to induce them.
We call \emph{LP-extremes} the alternatives remaining after the cleaning, and we denote them with $\NonDomSetDef\subseteq\InitSetDef$.
The name LP-extremes is borrowed from \citet[Section 11.2.1]{Kellerer2004}.

\begin{figure}[ht]
    \centering
    \includegraphics[width=0.3\textwidth]{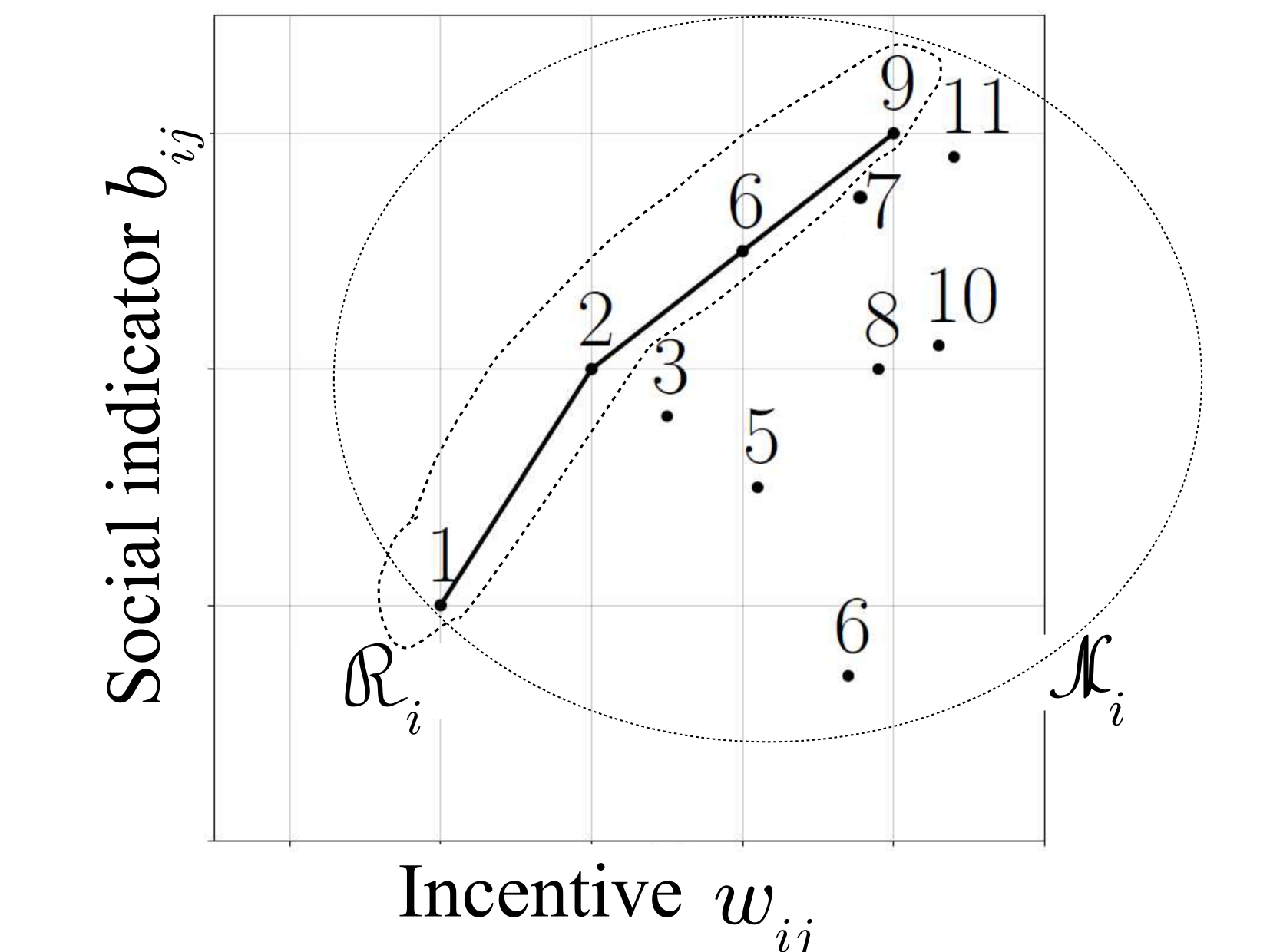}
    \caption{Alternative set $\InitSetDef$ of individual $\Ind$ and the subset $\NonDomSetDef$ of LP-extremes.}
    \label{fig:concavization}
\end{figure}

The process of constructing the set $\NonDomSetDef$ is called \emph{concavization} and is described in detail in Appendix \ref{sec:concavization}.
Here we just give the reader an intuition of it via Figure~\ref{fig:concavization}, which represent the incentive amount and social indicator for a set of alternatives $\InitSetDef$, of an individual $\Ind$.
In the figure, alternative $3$ is irrelevant since $2$ provides a larger social indicator, while requiring less incentive.
Alternative $7$ is irrelevant since it requires to spend more incentive than $6$, for a negligible gain in the social indicator.
It is much more convenient to make a slightly bigger investment to induce alternative $9$, which provides a significant social indicator improvement with respect to $6$.
More formally, we say that $7$ is LP-dominated by $6$ and $9$ (see Appendix \ref{sec:concavization} for more details).

We follow the Operations Research literature in the slight abuse of notation of denoting with $\WeightDef$ the incentive to be provided to the $j$-th alternative in $\NonDomSetDef$, where this is not ambiguous.
With no loss of generality, we can assume the ordering 
\begin{equation}
    \label{eq:ordering}
    \Weight{\Ind}{1} < \Weight{\Ind}{2} < \dots < \Weight{\Ind}{\NumOfNonDomDef}
\end{equation}
in the set $\NonDomSetDef$, where $\NumOfNonDomDef$ is its cardinality. 
The default alternative of any individual is neither dominated nor LP-dominated, since it requires no incentive ($\Weight{\Ind}{\DefAltDef}=0$).
Therefore, the default alternative is the first alternative in the set $\NonDomSetDef$ and $\Weight{\Ind}{1}=0$.

\begin{definition}[Efficiency and incremental efficiency]
    \label{def:efficiency}
    We define the \emph{efficiency} of an alternative $\Alt$ of individual $\Ind$ as
    \begin{equation*}
        \EffDef\defeq \frac{\SocUtDef - \SocUt{\Ind}{\DefAltDef} }{\WeightDef},
    \end{equation*}
    i.e., the gain in social indicator that we can gain via a unit of incentive allocated to that alternative.
    We define the incremental social indicator $\IncrSocUtDef$ and the incremental incentive $\IncrWeightDef$ required for each alternative $\Alt\in\NonDomSetDef$ as
    \begin{equation}
        \label{eq:incremental-weight-and-social}
        \begin{array}{rl}
            \IncrSocUtDef & \defeq \SocUtDef -\SocUt{\Ind}{\Alt-1} \\
            \IncrWeightDef  & \defeq \WeightDef -\Weight{\Ind}{\Alt-1}
        \end{array}, 
        \quad \Alt=2,\dots,\NumOfNonDomDef.
    \end{equation}
    The \emph{incremental efficiency} is then defined as
    \begin{equation}
        \label{eq:incremental_efficiency}
        \IncrEffDef \defeq \IncrSocUtDef / \IncrWeightDef.
    \end{equation}
\end{definition}
The incremental efficiency $\IncrEffDef$ can be interpreted as the increase in social welfare for each monetary unit spent, when individual $\Ind$ shifts from alternative $\Alt-1$ to alternative $\Alt$.

\subsection{Greedy Algorithm}

We want to find a curve $\Curve_\Budget=\left\{(\TotIncSym,\SocWelfareSym(\TotIncSym))\:|\:\TotIncSym\in[0,\Budget] \right\}$ that approximates the Maximum Social Welfare Curve $\Curve^*_\Budget$ \eqref{eq:max-soc-ut-curve}, i.e., such that $\SocWelfareSym(\TotIncSym)$ is close to $\MaxSocUtDeff$ for any value of budget $\TotIncSym\geq0$.

Very efficient algorithms, like the Dyer-Zemel algorithm \citep[Section~11.2.1]{Kellerer2004} are known to solve problem~\eqref{eq:max-soc-ut}, i.e., to approximate the maximum social welfare for a fixed single value of budget $\Budget$.
However, to apply them to the Maximum Social Welfare Curve problem, in which we want to find the maximum social welfare for a range of budget values $\TotIncSym\in[0,\Budget]$, instead of just one, we would have to run those algorithms from scratch for every single value of budget.
For this reason, we build our solutions upon a simpler greedy algorithm~\citep[Figure~11.2]{Kellerer2004}, which is less efficient to solve the Maximum Social Welfare problem (although still polynomial in time complexity), but easily extendable to also solve the Maximum Social Welfare Curve problem.
The other advantage deriving from such choice is that this greedy algorithm has interesting properties that increase its practical application and economic interpretability, as discussed in Section~\ref{sec:useful-properties-for-large-scale-applications}.
%The regulator willing to apply such algorithms could choose, for instance, to apply an incentive policy up to the point where the marginal efficiency (marginal welfare divided by marginal cost) becomes less than a certain threshold, fixed based on political or societal considerations, as we do in Section~\ref{sec:case_study}.

The pseudocode of the algorithm is in Algorithm~\ref{alg:greedy-max-soc-ut-curve}.
The notation $\AltPairDef$ stands for `$\Alt$-th alternative of individual $\Ind$'.
The idea of the algorithm is simple.
First, the algorithm finds all the LP-extremes alternatives and sorts them by order of decreasing incremental efficiency.
Then, at each iteration, the next pair of individual and alternative $\AltPair{\Ind'}{\Alt'}$ with the highest incremental efficiency is picked (line~\ref{ln:incr-eff}).
The alternative induced to $\Ind'$ is set to $\Alt'$ (line~\ref{ln:choice-update}) and the budget is reduced by the amount of the incremental weight (equation~\eqref{eq:incr-weight}).
An additional piece of the approximation of the social welfare curve is computed (equation~\eqref{eq:AlgSocUt}).
%The algorithm stops when the budget is depleted.
%From the allocation $\{\Dec{\Ind}{\Alt}\}\OfIndAltDef$ returned by the algorithm, we can deduce a state $(\Alt_1, \dots, \Alt_\NumOfInd)$, where $\Dec{\Ind}{\Alt_\Ind} = 1$, for any individual $\Ind$.
%Then, for any individual $\Ind$, the regulator proposes an incentive $\Inc{\Ind}{\Alt_\Ind}$ for alternative $\Alt_\Ind$ and no incentive for any alternative $\Alt\neq\Alt_\Ind$.
%Some individuals might receive no incentive if $\Dec{\Ind}{\DefAltDef}=1$ (the allocation returned by the algorithm is such that they choose their default alternative).
The algorithm stops when the maximum budget $\Budget$ is depleted and it returns a policy $\IncPolicy$, which is such that any individual $\Ind$, for whom the algorithm selected an alternative $\Alt\in\NonDomSetDef$, effectively chooses this alternative $\Alt$.

\begin{algorithm}[ht!]
\caption{Greedy algorithm for the Maximum Social Welfare and Maximum Social Welfare Curve problems.}
\label{alg:greedy-max-soc-ut-curve}
\footnotesize
\SetKwInOut{Input}{Input}\SetKwInOut{Output}{Output}
\Input{Social indicators $\{\SocUtDef\}\OfIndAltDef$, intrinsic utilities $\{\IndUtDef\}\OfIndAltDef$, budget $\Budget$ }
Iteration index $\IterSym:=0$\\
$\TotInc\OfIterDef:=0$, Total incentive allocated so far.\\
$\TotSocUt\OfIterDef:=0$, Social welfare obtained in the current allocation.\\
Compute the ordered set $\NonDomSetDef$ of LP-extremes of each individual $\Ind$.\\
Sort all the alternatives $\AltPairDef$ according to decreasing incremental efficiency $\IncrEffDef$ and put them in a set $\TotNonDomSet$.\\
Initialize the alternatives chosen by the individuals $\{\DecDef\}\OfIndAltDef$ as follows
\begin{equation*}
    \left\{
    \begin{array}{ll}
        \Dec{\Ind}{1} = 1, & \text{(default alternative)} \\
        \DecDef = 0, & \text{for any alternative }\Alt>1
    \end{array}
    \right.
\end{equation*}
\\
\While{$\TotNonDomSet\neq\emptyset$ and $\TotInc\OfIterDef\le \Budget$}
{
	Take $\AltPair{\Ind'}{\Alt'}$, the next alternative with the highest incremental efficiency $\IncrEff{\Ind'}{\Alt'}$ from $\TotNonDomSet$. \label{ln:incr-eff}\\ 
	Add $\AltPair{\Ind'}{\Alt'}$ to the solution, i.e.:
	\begin{align}
	\TotNonDomSet & := \TotNonDomSet\setminus\{\AltPair{\Ind'}{\Alt'}\}, & \nonumber
	\\
	\TotInc\OfIter{\IterSym+1} 
	& := \TotInc\OfIterDef + \IncrWeight{\Ind'}{\Alt'}
        \label{eq:incr-weight}
	\\
	\IncrEffSym\OfIterDef
	& := \IncrEff{\Ind'}{\Alt'}
	\label{eq:incr-eff-iter}
	\\
	\AlgSocUtDef
	& :=\TotSocUt\OfIterDef, 
	& \forall \TotIncSym\in [\TotInc\OfIterDef,\TotInc\OfIter{\IterSym+1})
	\label{eq:AlgSocUt}
	\\
	\TotSocUt\OfIter{\IterSym+1} 
	&:= \TotSocUt\OfIterDef + \IncrSocUt{\Ind'}{\Alt'} 
	& \nonumber \\
	\IterSym 
	& := \IterSym +1 \nonumber
	&
	\end{align}
	\label{ln:iteration-wise}
	\\
	Update the selected alternative for individual $\Ind'$, i.e.,
	\label{ln:choice-update}
\begin{equation*}
    \left\{
        \begin{array}{rll}
            \Dec{\Ind'}{\Alt'} & = 1, &  \\
            \Dec{\Ind'}{\Alt} & = 0, & \text{ for any other alternative }\Alt\neq \Alt'
        \end{array}
    \right.
\end{equation*}
}
%
%\rev{aa}{$\SocWelfareSym(\Budget) := \TotSocUt\OfIter{\IterSym+1}, \forall \Budget\in[\TotInc\OfIter{\IterSym+1},+\infty[$}{} \\
%
\Output{
Curve $\Curve_\Budget=\left\{(\TotIncSym,\AlgSocUtDef)\:|\:\TotIncSym\in[0,\Budget] \right\}$\\
Chosen alternatives $\{\DecDef\}\OfIndAltDef$\\
Incentive policy $\IncPolicy = \{\IncDef\}\OfIndAltDef$, where $\IncDef=\DecDef\cdot\WeightDef$\\
Split item $\AltPair{\SplitInd}{\SplitAlt}:=\AltPair{\Ind'}{\Alt'}$\\
Incremental efficiency of the split item $\IncrEff{\SplitInd}{\SplitAlt}$\\
Budget actually used $\BudgetUsed:=\TotInc\OfIter{\IterSym-1}$
}
\end{algorithm}

Observe that the curve $\Curve_\Budget$ given as output by the algorithm is an approximation of the solution $\Curve^*_\Budget$ of the Maximum Social Welfare Curve Problem (Section~\ref{sec:maximum-social-welfare-curve-problem}).
Moreover, given any maximum budget $\Budget$, the algorithm returns an approximation $\AlgSocUt{\Budget}$ to the solution $\MaxSocUtDef$ of the Maximum Social Welfare Problem~\eqref{eq:max-soc-ut}.
Note that, in order to achieve $\AlgSocUt{\Budget}$, the policy issued by the algorithm does not spend the entire maximum budget $\Budget$, but only $\BudgetUsed\le\Budget$.

The algorithm also gives as output the incremental efficiency of the `split item', denoted with $\IncrEff{\SplitInd}{\SplitAlt}$, useful to compute the optimality gap of the algorithm (Theorem~\ref{thm:bound} below).
The name \emph{split item}, which we borrow from~\cite{Kellerer2004}, reminds of the fact that, when we allocate the budget $\Budget$, we add to the solution all the LP-extreme alternatives, in decreasing order of incremental efficiency, up to $\AltPair{\SplitInd}{\SplitAlt}$.
In other words, such alternative $\AltPair{\SplitInd}{\SplitAlt}$ splits the set $\TotNonDomSet$ of all the LP-extremes in two parts: the first part consists of the alternatives we include in our solution, while we do not include the LP-extremes from the second part.

The distance to the optimum, in terms of social welfare, is bounded from above.
\begin{theorem}[Upper bound]
    \label{thm:bound}
    Let us run Algorithm~\ref{alg:greedy-max-soc-ut-curve} with budget $\Budget$, and let $\BudgetUsed$ be the budget actually used and $\IncrEff{\SplitInd}{\SplitAlt}$ be the incremental efficiency of the split item.
    The social welfare $\AlgSocUt{\Budget}$ we obtain is boundedly close to the social welfare $\MaxSocUtDef$ of any optimal personalized-incentive policy (Definition~\ref{def:optimal_incentive_policy}).
    In particular,
    \begin{align}
        \label{eq:bound}
        \MaxSocUtDef - \AlgSocUt{\Budget}
        \le \IncrEff{\SplitInd}{\SplitAlt} \cdot (\Budget-\BudgetUsed).
    \end{align}
\end{theorem}

\begin{corollary}
    \label{prop:curve_bound}
    The curve $\Curve_\Budget$ obtained via Algorithm~\ref{alg:greedy-max-soc-ut-curve} is boundedly close to the Maximum Social Welfare Curve $\Curve^*_\Budget$ from equation~\eqref{eq:max-soc-ut-curve} and the gap is given by Theorem~\ref{thm:bound}.
\end{corollary}

The next corollary says that, for any budget $\Budget$, the curve $\Curve_\Budget$ returned by Algorithm~\ref{alg:greedy-max-soc-ut-curve} and the Maximum Social Welfare Curve $\Curve^*_\Budget$ `touch each other'.
This ensures that the allocation computed by Algorithm \ref{alg:greedy-max-soc-ut-curve} at every iteration is optimal.
It is a direct consequence of Theorem~\ref{thm:bound}.

\begin{corollary}
    \label{prop:touch}
    The curve $\Curve_\Budget$ obtained via Algorithm~\ref{alg:greedy-max-soc-ut-curve} and the Maximum Social Welfare Curve $\Curve^*_\Budget$ from equation~\eqref{eq:max-soc-ut-curve} are such that $\SocWelfareSym(\TotInc\OfIterDef)=\MaxSocUt{\TotInc\OfIterDef}$ in all the values $\TotInc\OfIterDef$, $\IterSym=0,1,\dots$
\end{corollary}

\begin{figure}[ht]
    \centering
    \includegraphics[width=0.5\textwidth]{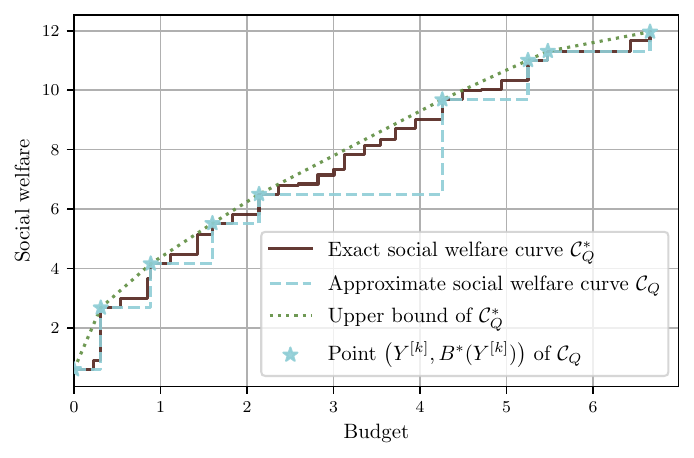}
    \caption{Distance between the social welfare curve $\Curve_\Budget$ computed by Algorithm~\ref{alg:greedy-max-soc-ut-curve}, the maximum social welfare curve $\Curve^*_\Budget$ (Section~\ref{sec:maximum-social-welfare-curve-problem}) and the upper bound of Theorem~\ref{thm:bound}. 
    The stars represent the incentive spent $\TotInc\OfIterDef$ and social welfare $\TotInc\OfIterDef = \MaxSocUt{\TotInc\OfIterDef}$ at each iteration $\IterSym=1,\dots,8$ of the algorithm (line~\ref{ln:iteration-wise}).}%
    \label{fig:distance_optimum}
\end{figure}

Figure~\ref{fig:distance_optimum} illustrates this property.
The continuous curve represents the Maximum Social Welfare Curve $\Curve^*_\Budget$ and the dashed curve represents the curve $\Curve_\Budget$ obtained via Algorithm~\ref{alg:greedy-max-soc-ut-curve}.
These two curves are step-functions because of the discreteness of the problem.
From Corollary \ref{prop:touch}, the curve $\Curve_\Budget$ intersects the curve $\Curve^*_\Budget$ at each iteration of the algorithm (represented by the stars).
The dotted curve represents the upper bound of $\Curve^*_\Budget$, computed from Theorem~\ref{thm:bound}.

\subsection{Useful Properties for Large-Scale Applications}
\label{sec:useful-properties-for-large-scale-applications}

Our aim is to compute a personalized-incentive policy in large scenarios in a small amount of time.
It is therefore crucial to show that our algorithm is computationally efficient.
\begin{proposition}
\label{prop:complexity}
The computational complexity of Algorithm \ref{alg:greedy-max-soc-ut-curve} is $O(\sum_{\Ind=1}^\NumOfInd \NumOfAltDef \cdot \log \NumOfConvHullDef + \TotNumOfConvHull \cdot \log m)$, where $\NumOfInd$ is the number of individuals, $\NumOfAltDef$ is the number of alternatives of individual $\Ind$, $\NumOfConvHullDef$ is the number of LP-extremes of individual $\Ind$ and $\TotNumOfConvHull\defeq\sum_{\Ind=1}^\NumOfInd \NumOfConvHullDef$.
\end{proposition}
%Note that the component $\sum_{\Ind=1}^\NumOfInd \NumOfAltDef \cdot \log \NumOfConvHullDef$ is for computing the LP-extremes, using the method of \citet{Kirkpatrick1986}.
Note that, since the alternatives of each individual are independent of the others, the sets $\NonDomSetDef$ can be computed in parallel, thus reducing even further the computation time.
%Therefore, the computation performance shown in the numerical results, in which we do not use any parallelization, should be considered as a pessimistic bound.

Despite our algorithm being computationally efficient, there might be cases in which it is desirable to stop it prematurely, without waiting for it to completely terminate.
This can be the case when a personalized-incentive policy must be computed on-the-fly, within tight time-constraints.
The following properties ensure that our algorithm is suitable to this situation, which eases its practical adoption.

\begin{remark}[Anytime algorithm]
\label{rem:anytime}
Algorithm~\ref{alg:greedy-max-soc-ut-curve} is \emph{anytime}: if we stop it prematurely at any iteration $\IterSym$, we get a valid solution for the Maximum Social Welfare and the Maximum Social Welfare Curve problems, with budget $\Budget'=\TotInc\OfIterDef$.
\end{remark}

%The property of Prop. \ref{prop:incr_optimality} ensures that all the solutions that the algorithm incrementally builds are optimal, with respect to the expenses of that allocation.
\begin{remark}[Incremental use]
Another desirable property of Algorithm~\ref{alg:greedy-max-soc-ut-curve} is that we can build on a previously computed incentive allocation whenever new available budget becomes available, instead of recomputing the entire allocation from scratch.
To explain this, let us suppose that we have a certain budget $\Budget$ and the algorithm returns the allocation $\{\DecDef\}\OfIndAltDef$, spending the corresponding incentive amount $\BudgetUsed$.
Suppose now that the available budget increases to $\Budget'>\Budget$. In this case, in order to exploit the new additional budget, we can simply resume the algorithm from its last iteration and continue up to the furthest iteration such that $\TotInc\OfIter{\IterSym+1}\le\Budget'$.
This is, per-se, a computational advantage with respect to algorithms that need to run from scratch every time new resources (budget) are available.
\end{remark}

In order to describe the diminishing return property of Algorithm~\ref{alg:greedy-max-soc-ut-curve}, we need the following definition.

\begin{definition}[Incremental and overall efficiency]
\label{def:marginal-and-overall-efficiency}

The \emph{incremental efficiency} provided by the algorithm at iteration $\IterSym$ is $\IncrEffSym\OfIterDef$, defined in equation \eqref{eq:incr-eff-iter}. 
We define the \emph{overall efficiency} of a personalized-incentive policy spending budget $\TotIncSym$ and achieving social welfare $\SocWelfareSym$ as $\EffSym=\SocWelfareSym / \TotIncSym$.
We denote with $\EffSym\OfIterDef$ the overall efficiency of the policy obtained by stopping Algorithm~\ref{alg:greedy-max-soc-ut-curve} at iteration~$\IterSym$, i.e., $\EffSym\OfIterDef = \SocWelfareSym\OfIterDef / \TotIncSym\OfIterDef $.
\end{definition}

The following proposition illustrates that, by spending more and more budget and allocating it as the algorithm dictates, we increase social welfare, but the marginal gain per unit of budget spent decreases.
\begin{proposition}[Diminishing returns]
    \label{prop:non-inc-eff}
    The incremental efficiency $\IncrEffSym\OfIterDef$ and the overall efficiency $\IncrEffSym\OfIterDef$ of the alternative added by Algorithm~\ref{alg:greedy-max-soc-ut-curve} at every iteration $\IterSym$ are both monotonically non-increasing.
\end{proposition}
%For the first iterations, the algorithm can find individuals who need only a small incentive to accept to switch to an alternative with a much larger social indicator.
%As the number of iterations increases, all good opportunities are depleted and the algorithm must resort to switch individuals who require a large incentive or who only increase social welfare by a small amount.

The following corollary is a consequence of Proposition~\ref{prop:non-inc-eff}.
\begin{corollary}
\label{cor:non-inc-eff}
At any iteration $\IterSym$, we can compute an upper bound $\SocWelfareSym^\text{ub}\ge \AlgSocUt{\Budget}$ to the social welfare we would get if we continue the algorithm until the end. Such an upper bound is $\SocWelfareSym^\text{ub} = \AlgSocUt{\TotIncSym\OfIterDef}+ \IncrEffSym\OfIterDef \cdot (\Budget - \TotIncSym\OfIterDef)$.
\end{corollary}
Therefore, if we notice that $\AlgSocUt{\TotIncSym\OfIterDef}$ is already sufficiently close to $\SocWelfareSym^\text{ub}$, then it is not worth continuing the algorithm, as we would not get  much additional social welfare. In this case, we can safely stop the algorithm, without waiting for it to end, thus saving time.

In some cases, the regulator would be willing to maximize social welfare under the constraints that the \emph{overall inverse efficiency} $\EffSym^{-1}$ is below a certain target.
For instance, in Section~\ref{sec:calculation-of-the-incentive-policy} the regulator does not want to spend more than 100 euros per ton of \cotwo{} saved, which is considered to be the carbon price.
In such cases, it is useful to observe that, thanks to Proposition~\ref{prop:non-inc-eff}, $(\EffSym\OfIterDef)^{-1}$ is non-decreasing.
Therefore, the regulator could run the algorithm and stop at the iteration where  $(\EffSym\OfIterDef)^{-1}$ goes above the target inverse efficiency.

We close this section with a definition that will be useful in Section~\ref{sec:policies_comparison}.
\begin{definition}[Maximum Step Size and Characteristic Incremental Efficiency]
\label{def:maximum-step-size}
Let us run Algorithm~\ref{alg:greedy-max-soc-ut-curve} with a certain budget $\Budget$ and record the values $\TotIncSym\OfIterDef$ calculated therein, as well as the incremental efficiency of the split item $\IncrEff{\SplitInd}{\SplitAlt}$. The maximum step size is defined as $\gamma_\Budget \defeq \max_{k=1,2,\dots} (\TotIncSym\OfIterDef - \TotIncSym\OfIter{\IterSym-1})$. The characteristic incremental efficiency of budget $\Budget$ is defined as $\IncrEffSym_\Budget \defeq \IncrEff{\SplitInd}{\SplitAlt}$.
\end{definition}

The properties presented in this section have shown that the proposed Algorithm~\ref{alg:greedy-max-soc-ut-curve} is computationally efficient and able to return an allocation providing a welfare close to the optimum.
Moreover, it has some features that make its adoption easier in practical large-scale scenarios.

\section{Comparison with Other Policies}
\label{sec:policies_comparison}

So far, we have considered that the regulator uses personalized incentives to increase social welfare.
In particular, the policy proposed is such that the loss in individual utility, due to the shift to another alternative, is compensated exactly by the incentive.
In this section, to frame our proposed personalized-incentive policy into a more general set of feasible policies, we generalize the formulation to include not only personalized incentives, but also enforcement policies, taxation, and non-personalized-incentive policies.
In Section \ref{sec:general-policy}, we show that any optimal personalized-incentive policy (Definition~\ref{def:optimal_incentive_policy}) is optimal in this more general class of policies.
In Section \ref{sec:alg-policy}, we show that Algorithm~\ref{alg:greedy-max-soc-ut-curve} can be used to compute an enforcement policy and a proportional tax-subsidy policy, which are both boundedly close to the optimal general policy.
We finally analytically show that non-personalized-incentive policies, like Tripod \citep{Araldo2019}, achieve by construction less social welfare than our personalized-incentive policy, and we provide a lower bound to this social welfare gap.

\subsection{Optimality of Personalized-Incentives Policies among General Policies}
\label{sec:general-policy}

We now consider a more general space of policies, including incentives, enforcement and taxation policies, and we define a criteria of optimality in this space.
In order to do so, we need to define some new quantities.

%\todoi{aa: I think the reasoning is easier to follow if we interpret $\VarUt{\Policy}$ in terms of `reduction'. I changed the sign accordingly.}
%\todoi{lj: I am not so sure about that because (i) the sign $\Delta$ represents more often variations than reductions and (ii) we are more discussing subsidies than taxation so, usually, the reduction in utility is negative, which is not natural. I leave it like that for now. To be discussed.}
The total loss in individual utility, of a policy $\Policy$, is
\begin{equation}
    \label{eq:utility_loss}
    \VarUt{\Policy} \equiv
     \sum^{\NumOfInd}_{\Ind=1} \left(
     	\IndUt{\Ind}{\DefAltDef}
     	-\IndUtAfterFunc{\Ind}{\AltEq{\Ind}{\Policy}}{\Policy}
     \right).
\end{equation}
With a taxation policy, i.e., a policy $\Policy$ such that $\TrDef\leq0$, $\forall\Ind,\Alt$, the loss in individual utility is non-negative, i.e., $\VarUt{\Policy}\geq0$.
With an incentive policy, i.e., a  policy $\Policy$ such that $\TrDef\geq0$, $\forall \Ind, \Alt$, the loss in individual utility is non-positive, i.e., $\VarUt{\Policy}\leq0$.
In particular, in any personalized-incentive policy obeying to Proposition~\ref{propo:incentive_amount}, the individuals are perfectly compensated for their loss in utility, and thus $\VarUt{\Policy} = 0$, in accordance with Remark~\ref{rem:unchanged-utility}.
Keeping everything else fixed, it is obvious that, the smaller $\Delta U$, the better.

The disutility, or cost, of a policy is measured by combining the loss in individual utilities and the expenses for the regulator, defined in equation \eqref{eq:expenses}.
\begin{definition}[Disutility] 
    The disutility $\Disut{\Policy}$ of a policy $\Policy$ is defined as the expenses of the regulator $\Exp{\Policy}$ plus the total loss in individual utilities $\VarUt{\Policy}$, i.e.,
    \begin{equation}
        \label{eq:disutility}
        \Disut{\Policy} \equiv \Exp{\Policy} + \VarUt{\Policy}.
    \end{equation}
\end{definition}

The following proposition shows that the disutility of a policy is always non-negative, which means that it is not possible that both the individuals increase their utility and the regulator collects revenues.
It also shows that, if two policies imply the same alternatives chosen, then they have the same disutility.
\begin{proposition}
\label{prop:non-negative-disutility}
Every policy $\Policy$ has a non-negative disutility that only depends on the alternative chosen by the individuals, rather than the actual incentive or taxation proposed. In particular:
\[
	\DisutDef = \sum_{\Ind=1}^{\NumOfInd} \left(
            \IndUt{\Ind}{\DefAltDef}
            - \IndUt{\Ind}{\AltEqDef}
        \right)
         \ge 0
\]
\end{proposition}

We can now define an optimal general policy, whose definition includes incentive, taxation and enforcement policies.
\begin{definition}[Optimal General Policy]
\label{def:optimal_general_policy}
An optimal general policy $\Policy$ with disutility threshold $\Budget\ge 0$ is the solution of the following problem:
\begin{equation}
    \label{eq:general-optimization-problem}
    \left\{
        \begin{array}{cl}
            \displaystyle\max_{\Policy} & \SocWelfare{\Policy}\\
            \text{s.t.} & \DisutDef \leq \Budget \\
                         %& \displaystyle\AltEqDef = \argmax_{\Altt\in\argmax_j \IndUtAfterFuncDef} \SocUt{\Ind}{\Altt}
        \end{array}
    \right..
\end{equation}
\end{definition}

Note that, problem~\eqref{eq:general-optimization-problem} is a generalization of~\eqref{eq:optimization_problem}.
Indeed, we  obtain the latter from the former by (i) constraining the policy to be a personalized-incentive policy, i.e., $\TrDef\ge 0,\forall \Ind\in\IndSet,\Alt\in\InitSetDef$ and (ii) imposing no change in individual utility, i.e., $\VarUt{\Policy}=0$.

The two next propositions characterize the optimal general policy.
The first one implies that finding an optimal general policy is equivalent to finding a chosen-alternative set which maximizes social welfare, subject to a disutility constraint.
\begin{proposition}
\label{prop:equivalence}
Let $\Policy$ be an optimal general policy with disutility threshold $\Budget$. Any other policy $\Policy'$ inducing the same alternatives is also an optimal general policy, independent of the actual value of the single incentives or taxes proposed.
\end{proposition}

The following proposition shows that the personalized-incentive policy, considered previously, is still relevant in this more general framework.
The proposition states that, for any disutility threshold $\Budget$, it is possible to find an optimal policy which is a personalized-incentive policy.
\begin{proposition}
\label{prop:optimal-incentive-general-policy}
For any $\Budget\geq0$, any optimal personalized-incentive policy $\IncPolicy$ with budget $\Budget$ (Definition~\ref{def:optimal_incentive_policy}) is also an optimal general policy with disutility threshold $\Budget$ (Definition~\ref{def:optimal_general_policy}).
\end{proposition}

The following corollary states that the social welfare bound for personalized-incentive policies (Theorem~\ref{thm:bound}) is equivalent for general policies.
\begin{corollary}
    \label{propo:bound-general}
    Let us run Algorithm~\ref{alg:greedy-max-soc-ut-curve} with budget $\Budget$ to construct a personalized-incentive policy.
    The social welfare $\AlgSocUt{\Budget}$ we obtain is boundedly close to the optimum $\MaxSocUtDef$, obtainable with an optimal general policy with disutiliy threshold $\Budget$.
    In particular,
    \begin{align*}
        \MaxSocUtDef - \AlgSocUt{\Budget}
        \le \IncrEff{\SplitInd}{\SplitAlt} \cdot (\Budget-\BudgetUsed).
    \end{align*} 

\end{corollary}

\subsection{Computing Optimal Enforcement and Proportional Tax-Subsidy Policy}\label{sec:alg-policy}

In this section, we show how Algorithm~\ref{alg:greedy-max-soc-ut-curve} can be used, in conjunction with Propositions \ref{prop:optimal-incentive-general-policy} and Corollary~\ref{propo:bound-general}, to compute an enforcement policy and a proportional tax-subsidy policy boundedly close to the optimum.

We provide a numerical comparison between these policies in Section~\ref{sec:comparison-with-other-policies}.

\subsubsection{Enforcement Policy}
\label{sec:enforcement}

With enforcement policies, the regulator constrains the individuals to choose an alternative among a subset of their choice set.
In the most extreme case, the individuals can choose only one alternative.
%Such policy is personalized in the sense that the alternative $\Alt_{\Ind}$ imposed to individual $\Ind$ can be different from the alternative $\Alt_{\Ind'}$ imposed to individual $\Ind'$.

Let $\IncPolicy$ be the personalized-incentive policy returned by Algorithm~\ref{alg:greedy-max-soc-ut-curve}, for a budget $\Budget$.
Consider now a policy $\Policy$ enforcing the individual to choose the same alternative that they would choose under the policy $\IncPolicy$, i.e.,
\begin{equation*}
    \left\{
        \begin{array}{ll}
            \TrDef = 0, & \text{if}\:\Alt = \AltEq{\Ind}{\IncPolicy} \\
            \TrDef = -\infty, & \text{otherwise} \\
        \end{array}
    \right., \quad \forall \Ind, \Alt.
\end{equation*}

\begin{proposition}
\label{prop:bound-enforcement}
The enforcement policy $\Policy$ constructed above is boundedly close to an optimal general policy with disutility constraint $\Budget$. The bound is the same as Corollary~\ref{propo:bound-general}.
\end{proposition}

\subsubsection{Proportional Tax-Subsidy Policy}
\label{sec:finding-the-optimal-tax-subsidy-policy}

We consider here policies $\Policy$ for which the monetary transfers are proportional to the social indicator of the alternatives, that is
\begin{equation}
    \label{eq:tax_def}
    \TrDef = \Tax \cdot (\SocUtDef - \Thr), \quad \forall \Ind, \Alt,
\end{equation}
where $\Tax>0$ is the \emph{tax-subsidy level} and $\Thr\in\mathbb{R}$ is an individual-specific \emph{baseline social-indicator}, set by the regulator.
We call these policies \emph{proportional tax-subsidy policies}.
Observe that, for any individual $\Ind$, her alternatives $\Alt\in\InitSetDef$ such that the social indicator is below the baseline are taxed (i.e., $\SocUtDef<\Thr\Rightarrow\TrDef<0$).
Conversely, alternatives $\Alt\in\InitSetDef$ having social indicator above the baseline are subsidized (i.e., $\SocUtDef>\Thr\Rightarrow\TrDef>0$).
The baseline social-indicators $\Thr$ can vary from individual to individual.
However, we impose that the tax-subsidy level $\Tax$ is the same for everyone.
In this sense, we consider that these policies are not personalized.

Observe from equations \eqref{eq:utility_def} and \eqref{eq:tax_def} that, considering any individual $\Ind$, if we vary the baseline $\Thr$ the variation of the utility $\IndUtAfterDef$ is the same for all alternatives $\Alt\in\InitSetDef$.
Hence, the value of $\Thr$ does not impact the choice of $\Ind$.
It simply represents a monetary transfer between the individual and the regulator.
More precisely, setting low $\Thr$ favours transfers from the regulator to individuals, thus increasing individual utilities, to the detriment of the regulator.
On the other hand, setting high $\Thr$, favours the revenue of the regulator, to the detriment of the utility of the individuals.

Note that, if $\SocUtDef$ represents a negative externality (as in Section~\ref{sec:case_study}), then the tax-subsidy policy defined above is equivalent to a Pigouvian tax if $\Tax$ is set to be equal to the external marginal cost of the externalities and $\Thr=0$.
For instance, it has been estimated \citep{quinet2009valeur}, that the social cost of 1 ton of \cotwo{} is 100 euros.
Then, if $\SocUtDef$ represents \cotwo{} emissions (in tons), the Pigouvian tax would be a proportional tax-subsidy policy with $\Tax=100$ euros.

The following theorem shows that we can use Algorithm~\ref{alg:greedy-max-soc-ut-curve} to compute a proportional tax-subsidy policy that is boundedly close to the theoretical optimum.

\begin{theorem}\label{theorem:taxation}
    We can construct a proportional tax-subsidy $\Policy$ under a certain disutility threshold $\Budget$ as follows.
    Run Algorithm~\ref{alg:greedy-max-soc-ut-curve} with budget constraint $\Budget$ and let $\IncrEff{\SplitInd}{\SplitAlt}$ be the incremental efficiency of the split item given as output.

    The proportional tax-subsidy policy $\Policy$, defined as in equation~\eqref{eq:tax_def}, with tax-subsidy level 
    \begin{align}
        \label{eq:tax_opt}
        \Tax=1/\IncrEff{\SplitInd}{\SplitAlt}
    \end{align}
    achieves a social welfare that is boundedly close to the optimal general policy with disutility threshold $\Budget$.
    The bound is the same as Corollary~\ref{propo:bound-general}.
\end{theorem}

\subsubsection{Comparison with Proportional-Incentive Policy and Tripod}
\label{sec:comparison-proportional-tripod}
A proportional-incentive policy $\Policy$ is a proportional tax-subsidy policy, as in~\eqref{eq:tax_def} where only subsidies and not taxes are distributed, i.e.
\begin{align}
    \label{eq:proportional-incentive}
    \Thr\le \SocUtDef, \quad \forall \Ind\in\IndSet,\Alt\in\InitSetDef.
\end{align}
An example of proportional-incentive policy is Tripod \citep{Araldo2019}. 

In this section we show that proportional-incentive policies are inefficient incentive policies, in the sense that, to achieve a certain social welfare level, they spend more incentives than needed.
We call `inefficiency gap' this additional incentive spent and we compute a lower bound for it in the following proposition.

\begin{proposition}
\label{prop:proportional-inefficiency}
Consider a proportional-incentive policy $\Policy$ as before.
There always exists a personalized-incentive policy $\IncPolicy$ that is able to achieve at least the same social welfare and provides the following savings in the amount of incentive spent:
\begin{align*}
	\TotInc(\Policy) - \TotInc(\IncPolicy) 
	\ge
	\frac{1}{\Tax}
	\cdot
	\sum_{\Ind\in\IndSet}
		\left( \IndUt{\Ind}{\AltEq{\Ind}{\Policy}}-\IndUt{\Ind}{\DefAltDef} \right)
		\cdot
		\Delta\Eff{\Ind}{\AltEq{\Ind}{\Policy}} 
		\defeq
		L(\Policy)
\end{align*}
where $\Delta\EffDef \defeq \EffDef - 1/\Tax$ 
is defined as \emph{efficiency loss}, $\forall \Ind\in\IndSet,\Alt\in\InitSetDef$.
The quantity $L(\Policy)$ defined above is a lower bound for the inefficiency gap.
\end{proposition}

Note that the \emph{efficiency loss} quantifies the fact that proportional-incentive policies are not able to exploit the inherent efficiency $\EffDef$ (see Definition~\ref{def:efficiency}) of the incentivized alternative. Indeed, instead of using such an alternative-dependent efficiency, they use a single value $1/\Tax$.

We compute in the following proposition a lower bound on the suboptimality gap of proportional-incentive policies.

\begin{proposition}
\label{prop:proportional-incentive-vs-algo}
Let us consider a proportional-incentive policy $\Policy$, achieving a social welfare $\AlgSocUt{\Policy}$ and spending an incentive amount $\TotInc(\Policy)$. 
Let us denote with $\MaxSocUt{\TotInc(\Policy)}$ the maximum social welfare achievable by an optimal personalized-incentive policy with that incentive amount.
The following lower bound holds on the suboptimality gap:
\[
	\MaxSocUt{\TotInc(\Policy)}-\AlgSocUt{\Policy} 
	\ge
	\max\left\lbrace 
		0, 
		\IncrEffSym_{\TotInc(\Policy)}
		\cdot \left( L(\Policy) - 2 \gamma_{\TotInc(\Policy)} \right) \right\rbrace
\]
where $\gamma_{\TotInc(\Policy)}$ and $\IncrEffSym_{\TotInc(\Policy)}$ are the maximum step size and the characteristic incremental efficiency, as defined in Definition~\ref{def:maximum-step-size}. 

\end{proposition}

We now draw an interesting parallel with Tripod, a proportional incentive policy described in~\cite{Araldo2019}. In Tripod, social welfare is represented, in particular, by energy reduction. While our formulation is general and can encompass any type of social welfare (provided that the assumptions of Section~\ref{sec:policy} are valid) in our case study (Section~\ref{sec:case_study}) we consider \cotwo{} reduction. In both our case study and Tripod, individuals are travellers and alternatives are modal choices. An important aspect of Tripod is that it is dynamic, i.e., time is slotted and, in each time slot, the incentives for the individuals happening to depart in that time slot are calculated.
With the Tripod policy, that we denote $\Policy^{\text{Tr}}$, the incentives proposed to individuals are proportional to the gain in the social indicator with respect to the default alternative, i.e.
\begin{equation*}
    %\label{eq:tripod}
    \TrDef^\text{Tr}=(\SocUtDef-\SocUt{\Ind}{\DefAltDef})/\text{TEE}_t
\end{equation*}
where the constant $\text{TEE}_t$ is called Token Energy Efficiency and is fixed by the regulator in every time slot $t$.
In Tripod, the incentives are distributed to individuals under a First-Come First-Served discipline, until a certain budget is depleted.
Therefore, out of the entire population $\IndSet_t$ of individuals departing at time slot $t$ only a subset $\IndSet_t^\text{Tr}$ actually receive an incentive.
%Let us denote with $\Policy^\text{Tr}\defeq \{\TrDef^\text{Tr}\}_{\Ind\in\IndSet^\text{Tr}}$ the Tripod policy. 
In \citet{Araldo2019}, the value of $\text{TEE}_t$ is fixed empirically, with a grid search, trying several values of $\text{TEE}_t$ in simulation and choosing the one with maximum social welfare. The calculation is based on a Model-Predictive Control (MPC) setting, where at every time slot the value of $\text{TEE}_t$ is calculated taking into account not only the current time slot, but also a prediction of the system state (congestion, individual arrival) in the subsequent time-slots, which are called \emph{optimization horizon}. Note that the MPC setting of Tripod allows to take into account the impact of the incentive policies on congestion, which we instead neglect, based on Assumptions~\ref{ass:independant_indiv} and~\ref{ass:independant_social}. Therefore, for adopting our policy to a real scenario, care should be taken in checking that such Assumptions are reasonable, as we do in our case study.

Within our framework, Tripod can be defined as a proportional-incentive policy, with $\Tax=1/\text{TEE}_t$ and $\Thr=\SocUt{\Ind}{\DefAltDef}$, applied to population $\IndSet_t^\text{Tr}$.

As a consequence of Proposition~\ref{prop:proportional-inefficiency}, Tripod is an inefficient incentive policy, under the assumptions of Section~\ref{sec:model-and-assumptions}. In Corollary~\ref{prop:tripod}, we lower-bound the additional incentive spent with respect to the theoretical best incentive policy.
\begin{corollary}
\label{prop:tripod}
For any Tripod incentive policy $\Policy^\text{Tr}$, there always exists a personalized-incentive policy~$\Policy$ that is able to achieve at least the same social welfare while spending less incentives. The saving in the incentives is:
\begin{equation*}
    \TotInc(\Policy^\text{Tr}) - \TotInc(\Policy) 
    \ge
    \sum_t
    \text{TEE}_t
    \cdot
    \sum_{\Ind\in\IndSet_t^\text{Tr}}
    \left( 
        \IndUt{\Ind}{\AltEq{\Ind}{\Policy^{\text{Tr}}}}-\IndUt{\Ind}{\DefAltDef} 
    \right)
    \cdot
    \Delta\Eff{\Ind}{\AltEq{\Ind}{\Policy^{\text{Tr}}}} 
\end{equation*}
where $\Delta\EffDef \defeq \EffDef - \text{TEE}_t$ 
is defined as \emph{efficiency loss}, $\forall \Ind\in\IndSet_t^\text{Tr},\Alt\in\InitSetDef$ and $\IndSet_t^\text{Tr}$ is the set of individuals getting incentives in Tripod in time-slot $t$.
\end{corollary}

Corollary~\ref{prop:tripod} shows that Tripod is far from minimizing the incentives needed to obtain a certain social welfare, while the policy issued by Algorithm~\ref{alg:greedy-max-soc-ut-curve} is generally close to using minimal incentives. This is confirmed by our numerical results in Section~\ref{sec:comparison-with-other-policies}.
An interpretation of the inefficiency suffered by Tripod follows.

\begin{remark}
\label{rem:tripod-inefficiency}
Tripod uses a single value $\text{TEE}_t=1/\Tax$ to compute incentives for all alternatives $\Alt$ of all users $\Ind\in\IndSet_t$ and gets $1/\text{TEE}_t$ additional units of social welfare per additional unit of incentive spent. The only incentivized alternatives are the ones for which $\EffDef\ge \text{TEE}_t$ (otherwise the proposed incentive would not be accepted by the individual).
In other words, Tripod gets always an efficiency (unit of social welfare improvement over unit of incentive spend) that is lower than the intrinsic efficiency of the incentivized alternatives.
By contrast, our personalized policy always entirely exploits the intrinsic efficiency of the incentivized alternatives.
\end{remark}

Since Tripod is a proportional-incentive policy, the same lower bound on the suboptimality gap as in Proposition~\ref{prop:proportional-incentive-vs-algo} also holds, which we do not write for the sake of space.

\section{Imperfect Information}
\label{sec:imperfect-information}

The assumption that the regulator knows perfectly the utility of the individuals may seem restrictive.
In this section, we show that the algorithm is still relevant when the utility is imperfectly known.
From discrete-choice theory \citep{Anderson1992}, we assume that intrinsic utility of alternative $\Alt$ of individual $\Ind$ is composed of a deterministic part $\IndUtDetDef$ and a random part $\RandomDef$:
\begin{equation*}
    %\label{eq:random-ind-ut}
    \IndUtDef = \IndUtDetDef + \RandomDef.
\end{equation*}
We assume that the regulator knows the deterministic part $\IndUtDetDef$ of the utility but not the random part $\RandomDef$.

Under this assumption, the regulator cannot compute the minimum incentive amount needed to induce individual $\Ind$ to shift from her default alternative $\DefAltDef$ to another alternative $\Alt$, using directly equation \eqref{eq:incentive_amount}.
A heuristic solution would be to set the incentive amount equal to the expectation of the utility difference between the two alternatives, given that $\DefAltDef$ is the default alternative chosen when there is no incentive.
%Observe that, by construction (see equation~\eqref{eq:default_alt}), $\IndUt{\Ind}{\DefAltDef} > \IndUtDef$.
In this case, the incentives $\{\Inc{\Ind}{\Alt}\}_{\Alt\in\InitSetDef}$ proposed by the regulator to individual~$\Ind$, to convince her to shift to alternative~$\Alt$, are such that $\Inc{\Ind}{\Altt}=0$, for any $\Altt\neq\Alt$, and
\begin{equation}
    \label{eq:incentive_amount_imperfect_0}
    \IncDef = \ExpSym(\IndUt{\Ind}{\DefAltDef} - \IndUtDef |\IndUt{\Ind}{\DefAltDef} > \IndUtDef) = \IncDetDef + \ExpSym(\Random{\Ind}{\DefAltDef} - \RandomDef | \Random{\Ind}{\DefAltDef} - \RandomDef > -\IncDetDef),
\end{equation}
where $\IncDetDef = \IndUtDet{\Ind}{\DefAltDef}-\IndUtDetDef$ is the difference in the deterministic part of the utility, known to the regulator.

Given an individual $\Ind$ and an alternative $\Alt\in\InitSetDef$, if the regulator proposes the incentive $\IncDef$, as defined by equation \eqref{eq:incentive_amount_imperfect_0}, then individual $\Ind$ has a positive probability to refuse the incentive.
Hence, the expenses of the regulator may be smaller than the total incentive amount proposed.

Algorithm~\ref{alg:greedy-max-soc-ut-curve} can be used to compute a personalized-incentive policy under imperfect information, by defining new weights
\begin{equation*}
    %\WeightDef = \mu\frac{ 1 + e^{\IncDetDef/\mu} }{ e^{\IncDetDef/\mu} } \ln\left( 1 + e^{\IncDetDef/\mu} \right), \quad \forall\Ind, \Alt.
    \WeightDef = \ExpSym(\IndUt{\Ind}{\DefAltDef} - \IndUtDef |\IndUt{\Ind}{\DefAltDef} > \IndUtDef).
\end{equation*}

At each iteration of the algorithm, the regulator proposes the incentive $\Weight{\Ind'}{\Alt'}$ to individual $\Ind'$ for alternative $\Alt'$, where $[\Ind', \Alt']$ is the pair of individual and alternative selected by the algorithm.
The regulator observes the response of the individual to the incentive.
If the individual accepts the incentive, it decreases the budget by the incentive amount.
The regulator keeps proposing incentives one by one until his budget is depleted.

Note that, if an individual $\Ind$ accepts an incentive $\IncDef$ for alternative $\Alt\in\InitSetDef$, the regulator can still propose  her, later, an incentive $\Inc{\Ind}{\Altt}$ for another alternative $\Altt\in\InitSetDef$.
If the individual refuses the second incentive $\Inc{\Ind}{\Altt}$, she still receives the first incentive $\IncDef$. 

In Section~\ref{sec:modal-choice-imperfect-information}, we apply the policy presented above to our case study and compare it to the case with perfect information, assuming that random terms are Gumbel-distributed.
The following proposition gives the exact expression of the incentives~\eqref{eq:incentive_amount_imperfect_0}, in case of Gumbel-distributed random terms.
\begin{proposition}
\label{prop:gumbel}
Let us assume that the random terms are i.i.d.~and follow a Gumbel distribution with scale parameter $\mu$ (i.e., $\RandomDef/\mu$ follows a standard Gumbel distribution).
%It is well known \citep{Nadarajah2005} that the difference of two i.i.d. Gumbel-distributed random variables is a standard logistic-distributed random variable, whose probability density function is $f(x)=\frac{e^x}{(e^x+1)^2}$.
Then, the incentive amount from equation \eqref{eq:incentive_amount_imperfect_0} can be written as
\begin{equation*}
    %\label{eq:incentive_amount_imperfect}
    \IncDef = \mu \frac{ 1 + e^{\IncDetDef/\mu} }{ e^{\IncDetDef/\mu} } \ln\left( 1 + e^{\IncDetDef/\mu} \right) \geq 0.
\end{equation*}
\end{proposition}

 \iffalse

 %==========
 % Probability of acceptance
 %==========

 The probability that individual $\Ind$ accepts the incentive $\IncDef$ proposed to her is
 \begin{equation*}
     \AccProbDef = \ProbSym(\IndUtDef + \IncDef \geq \IndUt{\Ind}{\DefAltDef} | \IndUt{\Ind}{\DefAltDef} > \IndUtDef).
 \end{equation*}
 Using equation \eqref{eq:random-ind-ut}, we get
 \begin{equation*}
     \AccProbDef = \ProbSym(\Random{\Ind}{\DefAltDef} - \RandomDef \leq \IncDef - \IncDetDef | \Random{\Ind}{\DefAltDef} - \RandomDef > -\IncDetDef).
 \end{equation*}
 With $\xi=\Random{\Ind}{\DefAltDef} - \RandomDef$,
 \begin{equation*}
     \AccProbDef = 1 - \ProbSym(\xi > \IncDef - \IncDetDef | \xi > -\IncDetDef) = 1 - \frac{1-F(\IncDef - \IncDetDef)}{1 - F(-\IncDetDef)}.
 \end{equation*}
 Then, using the cumulative distribution function of $\xi$,
 \begin{equation*}
     \AccProbDef = 1 - \frac{1+e^{-\IncDetDef/\mu}}{1+e^{(\IncDef-\IncDetDef)/\mu}} = \frac{e^{(\IncDef-\IncDetDef)/\mu}-e^{-\IncDetDef/\mu}}{1+e^{(\IncDef-\IncDetDef)/\mu}}.
 \end{equation*}
 Rearranging the terms yields
 \begin{equation}
     \label{eq:acceptance_prob}
     \AccProbDef = \frac{1-e^{-\IncDef/\mu}}{1+e^{(\IncDetDef-\IncDef)/\mu}}.
 \end{equation}

 \fi

\section{Numerical Results in an Application to Mode Choice}\label{sec:case_study}

In this section, we simulate an application of our personalized incentive policy to a scenario related to mode choice of individuals commuting to their workplace.
We consider a regulator willing to employ a limited monetary budget in order to promote eco-friendly modes of transportation.
The goal of the regulator is to reduce \cotwo{} emissions.
We compute the reduction in \cotwo{} emissions achieved via the personalized-incentive policy of Algorithm~\ref{alg:greedy-max-soc-ut-curve} and compare it with enforcement, proportional taxation and non-personalized-incentive policies.

Our approach is as follows.
After describing the census data used to build the simulation scenario (Section~\ref{sub:data}), we estimate a Multinomial Logit model for mode choice (Section~\ref{sec:multinomial-logit-model}).
Then, using the previous estimates, we simulate the utility of a home-work trip for a group of individuals, for all the modes of transportation considered (Section~\ref{sec:simulating-utilities}).
We then approximate the \cotwo{} emissions for these same trips (Section~\ref{computing-the-social-indicator}) and approximate the optimal personalized-incentive policy using Algorithm~\ref{alg:greedy-max-soc-ut-curve} (Section~\ref{sec:calculation-of-the-incentive-policy}).
We then study the modal shifts induced by such policy and the gain in \cotwo{} emissions achieved.
We conclude the numerical results by comparing our personalized incentive policy with other policies (enforcement, taxation, flat incentives -- Section~\ref{sec:comparison-with-other-policies}) and by evaluating its performance in case of imperfect information (Section~\ref{sec:modal-choice-imperfect-information}).

\subsection{Data}%
\label{sub:data}

We use census data from the French statistics institute INSEE, regarding households surveyed between 2015 and 2019.
We restrict the dataset to households whose home and workplace are in the Rhône department, which includes Lyon and its suburbs (about \num{222000} households in total).
Observed variables include city- or district-level home and work location, main mode of transportation used for commuting, and some socio-demographic variables.
The modes of transportation are divided in five categories: car, public transit, walking, cycling and motorcycle.
Appendix \ref{sec:data_appendix} provides a detailed description of the data.

%Travel times for home-work trips are computed with the routing engines OpenTripPlanner and GraphHopper, using data from \textcopyright{} OpenStreetMap contributors.
%Open-data public transit schedules are also used for the public-transit travel times.

%To compute \cotwo{} emissions, we use open-sourced data from the French agency ADEME.

%We consider four different modes of transportation for commuting: car, public transportation, walk and bike.
%For some individuals, public transportation is not an available alternative because their home and workplace are not connected.
%For others, commuting by car is not possible because the household does not own a vehicle.
%Therefore, the set of alternatives $\InitSetDef$ varies from individual to individual, with $2 \leq \NumOfAltDef \leq 4$.

\subsection{Multinomial Logit Model}
\label{sec:multinomial-logit-model}

Using the census data, we estimate a Multinomial Logit model for the mode choice of the individuals.
We consider five exogenous variables specific to the individual (age, sex, number of cars owned per employee in the household and professional occupation) and one exogenous variable which is specific to both the mode of transportation and the individual (travel time).
The number of cars owned is supposed to only impact the utility of commuting by car.
Following \citet{inoa2015}, we also include interaction variables between travel time and socio-demographic variables.
Details on how travel time is computed are provided on Appendix \ref{sec:travel_times}.
To estimate the utility of the round trip to work, travel times are doubled (we assume that the modes of transportation for the trip back and forth are the same).

Note that public transit is excluded from the choice set of the individuals whose commute to work cannot be performed by public transit ($\simeq$\num{16000} individuals, see Appendix \ref{sec:travel_times} for more details).
%\num{16161} individuals, representing \SI{10.87}{\%} of total sample weight

The four other modes of transportation (car, walking, cycling and motorcycle) are assumed to be in the choice set of all the individuals.
This is a strong assumption.
A regulator willing to deploy the personalized-incentive policy in practice could improve the precision of the model by using individual-specific data for vehicle ownership in order to remove some modes of transportation from the choice set of an individual, if she does not own the corresponding vehicle.

Table \ref{tab:mult_log_estimation} provides the results of the Multinomial Logit model, estimated with the R package \texttt{mlogit}.
The most frequent categories are used as reference category (car for the mode of transportation, man for the sex and employee for the occupation).

\begin{table}
    \centering
    \caption{Multinomial Logit model of mode choice}
    \label{tab:mult_log_estimation}
    \begin{footnotesize}
        \bgroup
        \def\arraystretch{.8}
        \begin{tabular}{lccccc} \toprule
             & (1) & (2) & (3) & (4) & (5) \\
             & car & public\_transit & walking & cycling & motorcycle \\ \midrule
             &  &  &  &  &  \\
            constant &  & 2.7709*** & 2.8659*** & 1.1340*** & -0.7284*** \\
                     &  & (0.0395) & (0.0488) & (0.0509) & (0.0773) \\
            age &  & -0.0150*** & -0.0026*** & -0.0139*** & -0.0019\\
                &  & (0.0008) & (0.0009) & (0.0010) & (0.0015) \\
            woman &  & 0.5349*** & 0.4361*** & -0.3882*** & -1.6909*** \\
                       &  & (0.0194) & (0.0248) & (0.0242) & (0.0527) \\
            car\_per\_indiv & 1.2138*** &  &  &  &  \\
                            & (0.0161) &  &  &  &  \\
            car\_per\_indiv$>$0 & 1.5604*** &  &  &  &  \\
                              & (0.0245) &  &  &  &  \\
            occupation: farmer &  & -3.9054*** & -1.0434*** & -2.3653*** & -0.8798** \\
                              &  & (0.4140) & (0.2012) & (0.5073) & (0.4400) \\
            occupation: artisan &  & -1.7023*** & -1.2153*** & -0.7848*** & -0.2261*** \\
                               &  & (0.0525) & (0.0566) & (0.0651) & (0.0841) \\
            occupation: executive &  & 0.1522*** & 0.2031*** & 1.1710*** & 0.2986*** \\
                                 &  & (0.0255) & (0.0327) & (0.0337) & (0.0575) \\
            occupation: intermediate &  & -0.2283*** & -0.1447*** & 0.4259*** & -0.0060 \\
                                    &  & (0.0242) & (0.0311) & (0.0349) & (0.0584) \\
            occupation: blue-collar &  & -0.7579*** & -0.9691*** & -0.4808*** & -0.0259 \\
                              &  & (0.0318) & (0.0413) & (0.0467) & (0.0616) \\
            travel\_time & -1.6281*** & -1.1746*** & -2.1032*** & -2.8474*** & -3.2075*** \\
                         & (0.0530) & (0.0480) & (0.0492) & (0.0581) & (0.0968) \\
            %travel\_time\_squared & -7.6667*** & -2.2365*** & -1.3315*** & -0.1602 & -6.5736*** \\
                                  %& (0.5511) & (0.0522) & (0.1046) & (0.2072) & (0.7575) \\
            travel\_time $\times$ age  & \multicolumn{5}{c}{-0.0026**} \\
                                             & \multicolumn{5}{c}{(0.0010)} \\
            travel\_time $\times$ woman  & \multicolumn{5}{c}{-0.1134***} \\
                                               & \multicolumn{5}{c}{(0.0266)} \\
            travel\_time $\times$ occupation: farmer  & \multicolumn{5}{c}{1.1027***} \\
                                                           & \multicolumn{5}{c}{(0.3621)} \\
            travel\_time $\times$ occupation: artisan  & \multicolumn{5}{c}{-0.0763} \\
                                                            & \multicolumn{5}{c}{(0.0918)} \\
            travel\_time $\times$ occupation: executive  & \multicolumn{5}{c}{-0.3671***} \\
                                                              & \multicolumn{5}{c}{(0.0354)} \\
            travel\_time $\times$ occupation: intermediate  & \multicolumn{5}{c}{-0.1986***} \\
                                                                 & \multicolumn{5}{c}{(0.0330)} \\
            travel\_time $\times$ occupation: blue-collar  & \multicolumn{5}{c}{0.2623***} \\
                                                                & \multicolumn{5}{c}{(0.0403)} \\
                                                    &  &  &  &  &  \\
            %Observations & 20,162,292 & 20,162,292 & 20,162,292 & 20,162,292 & \\ \hline
            \multicolumn{6}{c}{Reference category is male employee} \\
            \multicolumn{6}{c}{Travel time is expressed in hours} \\
            \multicolumn{6}{c}{Standard errors are reported in parentheses} \\
            \multicolumn{6}{c}{*** p$<$0.01, ** p$<$0.05, * p$<$0.1} \\
            \bottomrule
        \end{tabular}
        \egroup
    \end{footnotesize}
\end{table}

The results are consistent with the literature on commute mode choice.
For example, we find that being young and male increases the probability to commute by cycling, consistently with the literature review of cycling mode choice from \citet{Munoz2016}.
We also find that the coefficient of travel time is larger for public transit than for car which suggests that the value of time for commutes by public transit is slightly smaller than for commutes by car.
This is coherent with the meta-analysis on the value of travel-time in France from \citet{wardman2012european}.

\subsection{Simulating Utilities}
\label{sec:simulating-utilities}

We consider a regulator whose goal is to reduce the \cotwo{} emissions due to commute trips, by distributing incentives to the population described in the data (about \num{222000} individuals).

To apply Algorithm~\ref{alg:greedy-max-soc-ut-curve}, the regulator needs to know the utility and the \cotwo{} emissions of each individual, for each mode of transportation.
We describe in this subsection how we estimate them. Note that our estimations are individual specific.

\begin{remark}
\label{rem:exogeneity}
Recall that Assumption~\ref{ass:independant_indiv} implies that the utility of an individual when commuting by car or public transit does not depend on how many other individuals commute by car or by public transit.
Such an assumption is reasonable when the congestion on the road and transit occupation rate are approximately exogenous, i.e., they do not depend on the incentive policy.
This approximation is legitimate if the number of modal shifts induced by the policy is low, so that their impact on congestion and occupation is negligible.
A posteriori, we check that this latter assumption is verified in our case, since less than \num{1.60}\% of individuals shifted mode due to the personalized-incentive policy.
\end{remark}

Following the Multinomial Logit theory, we assume that utility of alternative $\Alt$ of individual $\Ind$ is composed of a deterministic part $\IndUtDetDef$ and a random part $\RandomDef$:
\begin{equation}
    \label{eq:logit_utility}
    \IndUtDef = \IndUtDetDef + \RandomDef.
\end{equation}
The deterministic part of the utility can be computed using the estimates from Table \ref{tab:mult_log_estimation}.
As for the random part, we simulate random draws from a random variable with standard Gumbel distribution (see Appendix \ref{sec:gumbel}).
In accordance with Assumption~\ref{ass:information}, the regulator is assumed to know perfectly both the estimates and the draws and thus the utilities.
We relax this assumption in Section~\ref{sec:modal-choice-imperfect-information}, where we provide results where the random draws are unknown to the regulator.

To normalize the utility in monetary units, we compare the value of travel time by car from our regression (expressed in utility units) with the value of travel time by car in France from the literature (expressed in euros).
We compute the value of travel time by taking the opposite of the average marginal effect on utility of increasing the travel time of the individuals by one hour.%
\footnote{For example, referring to the coefficients of Table~\ref{tab:mult_log_estimation}, for a 40-year old male employee, the value of travel time by car is 
\begin{equation*}
    - \big(\vGroup{-1.6281}{\text{coef.\ of travel\_time (car)}} + \vGroup{-0.0026}{\text{coef.\ of travel\_time$\times$age}} \cdot \vGroup{40}{\text{age} }\big) = 1.7321\text{ utility units}/\text{hour}.
\end{equation*}}
We find an average value of time of \num{1.88} utility units per hour.

Previous studies \citep{wardman2012european} have shown that the value of travel time, for car commuters, in France, is about \num{9.17} euros per hour.
This would imply that, in our estimates, one utility unit corresponds to $\mu=9.17/1.88=4.88$ euros.
In the following, we assume that the utility is normalized in monetary units, i.e., the values in equation \eqref{eq:logit_utility} are multiplied by $\mu$.
Note that it implies that the random variables $\RandomDef$ follows a Gumbel distribution with scale parameter $\mu$.

\subsection{Computing the Social Indicator}
\label{computing-the-social-indicator}

The regulator wants to reduce greenhouse gas emissions.
The social indicator associated to the mode of transportation $\Alt$ of individual $\Ind$ is the reduction in \cotwo{} equivalent of greenhouse gas emissions generated during the trip of $\Ind$ performed with mode $\Alt$, with respect to the emissions of the default mode.
%\todoi{lj: If we say that the social indicator is the reduction in \cotwo{} compared to the default mode, then the social welfare is incorrectly defined below.}
To compute \cotwo{} emissions for each individual and each mode of transportation, we take the distance of the fastest path between the individual's home and workplace and we multiply this distance with the \cotwo{} emissions equivalent per kilometre for the mode of transportation, using open-sourced data from the French agency ADEME (Agence de l'Environnement et de la Maîtrise de l'Énergie).%
\footnote{\url{https://www.bilans-ges.ademe.fr/en/}}

For car, we use the \cotwo{} emissions of a passenger car with average motorization (\num{0.193} kilogram of \cotwo{} per kilometre).
That is, we assume that the \cotwo{} emissions per kilometre are the same for everyone.
We pinpoint that this assumption may lead to some imprecision in the calculation of the actual \cotwo{} reduction.
%However, it is reasonable to optimistically assume that the imprecisions summed over all the individuals cancels out.
The application could be improved by using detailed data on the characteristics of the vehicle used by each individual.
%However, in that case, an individual with a fuel-inefficient vehicle would be more likely to receive incentive than an individual with an electric car.
%Paradoxically, this would encourage travellers to own more polluting cars, in order to receive larger compensation for each daily trip.
%\todoi{lj: I removed the comment about the imprecisions canceling out as it is not true: with true \cotwo{} emissions, there would be some individuals with fuel-inefficient cars that would be worse incentivizing first, which means that our results underestimate social welfare achievable in reality.}
%\todoi{lj: I removed the comment about fuel-inefficient cars as it contradicts the fact that our policy compensates exactly the individuals for their loss in utility.}

For Assumption~\ref{ass:independant_social} to be valid, \cotwo{} emissions due to the commuting trip of an individual must be independent from the mode of transportation chosen by the other commuters.
For the same argument of Remark~\ref{rem:exogeneity}, we can claim that this approximately holds true in the scenario.

As \citet{Chester2010}, we adopt a disaggregated view of \cotwo{} emissions from public transit.
We consider that the overall \cotwo{} generated by transit vehicles is shared among all travellers making trips within transit, proportionally to the kilometres travelled.
In other words, each trip on transit produces a quantity of \cotwo{} emissions equal to the number of kilometres travelled multiplied by the average \cotwo{} emissions per kilometre per passenger, assuming average and constant occupancy rate.
Observe that it is reasonable to assume an average occupancy rate that is constant over time from the argument of Remark~\ref{rem:exogeneity}.
The average \cotwo{} emissions per kilometre per passenger vary according to the mode of transportation used (e.g., bus, tramway or metro).
The mode of transportation taken for the fastest path are used to compute \cotwo{} emissions.
For multi-modal public-transit trips (e.g., bus then tramway), the \cotwo{} emissions are computed according to the distance travelled by each mode of transportation.

\cotwo{} emissions for walking and cycling trips are set to zero.
Hence, for each individual, the two alternatives corresponding to walking and cycling differ only in the intrinsic utility.
As a consequence, the alternative with smaller intrinsic utility can be neglected, thanks to Proposition~\ref{prop:pareto}.

\begin{table}
    \centering
    \caption{Our calculations of \cotwo{} emissions in the Rhône department.}%
    \label{tab:emissions}
    \begin{tabular}{lr}
        \toprule
        Daily \cotwo{} emissions (all home-work and work-home trips) & \num{595.26} tons of \cotwo{} \\
        Total \cotwo{} emissions in one year (200 working days) & \num{119050} tons of \cotwo{}\\
        Average yearly individual \cotwo{} emissions & 0.54 tons of \cotwo{}\\
        \bottomrule
    \end{tabular}
\end{table}

Recall that Assumption~\ref{ass:information} implies that the regulator knows perfectly the \cotwo{} emissions of the trips.
This is more realistic than for utility. 
In any case, measurement errors for \cotwo{} emissions are not as worrying as measurement errors for utility as we can assume that, if such errors are unbiased, they cancel out.
We will observe in Section~\ref{sec:modal-choice-imperfect-information} that the errors are much more severe when utilities are imperfectly known, as some individuals might reject the incentives, which leads to a suboptimal allocation.

Under the previous assumptions, we calculate the \cotwo{} emissions reported in Table~\ref{tab:emissions}, which results in 0.54 ton of \cotwo{} yearly per individual in the Rhône department.
This number is close to the publicly known estimation for the entire France: in 2007, the average French worker emitted 0.64 ton per year because of his/her home-work trips \citep{levy2011habitant}.

\subsection{Calculation of the Personalized-Incentive Policy}
\label{sec:calculation-of-the-incentive-policy}
%\todo[inline]{aa: one thing over which the reviewer could attack us is `you did not consider electric vehicles'}

We consider a large-scale scenario with more than 200 thousands individuals and over 1 million alternatives (Appendix \ref{sec:data_appendix}).
%\todo[inline]{aa: one home-work trip includes also the trip to come back? Is it thus the CO2 produced in a day? Or in a second? Or in a century? How does it compare to other numbers declared by auhtorities or other papers? Please, comment this things in the paper. Also, are such tons of CO2 related to the entire France? Or Paris? Or the 13th arrondissement? Or Europe?}
% https://www.insee.fr/fr/statistiques/1287608
%The initial consumer surplus (sum of individual utilities) is \num{2275700} while the consumer surplus when all individuals are forced to choose their socially optimal alternative is \num{-33559825}.
%Therefore, the maximum budget necessary \todo{AA: necessary to do what?} in terms of utility units is \num{35835525}.
We consider a policy in which the regulator proposes, each day, incentives to the individuals before their home-work trip.
The incentives are given conditional on the mode of transportation chosen for the round trip to work, thus the social indicator of an alternative is the reduction in \cotwo{} emissions for the trip \emph{back and forth}, with respect to the default alternative.
The budget represents the daily amount available to the regulator for incentives.

\begin{figure}
    \centering
    \includegraphics{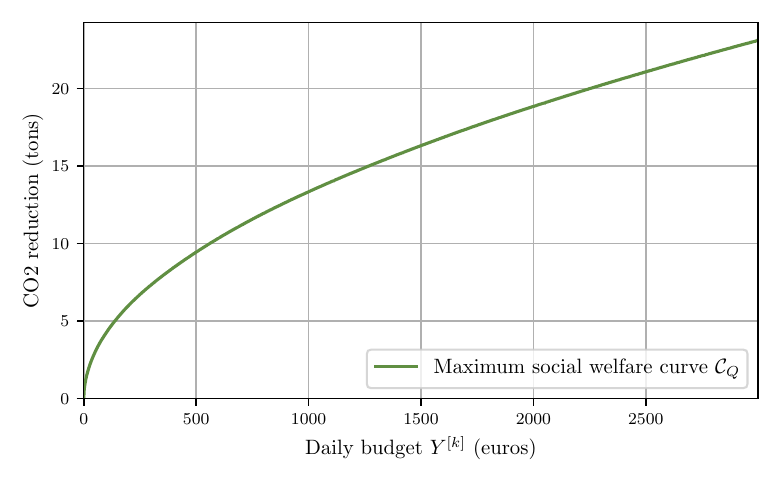}
    \caption{Maximum social welfare curve, up to a daily budget of 3000 euros. \textbf{Note}: The social welfare corresponds to the reduction in \cotwo{} emissions due to the personalized-incentive policy.}%
    \label{fig:max_soc_welfare_curve}
\end{figure}

First, we run Algorithm \ref{alg:greedy-max-soc-ut-curve} with a daily budget of \num{3000} euros and we plot the maximum social welfare curve (see Figure~\ref{fig:max_soc_welfare_curve}).
The maximum social welfare curve is an increasing step function (steps are small and thus not visible).
Consistently with Proposition~\ref{prop:non-inc-eff}, the slope of each step is non-increasing, which gives the curve a concave curvature.

\begin{figure}
    \centering
    \includegraphics{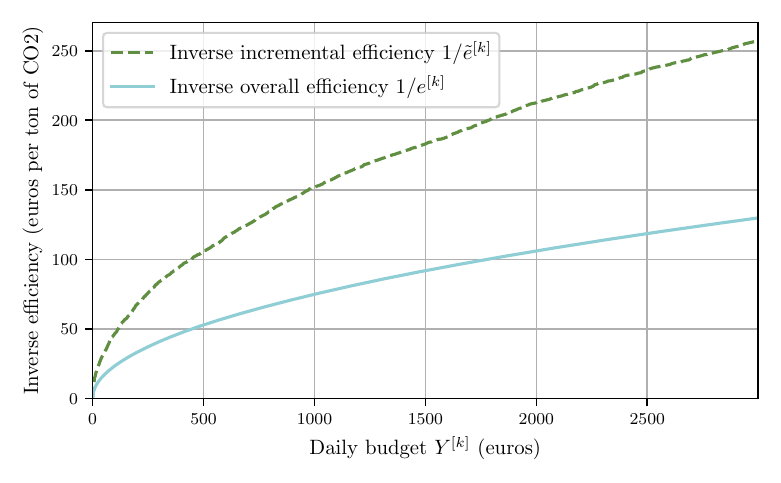}
    \caption{Cost of the policy in euros per ton of \cotwo{} prevented as a function of the daily budget.}%
    \label{fig:inverse_efficiency}
\end{figure}

\citet{quinet2009valeur} predict that the carbon price in France would be of 100 euros per ton of \cotwo{} in 2030.
It is thus reasonable to assume the regulator is interested in finding an incentive policy such that, for every 100 euros spent in incentives, pollution is reduced by at least a ton of \cotwo{}.
To this aim, the regulator can observe the curves of Figure~\ref{fig:inverse_efficiency}, which plots the inverse of the incremental and overall efficiency (the $\IncrEffSym\OfIterDef$ and $\EffSym\OfIterDef$ of Definition~\ref{def:marginal-and-overall-efficiency}), with respect to the budget $\TotIncSym\OfIterDef$ allocated by the algorithm at each iteration $\IterSym$.
Thanks to Proposition~\ref{prop:non-inc-eff}, $1/\EffSym\OfIterDef$ and $1/\IncrEffSym\OfIterDef$ increase with $\TotIncSym\OfIterDef$, as we proceed with the iterations of the algorithm.
Thanks to this monotonicity, the regulator can apply one of the following two criteria to fix the budget to invest.
It could run Algorithm~\ref{sec:algorithm} and stop it when $1/\EffSym\OfIterDef$ equals 100 euros per ton of \cotwo{}.
Alternatively, it can stop the Algorithm when $1/\IncrEffSym\OfIterDef$ equals 100 euros per ton of \cotwo{}.
From Figure~\ref{fig:inverse_efficiency}, we observe that with the first criterion the regulator would need to invest about 1800 euros per day, and about 500 euros with the second criterion.
In our opinion, both criteria would make sense, and the preference over one of them is a political choice.

We now set the budget of the regulator to $\Budget=\num{1800}$ euros.
Running Algorithm~\ref{alg:greedy-max-soc-ut-curve} with this budget required about \num{3500} iterations and took about 6 seconds (with Python, on a computer with an Intel i5-8350U 1.7GHz and 24GB of memory).
The algorithm allocates practically all the budget (\num{1798.59} euros).
We find that \num{1.57}\% of individuals received incentives and changed transportation mode, which results in a reduction of  \cotwo{} emission by \num{18} tons of \cotwo{} per day (\SI{3.00}{\%} of total \cotwo{} emissions).
Thus, this policy would cost on average \num{100.61} euros for each ton of \cotwo{} prevented.
%\todo[inline]{aa: to justify the polynomial time algorithm, we had discussed to compare its running time with that of a solver. Lucas, do you have such a comparison?\\
%lj: It runs in 500 seconds with SCIP solver (so roughly 50-100 times longer).\\
%aa: mmmh, 500 seconds would a sufficient time for the regulator to compute the incentives for the entire day! Therefore, in our studied case, the existence of our algorithm is not really justified. Such an algorithm would be more relevant at a bigger scale, when the solution from the solver would need about 10 hours or in an application with a much tighter time-constraints, for instance when decisions must be taken with a delay of 10 seconds.\\ Therefore, it is better to leave the information about the solver running time out of the paper and, if the reviewer ask, we will add bigger scenarios, multiplying by 2, 4, 8. 16 the population and plotting the computation time of our algorithm vs. the solver. But let's keep this for later.}

\begin{figure}
    \centering
    \includegraphics{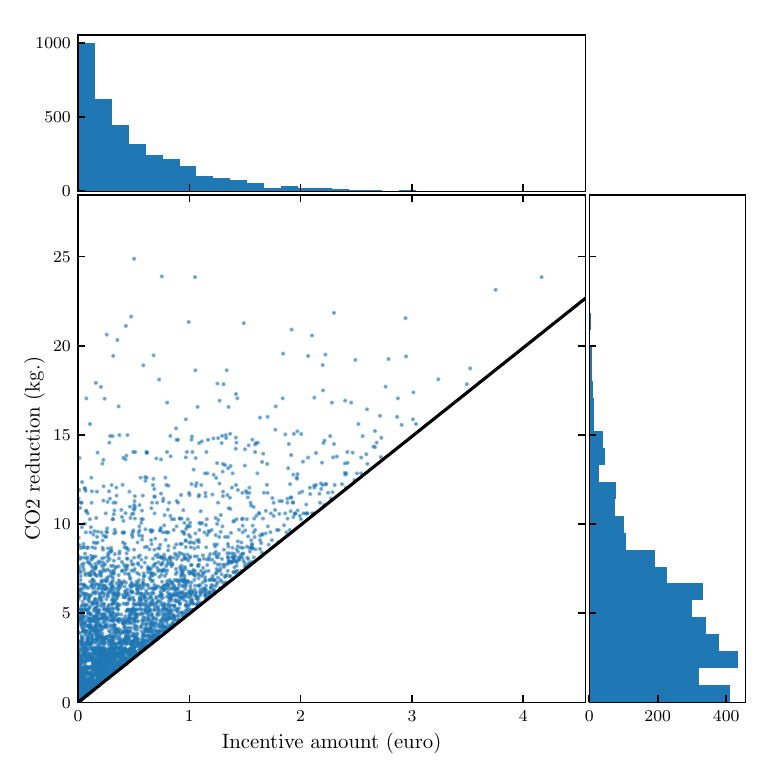}
    \caption{Distribution of incentive amount and \cotwo{} reduction for the incentives given in one day with budget $\Budget=1800$ euros.\\
    The slope of the black line represents the incremental efficiency of the split item returned by the algorithm, $\IncrEff{\SplitInd}{\SplitAlt}=\num{5}$ tons of \cotwo{} / euro.
    Note that all points are above the line because their incremental efficiency is larger.
    The histogram above represents the distribution of the incentive amounts.
    The histogram on the right represents the distribution of the \cotwo{} reduction for the incentives.}%
    \label{fig:jumps_scatter}
\end{figure}

Despite the small incentives, the reduction in \cotwo{} emissions is considerable.
Indeed, among the individuals who received incentives, the average amount of incentives is \num{0.52} euros per individual, for an average daily reduction in \cotwo{} emissions of \num{5} kilograms.
Recall that alternatives providing a large reduction in \cotwo{}, while requiring small incentive, have a high efficiency.
Hence, the algorithm selects first shifts achievable with a small incentive, i.e., where the individual is almost indifferent between the two alternatives, which however have a large difference in \cotwo{}.
Figure \ref{fig:jumps_scatter} shows the distribution of the incentive amount and the \cotwo{} reduction for the incentivized individuals.
For most incentives, the amount proposed to individuals is below \num{1} euro (incentives with a larger amount are not efficient enough, unless the \cotwo{} reduction is very high).

%On average, the reduction in \cotwo{} emissions from the incentives is \num{1.89} kilograms, while the average individual in the population emits only \num{1.28} kilograms per trip and the median is \num{0.65} kilogram.
%The share of car commutes decrease (from \SI{65}{\%} to \SI{45}{\%}) as the share of commutes by public transit (from \SI{18}{\%} to \SI{20}{\%}), by foot (from \SI{10}{\%} to \SI{13}{\%}) and by bike (from \SI{7}{\%} to \SI{22}{\%}) increase.
%Interestingly, the mode of transportation with the highest increase in use is the bike, which suggests most car drivers prefer to switch to the bike instead of walking.
%\todo[inline]{aa: I think this is not realistic. Nobody would expect that all of the sudden everyones gets on a bike. We limited the car possibility only to individuals owning a car. Maybe we should also limit the use of bike only to the individuals using the bike. Are we considering velib or similar? I am pretty sure that among the individuals switching to bikes, there are poor guys doing 30 Km by bike. Is there any way in the model to avoid this? Are we sure it does not happen. To control this does not happen, we should have a scatterplot like in Fig.~\ref{fig:transition}.
 %}
 %\begin{figure}
 %\includegraphics[width=0.7\textwidth]{graphs/andrea/transition}
 %\caption{In this plot, each point represents an individual and its transition, i.e., the mode she was choosing when 0 incentive budget is invested (her default choice) and the mode chosen when the incentive policy is implemented.}
 %\label{fig:transition}
 %\end{figure}

\begin{figure}
    \centering
    \includegraphics{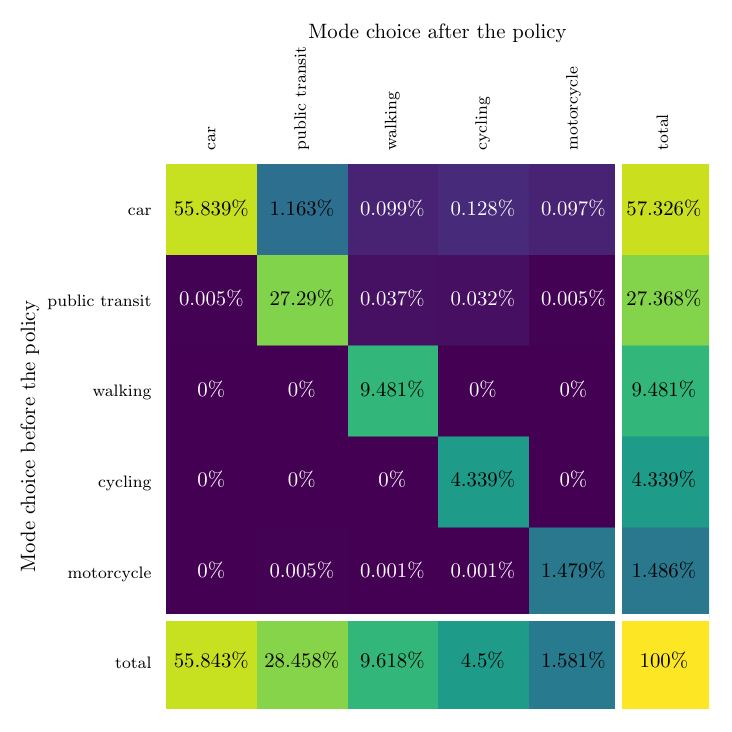}
    \caption{Evolution of mode share before and after the policy. \SI{1.163}{\%} of individuals were given incentives to shift from car to public transit, \SI{27.29}{\%} of individuals commuted by public transit before the policy and were not induced to shift.}%
    \label{fig:switches_heatmap}
\end{figure}

Figure \ref{fig:switches_heatmap} compares mode share before and after the policy.
Most individuals who received incentives are individuals who commuted by car and were induced to commute by public transit (\num{1.2}\% of all individuals, \num{74}\% of individuals who received incentives).
The share of individuals commuting by car decreased by 2.4\%, while public transit ridership increased by \num{4}\%.

%The incremental efficiency of the last iteration is \num{0.24} kilogram of \cotwo{} per euro.
%Hence, from Proposition \ref{prop:taxation}, the same results could be obtained with a proportional tax on the \cotwo{} emissions emitted by the individuals during their trips.
%The tax level should be set to $\Tax = 1/0.24 \approx 4.14$ euros per kilogram of \cotwo{} emitted, which is about 1.05 euros per kilometre by car.
%Such taxation would bring a revenue of \num{8390929.09} euros for the regulator but the sum of the individual utilities (expressed in euros) would decrease by \num{8390929.09} + \num{9999992.65} = \num{18390921.74} euros.
%%\todo[inline]{aa: What do you mean by `the same results'? They seem not to be the same, as you say that the individuals would loose more utility. Please, specify better.}

We now compute a bound of the optimality gap, i.e., the maximum additional \cotwo{} savings we would achieve if we could use a theoretical optimal policy instead of resorting to Algorithm~\ref{alg:greedy-max-soc-ut-curve}.
To do so, we apply Theorem~\ref{thm:bound}.
Since the incremental efficiency of the split item returned by the algorithm is $\IncrEff{\SplitInd}{\SplitAlt}\simeq 5$ kilograms of \cotwo{} per euro and the unused budget is $\Budget-\BudgetUsed=1.41$ euros, an optimal policy would reduce just $5\cdot 1.41 \simeq 7$ kilograms more than Algorithm~\ref{alg:greedy-max-soc-ut-curve}, which is negligible compared to the total \cotwo{} emissions reduction of \num{18} tons provided overall.

\subsection{Comparison with Other Policies}
\label{sec:comparison-with-other-policies}

In Section~\ref{sec:calculation-of-the-incentive-policy}, we evaluated the performance of the personalized incentive policy calculated by Algorithm~\ref{alg:greedy-max-soc-ut-curve}, which we denote with $\IncPolicy$.
We now compare it with three other policies from Section~\ref{sec:alg-policy}: an enforcement policy, a proportional taxation system and the Tripod incentive system from \citet{Araldo2019}.

Aggregate results for these policies are provided in Table~\ref{tab:policy}.
The policies are defined so that they induce the same choices for the individuals, using results from Section~\ref{sec:policy}.
Therefore, they provide the same reduction in \cotwo{} emissions and, from Proposition~\ref{prop:non-negative-disutility}, they have the same disutility.
However, they differ in their cost for the regulator and the variation in individual utilities implied.
It should be noted that the best policy to implement depends on social, political or juridical constraints.

In these results, we fix the disutility threshold to \num{1798.59} euros, which corresponds to the incentive $\BudgetUsed$ actually spent by Algorithm~\ref{alg:greedy-max-soc-ut-curve} when we set the budget to $\Budget=1800$ euros.
This means that if we run the algorithm setting a budget of \num{1798.59} euros, it spends it all and, thanks to Corollary~\ref{propo:bound-general},  the resulting policy is optimal under a budget constraint of \num{1798.59} euros.

\begin{table}
    \centering
    \caption{Summary of policies.}%
    \label{tab:policy}
    \footnotesize
    \begin{tabular}{lcccc}
        \toprule
        Policy $\Policy$ & Expenses (euros) & Ind.\ utility (euros) & Disutility (euros) & \cotwo{} reduction (tons) \\
                            & $\Exp{\Policy}$ & $\VarUt{\Policy}$ & $\Disut{\Policy}$ & $\SocWelfare{\Policy}$ \\
        \midrule
        Personalized incentives & \num{1798.59} & 0 & \num{1798.59} & \num{17.878} \\
        Enforcement & 0 & \num{-1798.59} & \num{1798.59} & \num{17.878} \\
        Proportional tax & \num{-116167.48} & \num{-114368.89} & \num{1798.59} & \num{17.878} \\
        Tripod incentives & \num{3596.97} & \num{1798.38} & \num{1798.59} & \num{17.878} \\
        \bottomrule
    \end{tabular}
\end{table}
% Note.
% Formula to compute the Tripod policy revenues: -(1 / efficiency of last jump) * total energy gains.

\paragraph{Enforcement policy}

Thanks to Proposition~\ref{prop:bound-enforcement}, the regulator can compute an enforcement policy $\Policy$ that is optimal for a disutility threshold of \num{1798.59} euros, by simply `imitating' the personalized incentive policy $\IncPolicy$, i.e., by inducing the same alternatives as $\IncPolicy$.
In order to do so, the regulator bans all the other alternatives, i.e., any alternative $\Alt$ such that $\Alt\neq\AltEq{\Ind}{\IncPolicy}$ is banned.
Obviously, it is not necessary to ban any alternative $\Alt$ if $\AltEq{\Ind}{\IncPolicy}$ is preferred to $\Alt$, in absence of policy, i.e., $\IndUtDef < \IndUt{\Ind}{\AltEq{\Ind}{\IncPolicy}}$.
Therefore, only the individuals receiving incentives under policy $\IncPolicy$ suffer bans with $\Policy$, which correspond to only 1.57\% of the population.

Contrarily to the personalized-incentive policy, the enforcement policy does not cost any money to the regulator (apart from eventual transaction costs) but it decreases individual utilities by \num{1798.59} euros.
Moreover, the \num{1.57}\% of individuals impacted by the ban may perceive that they are inequitably penalized with respect to the others. 
Hence, the enforcement policy might be less accepted by the population.
Still, this policy is well adapted to the context of imperfect information as it ensures that the individuals always choose the alternative wanted by the regulator.

\paragraph{Proportional tax}

The proportional tax policy is computed from equation \eqref{eq:tax_def}, using the tax level given by equation \eqref{eq:tax_opt} (Theorem~\ref{theorem:taxation}).
For each individual $\Ind$, the baseline social-indicators $\Thr$ is set to the \cotwo{} emissions of the default transportation mode, so that $\SocUtDef - \Thr$ is equal to the opposite of the \cotwo{} emissions of transportation mode~$\Alt$.
%This policy is referred to as a tax since the social indicators are always below the baseline, i.e., $\SocUtDef \leq \Thr$, $\forall \Ind, \Alt$.

Since the taxation provides revenues, the regulator is not constrained by his budget anymore.
However, taxation negatively impacts the utilities of the individuals, and is thus limited by political constraints, which we model by imposing that  the disutility of the policy must be below a threshold $\Budget=1798.59$ euros.

The tax must be paid by all individuals commuting either by car, public transit or motorcycle (about \SI{86}{\%} of individuals), even if the tax does not affect their choice.
This is different from the personalized-incentive policy, for which only \SI{1.57}{\%} of individuals are impacted.
This explains why the amount of taxes collected is about hundred times larger than the amount of incentive needed to reach the same reduction in \cotwo{} emissions (see Table~\ref{tab:policy}). 
A taxation policy is particularly penalizing for inelastic individuals who cannot shift to a more eco-friendly alternative, e.g., because they are living far from their workplace, or in a place with no transit offer.
Therefore, a taxation policy is much less acceptable than an incentive one.

On the other hand, the tax policy is not individual-specific, which means that  it requires less information (knowledge of individual utilities is required to compute the tax level from \eqref{eq:tax_opt} but the tax level does not change much under imperfect information and so the policy is still efficient).

\paragraph{Tripod policy}
We now compute a proportional-incentive policy as in Section~\ref{sec:comparison-proportional-tripod}.
Taking some additional assumption, we call such policy `Tripod' as in \citet{Araldo2019}.
In particular, we assume that the individuals described in the dataset are the first to log-in in the Tripod incentive system, such that budget $\Budget=1798.59$ euros is depleted after the Tripod system treats them.

For the sake of simplicity, we assume for both Tripod and our personalized policy that the entire population of the whole day is known in advance, as well as their alternatives. In this setting, we let Tripod calculate a TEE close to the best possible value, i.e., we set $\Tax$ (the inverse of the TEE) as in Theorem~\ref{theorem:taxation}. Observe that the possibility for Tripod to change the TEE from a time-slot to another might improve its efficiency with respect to what we observe here. On the other hand, the TEE calculation in Tripod is based on simulation-based prediction, which is imperfect by nature, and the TEE is never guaranteed to be close to the best one (which we are assuming). This would instead deteriorate the efficiency of Tripod with respect to what we observe here. We also set $\Thr = \SocUt{\Ind}{\DefAltDef}$, for any individual $\Ind$, and $\Budget=1798.59$ euros. 

The Tripod policy, like the personalized-incentive policy, is more adapted in cases where the regulator is endowed with a limited budget that he must use as efficiently as possible to increase social welfare.
In such cases, however, our personalized-incentive policy performs better than Tripod.
As explained in Remark~\ref{rem:tripod-inefficiency}, the reason is that we exploit the entire efficiency of the incentivized alternatives, thus getting the most additional social welfare out of every additional unit of incentive spent.
Tripod is instead limited to a fixed efficiency, generally smaller than the intrinsic efficiency of the incentivized alternatives. It is important to remark that, however, while our personalized policy needs exact information about individual utilities, Tripod does not, since it finds TEE empirically based on simulation-based prediction of its effects. In other words, Tripod needs a perfect simulation-based prediction instead of perfect information about individuals, which in many practical scenarios might more easily hold.

In this application, both policies reach the same social welfare but the Tripod policy require an incentive budget twice as large as the personalized-incentive policy.

However, it is important to remark that under our personalized incentive policy, two individuals providing the same social utility would receive a different incentive, based on their individual characteristics.
Although we ensure that all individuals keep their original individual utility, there is a risk that our policy may be perceived as discriminatory. In Tripod, instead, the incentive received by an individual only depends on the social utility she provides, which might be more easily accepted by citizens.

\subsection{Imperfect Information}
\label{sec:modal-choice-imperfect-information}

We show in this section the performance of our allocation policy when the regulator has imperfect information about individual utilities.
In this case, the allocation policy is computed as in Section~\ref{sec:imperfect-information}.
Using the values of the random variables $\RandomDef$ drawn previously, we can check whether individuals accept the incentives proposed to them.
The policy stops when the daily budget of \num{1800} euros is depleted.

\begin{table}
    \centering
    \caption{Comparison of the performance of the personalized-incentive policy for one day, with perfect and imperfect information.}%
    \label{tab:perfect-vs-imperfect}
    \begin{tabular}{lcc}
        \toprule
        & Perfect information (Sec.~\ref{sec:calculation-of-the-incentive-policy}) & Imperfect information \\
        \midrule
        Budget spent & \num{1798.59} euros & \num{1797.03} euros \\
        Incentives proposed & \num{3486} & \num{419}\\
        Incentives accepted & \num{3486} & \num{247}\\
        Acceptance rate & \SI{100}{\%} & \SI{59}{\%} \\
        \cotwo{} reduction & \num{17.9} tons & \num{3.8} tons \\
        \bottomrule
    \end{tabular}
\end{table}

Table~\ref{tab:perfect-vs-imperfect} compares the performance of our personalized-incentive policy under the perfect and imperfect information assumption.
Observe that, as expected, imperfect information decreases the efficacy of the policy.
Since the regulator does not exactly know the individual utilities, it may propose insufficient incentives, which are rejected by individuals (it happens \SI{41}{\%} of the times).
This results in a smaller reduction of \cotwo{} (\SI{21}{\%} compared with the perfect information case).
Note that less individuals are involved in the incentive program (only \SI{12}{\%} compared to the perfect information case) because incentive given to single individuals are on average larger, and thus the budget is depleted more quickly.

These results could be improved by learning from the responses of individual $\Ind$ to the incentives proposed earlier in order to compute the incentives that will be proposed to her for other alternatives.
For example, if the regulator observes that individual $\Ind$ refused the incentive to shift from car to walking, he learns information on the random term of the utility for car of individual $\Ind$.

Also, if it is not possible to propose incentives to individual $\Ind$ for different alternatives consecutively, the regulator could propose incentives for multiple alternatives simultaneously.

These extensions cannot be carried out with Algorithm~\ref{alg:greedy-max-soc-ut-curve}.
Future work could study the optimal personalized-incentive policy under imperfect information.

\section{Conclusion}\label{sec:conclusion}

% imperfect information
% multi-criteria problem
% multi-constraint problem
% strategic behaviour / unfairness (pollutants get more incentives)

This paper explores a new system of personalized incentives.
The agents face a discrete set of alternatives, and make independent discrete choices.
We consider situations where an individual utility for an alternative does not coincide with the social utility of this alternative.
Such situations call for regulation or State intervention.
The idea is to determine the optimal incentives to be provided to each individual to alter their choices in order to better align individual benefits and the Principal benefits (note that the Principal can be any regulator).
The regulator is assumed to have a fixed budget for the incentives.
Even if individuals make independent choices, the computation of the incentives to be provided has to consider all individuals’ preferences, so the problem is combinatorial.
We provided in this paper an algorithm to optimally distribute individual incentives given a budget constraint in order to maximize the social utility or the social welfare function.

In 2021, this incentive system may be somewhat in advance.
Nowadays, individual information is gathered via GPS, social networks and the Internet of things.
This is precious information, which can potentially be used to optimally compute the optimal set of incentives, and thus to better manage Society.
(Privacy issues are ignored here, which does not mean they are not important.)

Besides, humans remain unpredictable.
There is still (and hopefully for some time) some margin of freedom as far as to what people decide.
The recent pandemic shows that individuals or governments remain unpredictable \citep{Zhang2020} and that the right set of incentives remains hard to determine.
As a consequence, individual choices are described by the modeller as being probabilistic.
Incentives thus change choices up to some probability distribution.
%The specification of the nature of randomness (measurement / specification errors or idiosyncratic factors) added to the basic deterministic model, i.e., the full information model will trigger the specification of the Social Welfare Function to be optimized (the log sum formula when errors are purely idiosyncratic and choices are described by Logit probabilities).
%We have provided here some heuristic treatment, which can be seen as a first-pass analysis.
While we have just tackled imperfect information in the empirical application, the treatment of imperfect information appears to be tractable.
Preliminary computations, with the Logit, the workhorse of discrete choice models, suggest that such an extension is promising, including analytically.
Contrarily to the full information case, mainly envisaged in this paper, some incentives may be too large for some individuals (who could select the same choice with a smaller incentive), and this incentive is then inefficient; other incentives may be too small to modify individual choice as expected, and in such a case the incentive is ineffective.
The optimal solution makes a comprise between these two sources of imperfection.

In the empirical application, we have ignored congestion.
In our defence, let's recall that few commuters receive an incentive, which is a quality of our method.
In practice, congestion means that the utility of some individuals can change as other individuals are shifting, which renders the incentive amounts computed \emph{ex-ante} imprecise.
We have not solved the current problem with congestion because it is likely to be difficult.
But it is not impossible.
%One may slightly change to proposed algorithm by adding the social externality (difference between social and individual costs) when an incentive is provided to a commuter.
%This is not optimal because the externality is computed locally (for the current level of congestion), but such procedure will certainly improve the solution with respect to the current approach, which ignores congestion.
An iterative procedure alternating the incentive algorithm and the computation of the current level of congestion is promising.
Congestion can be treated as a static or dynamic (time of the day dependent) process.
Much work remains to be done along this line.

Finally, we have considered so far static choice, i.e., at a given point in historical time.
If we consider mode choice, it may be the case that incentive for public transport, for example, will have on the long run an impact on automobile ownership.
Moreover, in the medium run, a car left at home can be used by other family members for short trip.
Without any intervention, the trend could yield more trips and vehicle cold starts particularly on local roads, especially in places where vehicles continue to rely on internal combustion engines.
%Moreover, incentives to use electric vehicles may induce commuters to abandon public transport and to buy an electric car.
These examples show the need to also consider the medium and long-run impacts of incentives, by appending a predictive model to the incentive algorithm.
There are plenty of roads left to run.

\section*{Acknowledgments}

We are grateful to Moshe Ben-Akiva for his constructive and valuable feedback, which helped us to more deeply understand the limits and the implications of the personalized incentive policy proposed in this paper, in particular in a dynamic context and in terms of equity. His comments allowed us to better present and frame our approach within the state-of-the-art.
We thank Matthieu de Palma, Nathalie Picard and Ravi Seshadri for their helpful comments at various stages of this research.

\section*{Disclosure statement}

The authors report there are no competing interests to declare.

\section*{Funding}

This work was supported by NSFC-JPI UE under grant “MAAT” (project no. 18356856); ANR under grant ANR-11-LBX-0023-01; ANR under grant ANR-16-IDEX-0008.

\bibliographystyle{tfcad}
\bibliography{bibliography}

\newpage

\appendix

\section{Concavization}
\label{sec:concavization}
The process of concavization \citep[Figures~1 and~2]{Zoltners1979} consists in removing from the set of alternatives of any individual $\Ind$ some alternatives that we consider `irrelevant', as introduced in Section~\ref{sec:preliminary-steps}.

We introduce the concepts of dominance and LP-dominance and other definitions from \citet[Section 11.2]{Kellerer2004}.

\begin{definition}[Dominance]
    \label{def:dominance}
    Given an individual $\Ind$ and two of her alternatives $\Alt,\Alt'$, we say that $\Alt$ \emph{dominates} $\Alt'$ if it has a higher social indicator and requires less incentives to be adopted, i.e., $\SocUtDef \ge \SocUt{\Ind}{\Alt'}$ and $\WeightDef \le \Weight{\Ind}{\Alt'}$.
\end{definition}
Note that, from equation \eqref{eq:incentive_amount}, the condition $\WeightDef \le \Weight{\Ind}{\Alt'}$ is equivalent to $\IndUtDef \geq \IndUt{\Ind}{\Alt'}$ and thus the concept of dominance is equivalent to the concept of Pareto-dominance of Definition~\ref{def:dominance}.
Thanks to Assumption~\ref{ass:no-pareto}, we can assume they have been eliminated from our problem.

\begin{definition}[LP-dominance]
    Consider three alternatives $\Alt, \Alt',\Alt''$, such that $\SocUt{\Ind}{\Alt} < \SocUt{\Ind}{\Alt'} < \SocUt{\Ind}{\Alt''}$ and $\Weight{\Ind}{\Alt} < \Weight{\Ind}{\Alt'} < \Weight{\Ind}{\Alt''}$.
    We say that $\Alt'$ is \emph{LP-dominated} by $\Alt$ and $\Alt''$ if
    \[
        \frac{\SocUt{\Ind}{\Alt''}-\SocUt{\Ind}{\Alt'} }{\Weight{\Ind}{\Alt''}-\Weight{\Ind}{\Alt'} } \ge
        \frac{\SocUt{\Ind}{\Alt'}-\SocUt{\Ind}{\Alt} }{\Weight{\Ind}{\Alt'}-\Weight{\Ind}{\Alt} }.
    \]
\end{definition}

We denote with $\NonDomSetDef$ the set of alternatives of individual $\Ind$ that are neither dominated nor LP-dominated and $\NumOfNonDomDef$ its cardinality.
We call such alternatives \emph{LP-extremes}.
Note that this corresponds to the upper convex hull of $\InitSetDef$, as in \citet[Figure 11.1]{Kellerer2004}.

\newpage

\section{Proofs}
\label{sec:proofs}

\fakesubsec{Proofs of Section~\ref{sec:policy} }

\begin{proof}[Proof of Proposition~\ref{propo:incentive_amount}.]
Given any policy $\IncPolicy$, individual $\Ind$ chooses alternative $\Alt\in\InitSetDef$, with $\SocUtDef > \SocUt{\Ind}{\DefAltDef}$, if
%If the regulator wants to induce individual $\Ind$ to switch to alternative $\Alt\in\InitSetDef$, he must must propose her incentives $\{\IncDef\}_{\Alt\in\InitSetDef}$ such that her participation constraints are satisfied.
\begin{equation}
    \label{eq:particip_1}
    \IndUtDef + \IncDef \geq \IndUt{\Ind}{\Altt} + \Inc{\Ind}{\Altt},
    \quad\forall \Altt\in\InitSetDef,
\end{equation}
and
\begin{equation}
    \label{eq:particip_2}
    \IndUtDef + \IncDef > \IndUt{\Ind}{\Altt} + \Inc{\Ind}{\Altt}, 
    \quad\forall \Altt\in\InitSetDef\setminus\{\Alt\}: \SocUtDef \leq \SocUt{\Ind}{\Altt}.
\end{equation}
Indeed, equations~\eqref{eq:particip_1} and \eqref{eq:particip_2} ensure that~\eqref{eq:behavior} is satisfied.

Let $\Ind\in\IndSet$ and $\Alt\in\InitSetDef$, and consider a personalized-incentive policy $\IncPolicy$ such that $\Inc{\Ind}{\Altt}=0$, for any $\Altt\neq\Alt$ and $\IncDef = \IndUt{\Ind}{\DefAltDef} - \IndUtDef$.
Rewriting \eqref{eq:particip_1} and \eqref{eq:particip_2}, we can claim that individual $\Ind$ chooses alternative $\Alt$, if
\begin{equation}
    \label{eq:particip_3}
    \IndUt{\Ind}{\DefAltDef} \geq \IndUt{\Ind}{\Altt},
    \quad\forall \Altt\in\InitSetDef,
\end{equation}
and
\begin{equation}
    \label{eq:particip_4}
    \IndUt{\Ind}{\DefAltDef} > \IndUt{\Ind}{\Altt},
    \quad\forall \Altt\in\InitSetDef\setminus\{\Alt\}: \SocUtDef \leq \SocUt{\Ind}{\Altt}.
\end{equation}
Thanks to equation \eqref{eq:default_alt}, the personalized-incentive policy $\IncPolicy$ satisfies equation \eqref{eq:particip_3}.
It remains to prove that it always satisfies also equation~\eqref{eq:particip_4}.
Suppose by contradiction that there exists an alternative $\Altt\in\InitSetDef\setminus\{\Alt\}$ such that $\SocUtDef \leq \SocUt{\Ind}{\Altt}$, which does not satisfy equation~\eqref{eq:particip_4}.
Then we would have $\IndUt{\Ind}{\DefAltDef} \le \IndUt{\Ind}{\Altt}$.
By  construction $\SocUt{\Ind}{\Altt} \ge \SocUtDef > \SocUt{\Ind}{\DefAltDef}$.
This would contradict the definition of default alternative (equation~\eqref{eq:default_alt}).

At this point of the proof, we have demonstrated the first part of the Proposition, i.e., that, considering an option $\Alt$ such that $\SocUtDef > \SocUt{\Ind}{\DefAltDef}$, a personalized-incentive policy $\IncPolicy$  such that $\Inc{\Ind}{\Altt}=0$, for any $\Altt\neq\Alt$ and $\IncDef = \IndUt{\Ind}{\DefAltDef} - \IndUtDef$ successfully induces individual $\Ind$ to choose alternative $\Alt$.
We now prove the second part of the Proposition.

Observe that, if $\IncDef < \IndUt{\Ind}{\DefAltDef} - \IndUtDef$, then $\IndUtAfterFunc{\Ind}{\Alt}{\IncPolicy} = \IndUtDef + \IncDef < \IndUt{\Ind}{\DefAltDef}$ and individual $\Ind$ would never prefer $\Alt$ to $\DefAltDef$.
\end{proof}

\begin{proof}[Proof of Proposition~\ref{prop:pareto}.]

Consider an individual $\Ind$ and an alternative $\Alt$, Pareto-dominated by another alternative $\Altt$.
Suppose that the policy $\IncPolicy$ is such that $\Ind$ is induced to choose $\Alt$.
According to Assumption~\ref{ass:types-of-policies}, the incentive is $\IncDef = \IndUt{\Ind}{\DefAltDef} - \IndUtDef$ and the individual shifts from her default alternative $\DefAltDef$ to $\Alt$, increasing the social welfare by $\delta= \SocUtDef - \SocUt{\Ind}{\DefAltDef}$.

We can then construct a policy $\IncPolicy'$, which is identical to $\IncPolicy$, except for the incentive proposed to individual $\Ind$: she is incentivized to shift from her default alternative to $\Altt$, with an incentive $\Inc{\Ind}{\Altt}' = \IndUt{\Ind}{\DefAltDef} - \IndUt{\Ind}{\Altt}$.
The increase of social welfare is in this case $\delta'=\SocUt{\Ind}{\Altt} - \SocUt{\Ind}{\DefAltDef}$.

By the definition of Pareto-dominance, $\Inc{\Ind}{\Altt}' < \IncDef$ and $\delta'\ge\delta$.
Therefore, policy $\IncPolicy'$ obtains at least the same increase in social welfare than $\IncPolicy$ but spending less incentive budget. Therefore, it makes no sense to consider policy $\IncPolicy$.
\end{proof}

\fakesubsec{Proofs of Section~\ref{sec:algorithm}}

\begin{proof}[Proof of Theorem \ref{thm:bound}.]

If we run Algorithm~\ref{alg:greedy-max-soc-ut-curve} with budget $\Budget$, we practically make the same steps as the MCKP-Greedy algorithm \citep[equation (11.8) and Figure~11.2]{Kellerer2004}.
In line 5 of the aforementioned algorithm, the authors compute an upper bound to the solution of the Multiple Choice Knapsack Problem~\eqref{eq:max-soc-ut} as 
\[
    \textit{ub}= 
    \bar \SocWelfareSym\OfIterDef + \IncrSocUt{\SplitInd}{\SplitAlt} \cdot (\Budget - \TotInc\OfIterDef) / \IncrWeight{\SplitInd}{\SplitAlt}.
\]
where $\IterSym$ is the last iteration of the algorithm.

%\rev{aa}{In \cite{Kellerer2004}, an upper bound to the utility of problem~\eqref{eq:max-soc-ut} is obtained via the MCKP-Greedy algorithm \cite{Kellerer2004}.}{-}

%\footnote{
%\rev{aa}{
%Observe that $\textit{ub}$ is just a theoretical upper bound, and cannot be implemented in practice, as it assumes that an additional incentive $\Budget - \TotInc < \IncrWeight{\SplitInd}{\SplitAlt}$ given to individual $\SplitInd$, would generate an additional utility $\IncrSocUt{\SplitInd}{\SplitAlt} \cdot (\Budget - \TotInc) / \IncrWeight{\SplitInd}{\SplitAlt}$, which is in reality not true, as individual $\SplitInd$ would only jump to $\SplitAlt$ if the additional incentive is at least $IncrWeight{\SplitInd}{\SplitAlt}$.
%}{}
%}
Observing, by the definition of efficiency~\eqref{eq:incremental_efficiency}, that 
$\IncrEff{\SplitInd}{\SplitAlt}=\IncrSocUt{\SplitInd}{\SplitAlt} / \IncrWeight{\SplitInd}{\SplitAlt}$, we get $\textit{ub} - \bar \SocWelfareSym\OfIterDef = \IncrEff{\SplitInd}{\SplitAlt} \cdot (\Budget - \TotInc\OfIterDef)$.
By construction, the theoretical maximum social welfare $\MaxSocUtDef$ of problem~\eqref{eq:max-soc-ut} is less than or equal to the upper bound $ub$. Therefore:
\begin{equation*}
	\MaxSocUtDef-\bar \SocWelfareSym\OfIterDef \le 
	\textit{ub} - \bar \SocWelfareSym\OfIterDef = 
	\IncrEff{\SplitInd}{\SplitAlt} \cdot (\Budget - \TotInc\OfIterDef).
\end{equation*}
By construction, $\AlgSocUt{\Budget} = \bar\SocWelfareSym\OfIterDef$ and $\BudgetUsed = \TotInc\OfIterDef$, which gives the inequality~\eqref{eq:bound} that we want to prove.
Such inequality is illustrated in Figures~\ref{fig:distance_optimum} and \ref{fig:upper-bound}.
\begin{figure}[ht]
    \centering
    \includegraphics[width=0.35\textwidth]{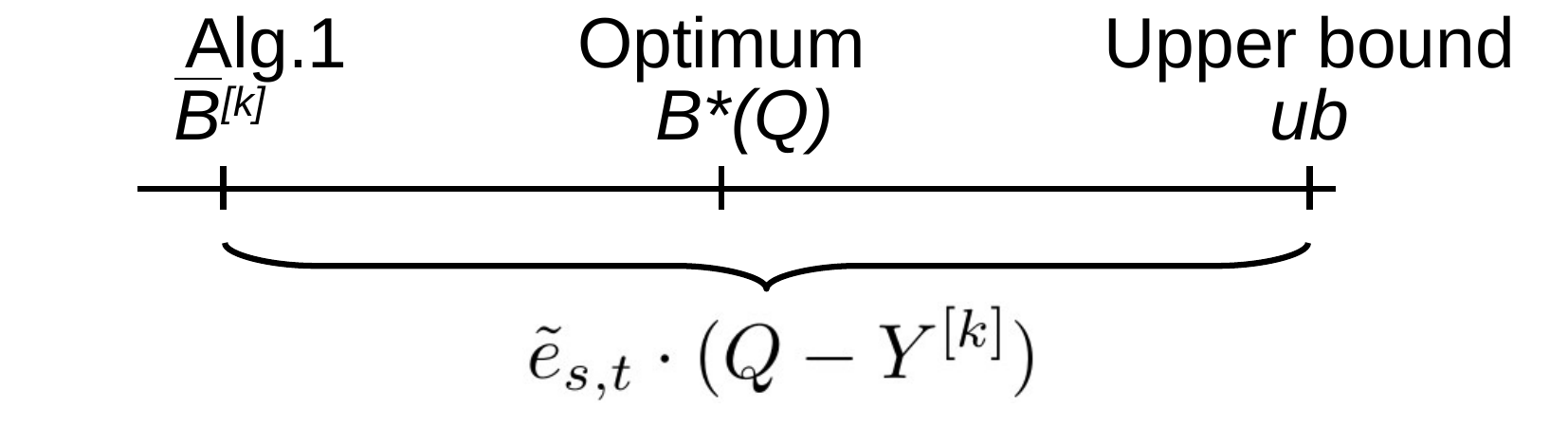}
    \caption{Illustration of the upper bound from Theorem~\ref{thm:bound}.}%
    \label{fig:upper-bound}
\end{figure}
\end{proof}

\begin{proof}[Proof of Proposition~\ref{prop:complexity}.]
To compute the ordered LP-extremes $\NonDomSetDef$ of the individual $\Ind$ we resort to the method of~\cite{Kirkpatrick1986} of complexity $O(\sum_{\Ind=1}^\NumOfInd \NumOfAltDef \cdot \log \NumOfConvHullDef)$.
To obtain the set $\TotNonDomSet$, we just need to merge these ordered sets into an aggregated ordered set.
This operation has complexity $O(\TotNumOfConvHull\cdot \log \NumOfInd)$.
The rest of the operations consists in adding to the solution the alternatives in $\TotNonDomSet$, one by one, which has complexity $O(\TotNumOfConvHull)$.
\end{proof}

\begin{proof}[Proof of Proposition~\ref{prop:non-inc-eff}.]
By construction, Algorithm.~\ref{alg:greedy-max-soc-ut-curve} gets at each iteration the alternative with the highest incremental efficiency (Line~\ref{ln:incr-eff}).
This proves the first part of the claim.

The second part of the claim can be shown geometrically.
In the figure above, we represent the total incentive and social welfare calculated by the algorithm at each iteration.

\centering
\includegraphics[width=0.5\textwidth]{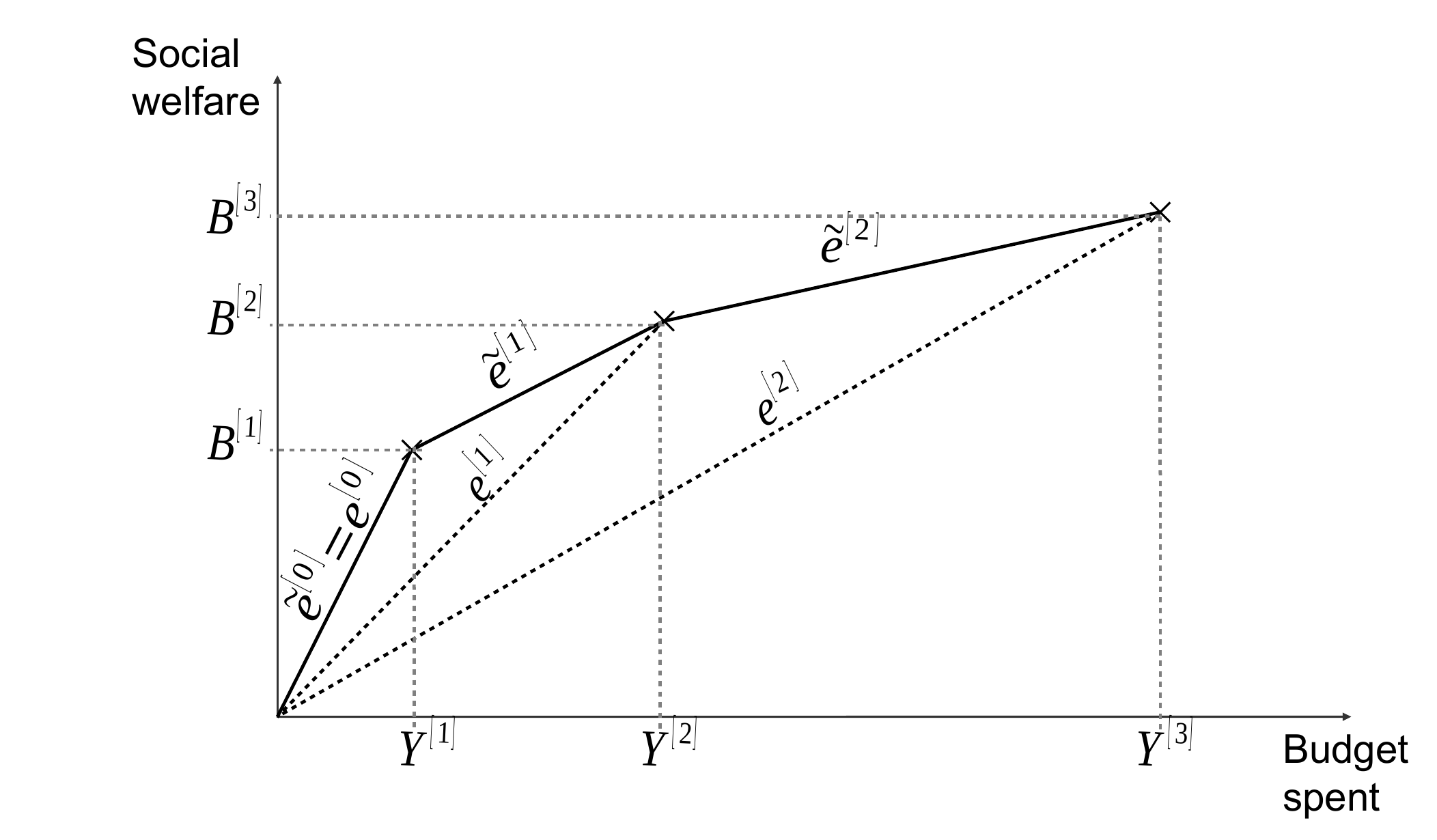}

Observe that the incremental efficiency $\IncrEffSym\OfIterDef$ is the inclination of the segment connecting $(\TotIncSym\OfIter{\IterSym-1}, \SocWelfareSym\OfIter{\IterSym-1})$ to $(\TotIncSym\OfIter{\IterSym}, \SocWelfareSym\OfIter{\IterSym})$ and that the efficiency $\EffSym\OfIterDef$ is the inclination of the segment connecting $(0,0)$ to $(\TotIncSym\OfIter{\IterSym},\SocWelfareSym\OfIter{\IterSym})$.
It becomes then evident that the monotonicity of $\IncrEffSym\OfIterDef$ implies also the monotonicity of $\EffSym\OfIterDef$.
\end{proof}

\begin{proof}[Proof of Corollary~\ref{cor:non-inc-eff}.]
Observe that at every iteration $\IterSym$ we increase the social welfare by $\IncrEffSym\OfIterDef\cdot (\TotIncSym\OfIter{\IterSym+1} - \TotIncSym\OfIterDef)$.
Therefore 
$\AlgSocUt{\Budget} = \AlgSocUt{\TotInc\OfIterDef} + \IncrEffSym\OfIterDef\cdot (\TotIncSym\OfIter{\IterSym+1} - \TotIncSym\OfIterDef) + \IncrEffSym\OfIter{\IterSym+1}\cdot (\TotIncSym\OfIter{\IterSym+2} - \TotIncSym\OfIter{\IterSym+1})+\dots$.
Observing that $\IncrEffSym\OfIterDef$ is non increasing, we get:
$\AlgSocUt{\Budget}
\le
\AlgSocUt{\TotInc\OfIterDef} +
\IncrEffSym\OfIterDef
\cdot  
\left[ 
	(\TotIncSym\OfIter{\IterSym+1} - \TotIncSym\OfIterDef) +  (\TotIncSym\OfIter{\IterSym+2} - \TotIncSym\OfIter{\IterSym+1})+\dots 
\right] = 
\AlgSocUt{\TotInc\OfIterDef} +
\IncrEffSym\OfIterDef
\cdot  
(\Budget - \TotIncSym\OfIterDef)$
\end{proof}

\fakesubsec{Proofs of Section~\ref{sec:policies_comparison} }

\begin{proof}[Proof of the Proposition~\ref{prop:non-negative-disutility}.]
From equations \eqref{eq:utility_def}, \eqref{eq:expenses}, \eqref{eq:utility_loss} and \eqref{eq:disutility}, observe that
    \begin{equation*}
        \DisutDef
        = \sum_{\Ind=1}^{\NumOfInd} \left(
            \IndUt{\Ind}{\DefAltDef}
            - \IndUt{\Ind}{\AltEqDef}
        \right)
        \geq0,
    \end{equation*}
\end{proof}

\begin{proof}[Proof of Proposition~\ref{prop:equivalence}]
Thanks to Proposition~\ref{prop:non-negative-disutility}, $\Disut{\Policy'} = \DisutDef \le \Budget$.
Moreover, the social welfare is also the same, i.e., $\AlgSocUt{\Policy} = \AlgSocUt{\Policy'}$, since it only depends on the alternative chosen.
This shows the proposition.
\end{proof}

\begin{proof}[Proof of Proposition~\ref{prop:optimal-incentive-general-policy}]
Assume, by contradiction, that the optimal personalized-incentive policy $\IncPolicy$ is not an optimal general policy.
This would imply the existence of a policy $\Policy$ such that $\SocWelfare{\Policy} > \SocWelfare{\IncPolicy}$ and $\Disut{\Policy} \leq \Budget$.

Consider now a personalized-incentive policy $\IncPolicy'$ such that
\begin{equation*}
    \left\{
        \begin{array}{ll}
            \IncDef' = \IndUt{\Ind}{\DefAlt{\Ind}} - \IndUt{\Ind}{\Alt}, & \text{if}\:\Alt = \AltEq{\Ind}{\Policy} \\
            \IncDef' = 0, & \text{otherwise} \\
        \end{array}
    \right..
\end{equation*}
Then, by construction, $\IncPolicy'$ is such that $\AltEq{\Ind}{\IncPolicy'} = \AltEq{\Ind}{\Policy}$, $\forall \Ind\in\IndSet$.

Moreover, observe that $\IncPolicy'$ is such that $\SocWelfare{\IncPolicy'} > \SocWelfare{\IncPolicy}$ and
$
    \Exp{\IncPolicy'} = \Disut{\IncPolicy'} \leq \Budget
$.
Therefore, $\IncPolicy'$ would be a better personalized-incentive policy than $\IncPolicy$, which is absurd, since by construction $\IncPolicy$ is an optimal personalized-incentive policy.
\end{proof}

\begin{proof}[Proof of Corollary~\ref{propo:bound-general}]

Let $\AlgSocUt{\Budget}$ be the social welfare returned by Algorithm~\ref{alg:greedy-max-soc-ut-curve} for a budget $\Budget$.
We know from Theorem~\ref{thm:bound} that it is boundedly close to the social welfare $\MaxSocUtDef$ obtained with an optimal personalized-incentive policy, with the following bound
\begin{align*}
    \MaxSocUtDef - \AlgSocUt{\Budget}
    \le \IncrEff{\SplitInd}{\SplitAlt} \cdot (\Budget-\BudgetUsed).
\end{align*}

Thanks to Proposition~\ref{prop:optimal-incentive-general-policy}, $\MaxSocUtDef$ is also the social welfare obtained via an optimal general policy with disutility threshold $\Budget$. This proves the Corollary.
\end{proof}

%\begin{proof}[Proof of Proposition~\ref{prop:pigouvian}.]

%The utility $\IndUtAfterFunc{\Ind}{\Alt}{\TaxPolicy}$ of individual $\Ind$ when choosing alternative $\Alt\in\InitSetDef$ can be written as $\IndUtAfterFunc{\Ind}{\Alt}{\TaxPolicy} = \IndUtDef + \SocUtDef$, by applying~\eqref{eq:tax_def} and~\eqref{eq:IndUtAfterDef}. Therefore, the utility maximization by the individuals coincide with the total surplus maximization.

%\end{proof}

\begin{proof}[Proof of Proposition~\ref{prop:bound-enforcement}.]
By construction, the enforcement policy $\Policy$ induce the same individual alternatives as the personalized-incentive policy $\IncPolicy$.
Then, thanks to Proposition~\ref{prop:non-negative-disutility}, they have the same disutility $\Disut{\Policy}=\Disut{\IncPolicy}\le\Budget$ and achieve the same social welfare $\AlgSocUt{\Policy}=\AlgSocUt{\IncPolicy}$.
Thanks to Corollary~\ref{propo:bound-general}, $\AlgSocUt{\IncPolicy}$ is boundedly close to the optimum $\AlgSocUt{\Budget}$, and so is $\AlgSocUt{\Policy}$.
\end{proof}

\begin{proof}[Proof of Theorem \ref{theorem:taxation}.]
Let $\Policy$ be a policy such that $\TrDef = \Tax(\SocUtDef - \Thr)$, with $\Tax$ given by equation \eqref{eq:tax_opt} and $\Thr\in\mathbb{R}$.

Let $\IncPolicy$ be the personalized-incentive policy obtained running the algorithm as explained in the statement of this theorem.
Thanks to Corollary~\ref{propo:bound-general}, we know that $\AlgSocUt{\IncPolicy}$ is such that 
\begin{align}
\label{eq:bound-to-exploit}
\MaxSocUtDef-\AlgSocUt{\IncPolicy} \le \IncrEff{\SplitInd}{\SplitAlt}\cdot (\Budget-\BudgetUsed).
\end{align}
If we prove that $\AltEq{\Ind}{\IncPolicy} = \AltEq{\Ind}{\Policy}$, $\forall \Ind\in\IndSet$, we could claim that $\AlgSocUt{\Policy}=\AlgSocUt{\IncPolicy}$ and also, thanks to Proposition~\ref{prop:non-negative-disutility}, that $\Disut{\Policy}=\Disut{\IncPolicy}\le\Budget$.
In this case, the bound~\eqref{eq:bound-to-exploit} would also hold for $\Policy$.

%For what said, to prove the theorem, we need to show that
%\begin{equation}
    %\label{eq:proof_goal}
    %\AltEqDef = \AltEq{\Ind}{\TaxPolicy}, \forall \Ind\in\IndSet
%\end{equation}
%where $\AltEqDef$ is the alternative induced to individual~$\Ind$ by the incentive policy $\IncPolicy$ returned by Algorithm~\ref{alg:greedy-max-soc-ut-curve}, i.e., $\Dec{\Ind}{\AltEqDef} = 1$ and $\DecDef = 0$, for any $\Alt\in\InitSetDef\setminus\{\AltEqDef\}$, and where
%\begin{equation}
    %\label{eq:goal_def}
    %\AltEq{\Ind}{\TaxPolicy} = \argmax_{k\in\argmax_j \IndUtAfterFunc{\Ind}{\Alt}{\TaxPolicy}} \SocUt{\Ind}{k}.
%\end{equation}

To do so, we show that (i) alternative $\AltEq{\Ind}{\Policy}$ is in the set $\NonDomSetDef$ of the LP-extremes alternatives and (ii) alternative $\AltEq{\Ind}{\IncPolicy}$ maximizes $\IndUtAfterFunc{\Ind}{\Alt}{\Policy} = \IndUtDef + \Tax(\SocUtDef - \Thr)$, over all alternatives $\Alt\in\NonDomSetDef$.

\paragraph{Proof of (i)}

Assume, by contradiction, that $\AltEq{\Ind}{\Policy}$ is LP dominated by alternatives $\Alt$ and $\Alt'$, i.e., $\SocUt{\Ind}{\Alt} < \SocUt{\Ind}{\AltEq{\Ind}{\Policy}} < \SocUt{\Ind}{\Alt'}$ and $\Inc{\Ind}{\Alt} < \Inc{\Ind}{\AltEq{\Ind}{\Policy}} < \Inc{\Ind}{\Alt'}$, and
\begin{equation*}
    \frac{\SocUt{\Ind}{\Alt'} - \SocUt{\Ind}{\AltEq{\Ind}{\Policy}}}{\Weight{\Ind}{\Alt'} - \Weight{\Ind}{\AltEq{\Ind}{\Policy}}} \geq \frac{\SocUt{\Ind}{\AltEq{\Ind}{\Policy}} - \SocUt{\Ind}{\Alt}}{\Weight{\Ind}{\AltEq{\Ind}{\Policy}} - \Weight{\Ind}{\Alt}}.
\end{equation*}
From equation \eqref{eq:weight_def}, $\Weight{\Ind}{\Alt} = \IndUt{\Ind}{\DefAltDef} - \IndUt{\Ind}{\Alt}$ and thus the previous condition can be written as
\begin{equation*}
    \frac{\SocUt{\Ind}{\Alt'} - \SocUt{\Ind}{\AltEq{\Ind}{\Policy}}}{\IndUt{\Ind}{\AltEq{\Ind}{\Policy}} - \IndUt{\Ind}{\Alt'}} \geq \frac{\SocUt{\Ind}{\AltEq{\Ind}{\Policy}} - \SocUt{\Ind}{\Alt}}{\IndUt{\Ind}{\Alt} - \IndUt{\Ind}{\AltEq{\Ind}{\Policy}}}.
\end{equation*}
Multiplying by $\Tax>0$ on both sides and adding and subtracting $\Thr$ yields
\begin{equation*}
    \Tax\frac{\SocUt{\Ind}{\Alt'} - \SocUt{\Ind}{\AltEq{\Ind}{\Policy}}-A+A}{\IndUt{\Ind}{\AltEq{\Ind}{\Policy}} - \IndUt{\Ind}{\Alt'}} \geq \Tax\frac{\SocUt{\Ind}{\AltEq{\Ind}{\Policy}} - \SocUt{\Ind}{\Alt}-A+A}{\IndUt{\Ind}{\Alt} - \IndUt{\Ind}{\AltEq{\Ind}{\Policy}}}.
\end{equation*}
Rearranging the terms and using equation \eqref{eq:tax_def} yields
\begin{equation*}
    \frac{\Tr{\Ind}{\Alt'}-\Tr{\Ind}{\AltEq{\Ind}{\Policy}}}{\IndUt{\Ind}{\AltEq{\Ind}{\Policy}} - \IndUt{\Ind}{\Alt'}} \geq \frac{\Tr{\Ind}{\AltEq{\Ind}{\Policy}}-\TrDef}{\IndUt{\Ind}{\Alt} - \IndUt{\Ind}{\AltEq{\Ind}{\Policy}}}.
\end{equation*}
%\[ \frac{\Tax(A-\SocUt{\Ind}{\AltEqDef}) - \Tax(A-\SocUt{\Ind}{\Alt'})}{\IndUt{\Ind}{\AltEqDef} - \IndUt{\Ind}{\Alt'}} -1 +1 \geq \frac{\Tax(A-\SocUt{\Ind}{\Alt}) - \Tax(A-\SocUt{\Ind}{\AltEqDeflt}})}{\IndUt{\Ind}{\Alt} - \IndUt{\Ind}{\AltEqDef}} -1 +1. \] 
Finally, using equation \eqref{eq:utility_def}, we get
\begin{equation*}
    \frac{\IndUtAfterFunc{\Ind}{\Alt'}{\Policy} - \IndUtAfterFunc{\Ind}{\AltEq{\Ind}{\Policy}}{\Policy}}{\IndUt{\Ind}{\AltEq{\Ind}{\Policy}} - \IndUt{\Ind}{\Alt'}} \geq \frac{\IndUtAfterFunc{\Ind}{\AltEq{\Ind}{\Policy}}{\Policy} - \IndUtAfterFunc{\Ind}{\Alt}{\Policy}}{\IndUt{\Ind}{\Alt} - \IndUt{\Ind}{\AltEq{\Ind}{\Policy}}}.
\end{equation*}
Let $\alpha = \IndUt{\Ind}{\Alt} - \IndUt{\Ind}{\AltEq{\Ind}{\Policy}}$ and $\alpha' = \IndUt{\Ind}{\AltEq{\Ind}{\Policy}} - \IndUt{\Ind}{\Alt'}$.
From $\Inc{\Ind}{\Alt} < \Inc{\Ind}{\AltEq{\Ind}{\Policy}} < \Inc{\Ind}{\Alt'}$, it follows that $\IndUtDef > \IndUt{\Ind}{\AltEq{\Ind}{\Policy}} > \IndUt{\Ind}{\Alt'}$ and thus $\alpha, \alpha' > 0$.
Then, we get
\begin{equation*}
    \frac{\IndUtAfterFunc{\Ind}{\Alt'}{\Policy} - \IndUtAfterFunc{\Ind}{\AltEq{\Ind}{\Policy}}{\Policy}}{\alpha'} \geq \frac{\IndUtAfterFunc{\Ind}{\AltEq{\Ind}{\Policy}}{\Policy} - \IndUtAfterFunc{\Ind}{\Alt}{\Policy}}{\alpha}.
\end{equation*}
%\[ \alpha \IndUtAfter{\Ind}{\Alt'} - \alpha \IndUtAfter{\Ind}{\AltEqDef} \geq \alpha'\IndUtAfter{\Ind}{\AltEqDef} - \alpha'\IndUtAfter{\Ind}{\Alt}. \] 
%\[ \alpha \IndUtAfter{\Ind}{\Alt'} + \alpha'\IndUtAfter{\Ind}{\Alt}  \geq \alpha\IndUtAfter{\Ind}{\AltEqDef} + \alpha' \IndUtAfter{\Ind}{\AltEqDef}. \] 
By simple arithmetic calculation, one can see that this is equivalent to
\begin{equation}
    \label{eq:weighted_average}
    \frac{\alpha \IndUtAfterFunc{\Ind}{\Alt'}{\Policy} + \alpha'\IndUtAfterFunc{\Ind}{\Alt}{\Policy}}{\alpha+\alpha'}  \geq \IndUtAfterFunc{\Ind}{\AltEq{\Ind}{\Policy}}{\Policy}.
\end{equation}
Equation \eqref{eq:weighted_average} means that the utility of $\AltEq{\Ind}{\Policy}$ is less than or equal to the weighted average of the utility of $\Alt$ and $\Alt'$.
Two cases could then hold:
\begin{itemize}
    \item Either $\Alt$ or $\Alt'$ is preferred to $\AltEq{\Ind}{\Policy}$, i.e., $\IndUtAfterFunc{\Ind}{\Alt}{\Policy} > \IndUtAfterFunc{\Ind}{\AltEq{\Ind}{\Policy}}{\Policy}$ or $\IndUtAfterFunc{\Ind}{\Alt'}{\Policy} > \IndUtAfterFunc{\Ind}{\AltEq{\Ind}{\Policy}}{\Policy}$.
    This would mean that $\AltEq{\Ind}{\Policy}$ does not maximizes utility and would contradict equation \eqref{eq:behavior}.
    \item The three alternatives are equivalent, i.e., $\IndUtAfterFunc{\Ind}{\Alt}{\Policy} = \IndUtAfterFunc{\Ind}{\AltEq{\Ind}{\Policy}}{\Policy} = \IndUtAfterFunc{\Ind}{\Alt'}{\Policy}$.
        This would also contradict equation \eqref{eq:behavior} because $\SocUt{\Ind}{\Alt'} > \SocUt{\Ind}{\AltEq{\Ind}{\Policy}}$, by assumption.
\end{itemize}

Therefore, $\AltEq{\Ind}{\Policy}$ is not a LP-dominated alternative.
Clearly, $\AltEq{\Ind}{\Policy}$ is not dominated either and thus $\AltEq{\Ind}{\Policy}\in\NonDomSetDef$.

\paragraph{Proof of (ii)}
The proof of (ii) requires the following lemmas.

\underline{Lemma A}
%\begin{lemma}
%    \label{lemma:increff}
\textit{
    If the alternatives in $\NonDomSetDef$ are ordered according to equation \eqref{eq:ordering}, i.e., they are ordered by increasing weight, then
    \begin{equation*}
        \IncrEff{\Ind}{1} > \IncrEff{\Ind}{2} > \dots > \IncrEff{\Ind}{\NumOfNonDomDef}.
    \end{equation*}
}

%\end{lemma}

To prove this lemma, show by contradiction that if $\IncrEffDef\leq \IncrEff{\Ind}{\Alt+1}$, then $\Alt$ would be LP-dominated by $\Alt-1$ and $\Alt+1$.

\underline{Lemma B}
    %\label{lemma:inequality}
    \textit{ If $\Alt\in\NonDomSetDef$ is such that $\IncrEffDef \geq 1/\Tax$, then $\IndUtAfterFunc{\Ind}{\Alt-1}{\Policy}\leq\IndUtAfterFunc{\Ind}{\Alt}{\Policy}$, where $\Alt-1$ denotes the alternative which comes just before $\Alt$ in the ordered set $\NonDomSetDef$.
}

To prove this lemma, note that, using equations \eqref{eq:weight_def}, \eqref{eq:incremental-weight-and-social} and \eqref{eq:incremental_efficiency}, the inequality $\IncrEffDef \geq 1/\Tax$ can be written as
\begin{equation*}
    \frac{\SocUt{\Ind}{\Alt}-\SocUt{\Ind}{\Alt-1}}{\IndUt{\Ind}{\Alt-1}-\IndUt{\Ind}{\Alt}} \geq 1/\Tax.
\end{equation*}
Multiplying by $\Tax(\IndUt{\Ind}{\Alt}-\IndUt{\Ind}{\Alt-1})>0$, subtracting $\Tax \cdot \Thr$ from both sides and rearranging the terms, we get $\IndUt{\Ind}{\Alt-1} - \tau(A - \SocUt{\Ind}{\Alt-1}) \leq \IndUt{\Ind}{\Alt} - \tau(A - \SocUt{\Ind}{\Alt})$.
Using equations \eqref{eq:tax_def} and~\eqref{eq:utility_def} yields $\IndUtAfterFunc{\Ind}{\Alt-1}{\Policy} \leq \IndUtAfterFunc{\Ind}{\Alt}{\Policy}$.

\underline{Lemma C}
\textit{
    %\label{lemma:inequality2}
    If $\Alt\in\NonDomSetDef$ is such that $\IncrEffDef<1/\Tax$, then $\IndUtAfterFunc{\Ind}{\Alt-1}{\Policy}>\IndUtAfterFunc{\Ind}{\Alt}{\Policy}$, where $\Alt-1$ denotes the alternative which comes just before $\Alt$ in the ordered set $\NonDomSetDef$.
}

This lemma can be proved with the same reasoning as Lemma B.

Let $\Alt\in\NonDomSetDef$ be such that
\begin{equation}
    \label{eq:def_max_alt}
    \IncrEffDef \ge 1/\Tax > \IncrEff{\Ind}{\Alt+1}.
\end{equation}
Then, Lemmas A and B imply that the alternatives in the set $\{\Altt\in\NonDomSetDef:\:\Altt\leq\Alt\}$ are ordered by non-decreasing utility.
Similarly, Lemmas A and C imply that the alternatives in the set $\{\Altt\in\NonDomSetDef:\:\Altt\geq\Alt\}$ are ordered by decreasing utility.
Hence, alternative $\Alt$, defined by equation \eqref{eq:def_max_alt}, is the alternative which maximizes the utility $\IndUtAfterFunc{\Ind}{\Alt}{\Policy}$, over all alternatives in $\NonDomSetDef$.

Observe that, by construction of Algorithm~\ref{alg:greedy-max-soc-ut-curve}, the alternative $\AltEq{\Ind}{\IncPolicy}$ satisfies equation \eqref{eq:def_max_alt}.
Hence, alternative $\AltEq{\Ind}{\IncPolicy}$ maximizes the utility $\IndUtAfterFunc{\Ind}{\Alt}{\Policy}$, over all alternatives $\Alt\in\NonDomSetDef$.
As we have shown that $\AltEq{\Ind}{\Policy}\in\NonDomSetDef$, it must be that $\AltEq{\Ind}{\IncPolicy} = \AltEq{\Ind}{\Policy}$.
\end{proof}

\begin{proof}[Proof of Proposition~\ref{prop:proportional-inefficiency}.]
For any individual $\Ind\in\IndSet$, let us consider the alternative $\AltEq{\Ind}{\Policy}$ she chooses under policy $\Policy$ and compute the respective incentive
\begin{equation*}
    \Tr{\Ind}{\AltEq{\Ind}{\Policy}}
    = 
    \Tax \cdot  
    (\SocUt{\Ind}{\AltEq{\Ind}{\Policy}} 	- \Thr )
    \ge
    \Tax \cdot  
    (\SocUt{\Ind}{\AltEq{\Ind}{\Policy}} 	- \SocUt{\Ind}{\DefAltDef} )
    \ge \IndUt{\Ind}{\AltEq{\Ind}{0} } - \IndUt{\Ind}{\AltEq{\Ind}{\Policy}} .
\end{equation*}
where the first inequality is a consequence of the definition of proportional-incentive policy -- see~\eqref{eq:proportional-incentive}, while the last inequality ensures that the incentive compensates for the loss in individual utility when shifting to alternative $\AltEq{\Ind}{\Policy}$, which is a necessary condition for the individual to accept the incentive and shift to $\AltEq{\Ind}{\Policy}$.

Therefore, recalling from Definition~\ref{def:efficiency} that the efficiency of a generic alternative $\Alt$ as 
$
	\EffDef\defeq 
	\frac
	{\SocUtDef-\SocUt{\Ind}{\DefAltDef}}
	{ \IndUt{\Ind}{\AltEq{\Ind}{0} } - \IndUtDef }
$, we can write:
\begin{equation*}
    \frac{1}{\Tax}
    \le \Eff{\Ind}{\AltEq{\Ind}{\Policy}}, \ \ \forall \Ind\in\IndSet
    \quad\Longrightarrow\quad
    \frac{1}{\Tax}
    \le 
    \min_{\Ind\in\IndSet}
    \Eff{\Ind}{\AltEq{\Ind}{\Policy}}
\end{equation*}

Observe that the smaller $\Tax$, the smaller the incentive spent by $\Policy$. Therefore, it is always best to choose 
\[
	\Tax = \frac{1}
		{
			\min_{\Ind\in\IndSet}
			\Eff{\Ind}{\AltEq{\Ind}{\Policy}}
		}
\]

Let us now consider a policy $\IncPolicy$ that incentivizes the same individuals $\Ind\in\IndSet$.
In particular, it incentivizes the same alternative $\AltEq{\Ind}{\Policy}$, with a quantity $\Inc{\Ind}{\AltEq{\Ind}{\Policy}} = \IndUt{\Ind}{\AltEq{\Ind}{0} } - \IndUt{\Ind}{\AltEq{\Ind}{\Policy}} $. This incentive is sufficient to induce each individual to choose such an alternative. Therefore, the social welfare of this new policy $\IncPolicy$ will be the same as $\Policy$, i.e., $\AlgSocUt{\IncPolicy} = \AlgSocUt{\Policy}$. However, the saving of incentive distributed is:

\begin{align*}
    \TotInc(\Policy) - \TotInc(\IncPolicy) 
        &
        = \sum_{\Ind\in\IndSet} 
        (
        \Tr{\Ind}{\AltEq{\Ind}{\Policy}} 
        - \Inc{\Ind}{\AltEq{\Ind}{\Policy}}
        )
        = \sum_{\Ind\in\IndSet} (
        \SocUt{\Ind}{\AltEq{\Ind}{\Policy}}-\SocUt{\Ind}{\DefAltDef}
        )
        \cdot 
        \left(
            \Tax
            - \frac{1}{ \Eff{\Ind}{ \AltEq{\Ind}{\Policy} } }
        \right)		
        \\
        & 
        = 
        \frac{1}{\Tax}
        \cdot
        \sum_{\Ind\in\IndSet}
        \frac{
            \SocUt{\Ind}{\AltEq{\Ind}{\Policy}} - \SocUt{\Ind}{\DefAltDef}
        }
        {
            \Eff{\Ind}{\AltEq{\Ind}{\Policy} }
        }
        \cdot
        \left(
            \Eff{\Ind}{ \AltEq{\Ind}{\Policy} } - \frac{1}{\Tax} 
        \right)
        = 
        \frac{1}{\Tax}
        \cdot
        \sum_{\Ind\in\IndSet}
        \left( 
            \IndUt{\Ind}{\AltEq{\Ind}{\Policy}}-\IndUt{\Ind}{\DefAltDef} 
        \right)
        \cdot
        \Delta\Eff{\Ind}{ \AltEq{\Ind}{\Policy} } 
    \end{align*}

\end{proof}

\begin{proof}[Proof of Proposition~\ref{prop:proportional-incentive-vs-algo}]

Let us run Algorithm~\ref{alg:greedy-max-soc-ut-curve} with budget $\Budget=\TotInc(\Policy)$, which allows to get the values of $\IncrEffSym_{\TotInc(\Policy)}$ and $\gamma_{\TotInc(\Policy)}$. Thanks to Proposition~\ref{prop:proportional-inefficiency}, there always exists a personalized-incentive policy policy $\IncPolicy$ that achieves at least the same social welfare of $\Policy$:
\begin{align}
	\label{eq:by-vs-bt}
	\AlgSocUt{\IncPolicy} \ge \AlgSocUt{\Policy}
\end{align}
 while providing incentive savings of at lest $L(\Policy)$.
Let $\IterSym'$ be the first iteration of the algorithm in which $\TotInc\OfIter{\IterSym'}\ge \TotInc(\IncPolicy)$ and $\IterSym''$ the last iteration in which $\TotInc\OfIter{\IterSym''}\le \TotInc(\Policy)$. 

First, suppose $L(\Policy)\ge 2 \gamma_{\TotInc(\Policy)}$. In this case, observe that
\begin{align*}
	\TotInc\OfIter{\IterSym'} - \TotInc(\IncPolicy) 
	& \le \gamma_{\TotInc(\Policy)}
	\\
	\TotInc(\Policy) - \TotInc\OfIter{\IterSym''}
	& \le \gamma_{\TotInc(\Policy)}
	\\
	\TotInc(\Policy) - \TotInc(\IncPolicy)  & \ge L(\Policy).
\end{align*}
This is shown, for the sake of understanding, in the following figure
\begin{center}
    \includegraphics[width=0.5\textwidth]{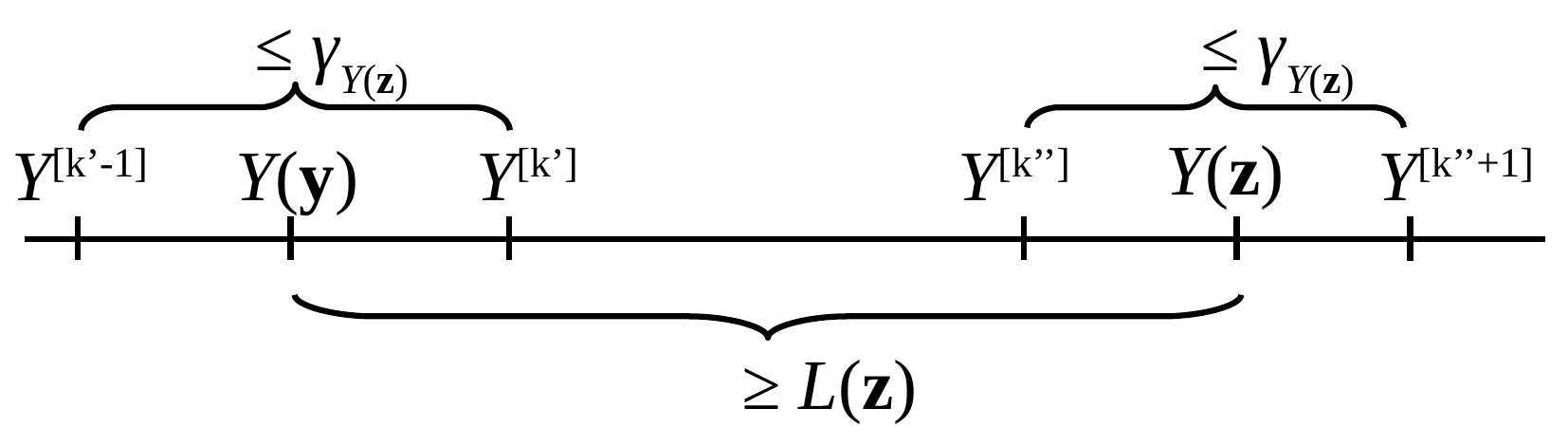}
\end{center}

Summing the first two of the inequalities above and then replacing $\TotInc(\Policy) - \TotInc(\IncPolicy)$ with the third inequality, we get
\[
	\TotInc\OfIter{\IterSym'} - \TotInc\OfIter{\IterSym''} + L(\Policy)
	\le
	\TotInc\OfIter{\IterSym'} - \TotInc(\IncPolicy)  
	+ \TotInc(\Policy) - \TotInc\OfIter{\IterSym''}
	\le 2 \gamma_{\TotInc(\Policy)}
\]
Rearranging the elements between the first and third terms, we get
\begin{align}
	\label{eq:ineq-L}
	\TotInc\OfIter{\IterSym''} - \TotInc\OfIter{\IterSym'} 
	\ge L(\Policy) - 2 \gamma_{\TotInc(\Policy)}.
\end{align}

Observe that:
\begin{align*}
	\SocWelfareSym\OfIter{\IterSym''} - \SocWelfareSym\OfIter{\IterSym'}
    &= \sum_{k=k'}^{k''-1} 
		( \SocWelfareSym\OfIter{\IterSym+1} - \SocWelfareSym\OfIter{\IterSym} ) \\
    &= \sum_{k=k'}^{k''-1} 
		\IncrEffSym\OfIterDef \cdot
			( \TotIncSym\OfIter{\IterSym+1} - \TotIncSym\OfIter{\IterSym} )
	\ge \IncrEffSym\OfIter{\IterSym''} \\
    &= \sum_{k=k'}^{k''-1} 
			( \TotIncSym\OfIter{\IterSym+1} - \TotIncSym\OfIter{\IterSym} ) \\
    &=	\IncrEffSym\OfIter{\IterSym''} \cdot 
		(\TotInc\OfIter{\IterSym''} - \TotInc\OfIter{\IterSym'} )
\end{align*}
where the inequality holds thanks to the monotonicity of incremental efficiencies (Proposition~\ref{prop:non-inc-eff}). Applying~\eqref{eq:ineq-L}:
\[
	\SocWelfareSym\OfIter{\IterSym''} - \SocWelfareSym\OfIter{\IterSym'}
	\ge \IncrEffSym\OfIter{\IterSym''} \cdot  
			(L(\Policy) -  2 \gamma_{\TotInc(\Policy)})
\]

By construction, $\IncrEffSym\OfIter{\IterSym''}=\IncrEffSym_{\TotInc(\Policy)}$. Moreover, $\MaxSocUt{\TotInc(\Policy)} \ge \MaxSocUt{\TotInc\OfIter{\IterSym''}} = \SocWelfareSym\OfIter{\IterSym''}$, where the inequality holds thanks to the monotonicity of the maximum social welfare curve and the equality holds thanks to Corollary~\ref{prop:touch} . We also know that $\SocWelfareSym\OfIter{\IterSym'} = \MaxSocUt{\TotInc\OfIter{\IterSym'}} \ge \MaxSocUt{\TotInc(\IncPolicy)}\ge \AlgSocUt{\IncPolicy}\ge \AlgSocUt{\Policy}$, where the first equality derives from Corollary~\ref{prop:touch}, the second inequality from the monotonicity of the maximum social welfare curve, the third inequality by the definition of optimal personalized-incentive policy and the last by~\eqref{eq:by-vs-bt}.

\end{proof}

\fakesubsec{Proofs of Section~\ref{sec:imperfect-information} }

\begin{proof}[Proof of Proposition~\ref{prop:gumbel}.]

Let $\xi = \Random{\Ind}{\DefAltDef} - \RandomDef$.
Then, $\xi$ is the difference of two i.i.d.~Gumbel-distributed random variables with scale parameter $\mu$, and thus it is a logistic-distributed random variable with scale parameter $\mu$ \citep{Nadarajah2005}.
Its probability density function is $f(x) = \frac{e^{x/\mu}}{\mu(e^{x/\mu}+1)^2}$ and its cumulative distribution function is $F(x) = \frac{e^{x/\mu}}{e^{x/\mu}+1}$.
For any $z\in\mathbb{R}$, we have
\begin{align*}
    \ExpSym(\xi|\xi>z) =& \frac{\int_{z}^{\infty} x f(x) dx}{1-F(z)} \\
    =& \frac{1}{1-e^{z/\mu}/(1+e^{z/\mu})} \int_{z}^{\infty} \frac{xe^{x/\mu}}{\mu(e^{x/\mu}+1)^2}dx \\
    =& \left(1 + e^{z/\mu}\right) \int_{z}^{\infty} \frac{xe^{x/\mu}}{\mu(e^{x/\mu}+1)^2}dx.
\end{align*}
Using
\begin{equation*}
    \frac{\partial}{\partial x} \left(- \frac{x}{e^{x/\mu}+1}\right) = \frac{xe^{x/\mu}}{\mu(e^{x/\mu}+1)^2} - \frac{1}{e^{x/\mu}+1},
\end{equation*}
and
\begin{equation*}
    \frac{\partial}{\partial x} \mu\ln(1+e^{-x/\mu}) = \frac{-e^{-x/\mu}}{1+e^{-x/\mu}} = \frac{-1}{e^{x/\mu}+1},
\end{equation*}
we get
\begin{align*}
    \int_{z}^{\infty} \frac{xe^{x/\mu}}{\mu(e^{x/\mu}+1)^2}dx &= \frac{z}{e^{z/\mu}+1} - \int_{z}^{\infty} \frac{-1}{e^{x/\mu}+1}dx. \\
                                                        &= \frac{z}{e^{z/\mu}+1} + \mu\ln(1+e^{-z/\mu}).
\end{align*}
Finally,
\begin{align*}
    \IncDef &= \IncDetDef + \ExpSym(\xi|\xi>-\IncDetDef) \\
            &= \IncDetDef - \IncDetDef + \mu(1+e^{-\IncDetDef/\mu})\ln(1+e^{\IncDetDef/\mu}) \\
            &= \mu\frac{1+e^{\IncDetDef/\mu}}{e^{\IncDetDef/\mu}}\ln(1+e^{\IncDetDef/\mu}).
\end{align*}
\end{proof}

\newpage

\section{Census Data Description}%
\label{sec:data_appendix}

We now describe the census data we use.%
\footnote{\url{https://www.insee.fr/fr/statistiques/4507890}}
They are published by INSEE and concern the period from 2015 to 2019.
The data contain observations for \num{7861201} households, representing \num{21810707} individuals (about a third of national population).
Only one individual is surveyed in each household, which means that, for example, the main mode of transportation is only observed for one individual in the household.
Hence, in each household, we consider only the surveyed individual.

We restrict our sample to workers living and working in the Rhône department, with a valid mode of transportation (i.e., unemployed and individuals working from home are excluded).
We remove some outliers, i.e., individuals travelling more than 90 minutes, which were about \num{2000}.
The final dataset contains \num{221571} individuals.
The total number of alternatives is \num{1092748}.

Note that census data do not represent an exhaustive sample of the population. Therefore, some categories of individuals might be over- or under-represented. To correct for such imbalances, INSEE computes a weight for each surveyed person.
To compute the statistics below and to perform the multinomial regression, we use these weights.

\paragraph{Home and Work Location.}

%The home and work location of the individuals is reported at the city-level, except for Paris, Lyon and Marseille where it is reported at the district-level.
The home and work location of the individuals is reported at the city-level, except for Lyon where it is reported at the district-level.
There are \num{275} unique home locations (an average of \num{812} individuals living at each location).
%The average area of home locations is \SI{11.18}{\square\kilo\meter}.

\paragraph{Mode of Transportation.}

The main mode of transportation used for commuting is in one of the following five categories: car, public transit, walking, cycling and motorcycle.
The share of each category are reported on Table \ref{tab:mode_share}.

\begin{table}[htbp]
    \centering
    \caption{Share of each mode of transportation reported.}%
    \label{tab:mode_share}
    \begin{tabular}{cc}
        \toprule
        Mode of transportation & Share \\
        \midrule
        Car & \SI{60.69}{\%} \\
        Public transit & \SI{25.07}{\%} \\
        Walking & \SI{8.83}{\%} \\
        Cycling & \SI{3.95}{\%} \\
        Motorcycle & \SI{1.47}{\%} \\
        \bottomrule
    \end{tabular}
    \\
    \footnotesize{Source: population census for Rhône department, INSEE.}
\end{table}

%\paragraph{Number of Cars Owned.}

%The data report the number of cars owned by the members of the household.
%This variable will be used to define the alternative set of the individuals (car is an alternative available to the individual only if there is at least one car in her household).
%In the dataset, the average number of cars owned is \num{1.57} and the share of households without a car is \SI{7.95}{\%}.

\paragraph{Socio-Demographic Variables.}

The data contain socio-demographic variables which are used to estimate a Multinomial Logit model for mode choice.
Table \ref{tab:numeric_variables} reports the list of numeric variables that we use, Table \ref{tab:categorical_variables} reports the list of categorical variables that we use.

\begin{table}[htbp]
    \centering
    \caption{Description of the numeric socio-demographic variables.}%
    \label{tab:numeric_variables}
    \small
    \begin{tabular}{ccc}
        \toprule
        Name & Description & Mean \\
        \midrule
        Age & Age of the individual, rounded to the nearest five-year age group & \num{38.49} \\
        Cars per individual & Number of cars owned divided by number of employed in the household & \num{0.84} \\
        \bottomrule
    \end{tabular}
    \\
    \footnotesize{Source: population census for Rhône department, INSEE.}
\end{table}

\begin{table}[htbp]
    \centering
    \caption{Description of the categorical socio-demographic variables.}%
    \label{tab:categorical_variables}
    \small
    \begin{tabular}{ccc}
        \toprule
        Name & Description & Most frequent category \\
        \midrule
        Sex & Sex of the individual & man (\SI{50.67}{\%}) \\
        Occupation & Occupation of the individual, using INSEE nomenclature & employee (\SI{24.93}{\%}) \\
        \bottomrule
    \end{tabular}
    \\
    \footnotesize{Source: population census for Rhône department, INSEE.}
\end{table}

\newpage

\section{Computation of Travel Times}%
\label{sec:travel_times}

For any individual, the origin point of her trips is set to the town hall of the city where she lives and the destination point is set to the town hall of the city where she works (for district-level home and workplace, the town hall of the district is used).
The coordinates of the town halls are retrieved from OpenStreetMap.

Travel time of each mode is set to the travel time of the fastest path, within that mode, which connects the two locations, computed from the open-source routing engine GraphHopper.
The road network for pedestrians, bicycles, motorcycles and cars is retrieved from OpenStreetMap data.
It is assumed that there is no congestion.
For public transit trips, the fastest path is computed using public transit timetables, retrieved from open-data GTFS files.
The departure time is assumed to be at 8 a.m.~on a weekday.

Note that, for some individuals, no path can be found to travel by public transit from their origin to their destination (\num{16161} individuals, representing 10.87\%  of total sample weight).
For these individuals, we exclude public transit from their choice set.

Some individuals are living and working in the same city (\num{61497} individuals, representing \SI{26.61}{\%} of total sample weight).
For these individuals, travel times are computed by supposing that trip distance is equal to the radius of the city (assuming cities are circular) and that speed is equal to the average speed of intercity trips.

%At this point, we discard all individuals with a commute time larger than 90 minutes, when using the mode of transportation they reported (\num{6656} individuals, representing \SI{3.28}{\%} of total sample weight).
%These observations probably represent individuals with two homes or two workplaces, or individuals telecommuting.

\begin{figure}
    \centering
    \includegraphics{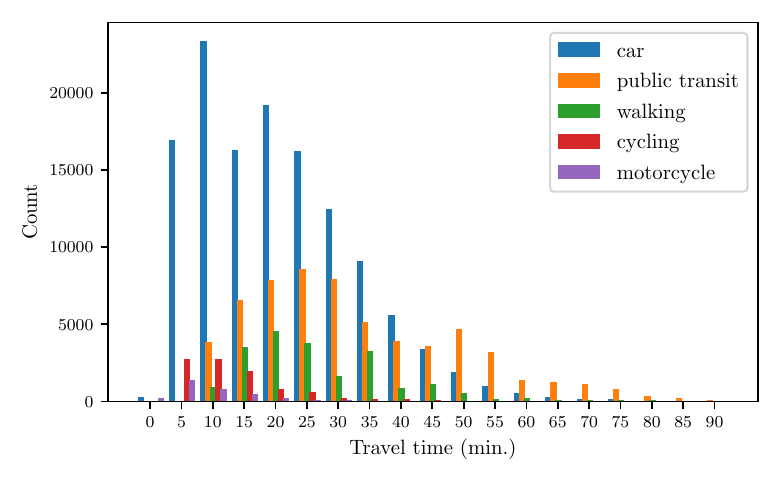}
    \caption{Distribution of travel times in the population (before the policy).}%
    \label{fig:tt_hist}
\end{figure}

Figure \ref{fig:tt_hist} shows the distribution of travel times in the population, for each mode of transportation.
Except for public transit trips, most trips last less than 30 minutes.

\newpage

\section{Simulating Utilities}%
\label{sec:gumbel}

In the Multinomial Logit model, the utility of individual $\Ind$ with mode of transportation $\Alt$ is
\begin{equation*}
    \IndUtDef = \IndUtDetDef + \RandomDef,
\end{equation*}
where $\IndUtDetDef$ is the deterministic part of the utility, which depends on the individual- and alternative-specific exogenous variables, and $\RandomDef$ is a random variable with standard Gumbel distribution.

The deterministic part $\IndUtDetDef$ are computed from the estimates of the Multinomial Logit model.
From the data, we know the alternative $\DefAltDef$ chosen by any individual $\Ind$ so we must have
\begin{equation}
    \label{eq:gumbel_constraint}
    \IndUt{\Ind}{\DefAltDef} > \IndUtDef, \quad \forall \Alt \neq \DefAltDef.
\end{equation}
To simulate draws of standard Gumbel variables conditional on equation \eqref{eq:gumbel_constraint}, we use the rejection sampling method, i.e., we draw values from the standard Gumbel distribution until the constraint of equation \eqref{eq:gumbel_constraint} is satisfied.

\end{document}